\documentstyle [fleqn,11pt] {report}
\newcommand{\FF}{{\cal F}}
\newcommand{\cd}{\cdot}

\newcommand{\al}{\alpha}
\newcommand{\ts}{\textstyle}
\renewcommand{\b}{\beta}
\newcommand{\de}{\delta}
\newcommand{\De}{\Delta}
\newcommand{\ep}{\epsilon}
\newcommand{\ga}{\gamma}
\newcommand{\Ga}{\Gamma}

\newcommand{\io}{\iota}
\newcommand{\La}{\Lambda}
\newcommand{\la}{\lambda}
\newcommand{\Om}{\Omega}
\newcommand{\om}{\omega}
\newcommand{\si}{\sigma}
\newcommand{\Si}{\Sigma}
\newcommand{\th}{\theta}
\newcommand{\vth}{\vartheta}

\newcommand{\ra}{\rightarrow}
\newcommand{\mm}{\mbox{$\cal M$}}
\newcommand{\tr}{\mbox{tr}}

\newcommand{\lap}{\triangle}
\newcommand{\arctg}{\mbox{arctg}}
\newcommand{\bm}[1]{\mbox{\boldmath $#1$}}
\newcommand{\be}{\begin{equation}}
\newcommand{\ee}{\end{equation}}
\newcommand{\bea}{\begin{eqnarray}}
\newcommand{\eea}{\end{eqnarray}}
\newcommand{\bean}{\begin{eqnarray*}}
\newcommand{\eean}{\end{eqnarray*}}
\newcommand{\dd}{\partial}

\begin{document}

   \begin{titlepage}
 \begin{flushright} ZU--TH14/92   \end{flushright}
 \vspace*{2cm}
 \begin{center}
   {\huge GAUGE INVARIANT \vspace{0.5cm} COSMOLOGICAL
      PERTURBATION  \vspace{0.5cm} THEORY}\\ {\LARGE A General
	 	Study and its application to the Texture
	Scenario of  Structure Formation}
   \vspace{2cm} \\
    {\bf \large Ruth Durrer}
    \vspace{1cm} {\large
   \\ Universit\"at Z\"urich, Institut f\"ur Theoretische Physik,
      Winterthurerstrasse ~190, \\ CH-8057 Z\"urich, Switzerland} \vspace{2cm}
 \end{center}

 \begin{abstract}
 After an introduction to the problem of cosmological structure formation,
 we develop gauge invariant cosmological perturbation theory. We derive the
 first order perturbation equations of Einstein's equations and energy
 momentum ``conservation''. Furthermore, the perturbations of Liouville's
 equation for collisionless particles and Boltzmann's equation for Compton
 scattering are worked out. We fully discuss the propagation of photons in a
 perturbed Friedmann universe, calculating the Sachs--Wolfe effect and
 light deflection.  The perturbation equations are extended to
 accommodate also perturbations  induced by seeds.

 With these general results we discuss some of the  main aspects of the
 texture model for the formation of large scale structure in the
 Universe (galaxies, clusters, sheets, voids).
 In this model, perturbations in the dark matter are induced by texture seeds.
 The gravitational effects of a spherically symmetric collapsing texture
 on dark matter, baryonic matter and photons are calculated
 in first order perturbation theory. We study
 the characteristic signature of the microwave background fluctuations
 induced in this scenario  and compare it with
 the  COBE observations.
 \end{abstract}
 \end{titlepage}

 \tableofcontents
  \newpage
 \addcontentsline{toc}{chapter}{Introduction}
 \noindent {\LARGE{\bf Introduction}}\vspace{0.2cm}\\
 Within standard cosmology, the formation of large scale structure remains
 one of the biggest unsolved problems, despite of great efforts.
 The most natural scenario, where structure forms by  growth of
 adiabatic perturbations in a baryon dominated universe is clearly ruled out.
 Other extensively worked out scenarios like isocurvature baryons
 or baryons and collisionless matter (cold dark matter or hot dark matter)
 face severe difficulties
 \cite{GH,MESL,SFR,GSS2} (for a  short review see \cite{PeS}).

 On the other hand, there is the interesting possibility to
 induce perturbations in the baryonic and dark matter by seeds.
  Seeds are  an inhomogeneously distributed form of energy which
  contributes only a small fraction of the
 total energy density of the universe. Thus, linear perturbation theory
  can be used to calculate the induced fluctuations and their time evolution.
  Gauge--invariant linear perturbation theory \cite{Ba} is
 superior to  gauge dependent methods, since it is not plagued
 by gauge modes, and it leads in all known cases to the simplest systems of
 equations.

 Examples of seeds are primordial black holes, boson stars, a first
 generation of stars, cosmic strings, global monopoles, or global texture.
 These last three are especially appealing since they  can
 originate in a natural way from phase transitions in the early universe
 \cite{K1}. If they indeed play
 an important role in structure formation they would  link
 the smallest microscopical scales never probed by particle accelerators
 (down to $10^{-16}GeV^{-1}$) and the largest structures (up to  $100$Mpc
 and maybe more)!  Another attractive feature of topological defects is that
 the gravitational effects of each class of them (cosmic strings are
 $\pi_1$ defects, monopoles are $\pi_2$ defects and texture are $\pi_3$
 defects)  are quite insensitive to
 the detailed symmetry breaking mechanism and only depend on the
 symmetry breaking scale $\eta$, which we cast in the dimensionless quantity
 $\ep = 16\pi G\eta^2$. This quantity   determines the amplitude but
  not the shape or the time evolution of the perturbations.

 Here, we mainly consider the $\pi_3$ defect global texture which
 was first proposed by  Turok [1989] as a seed for large scale structure.
 Several subsequent
  investigations of this scenario  gave promising results:
 The spatial and angular correlation functions, the large scale
 velocity fields
 and other statistical quantities obtained by numerical simulations
 agree roughly with  observations (see Pen et al. [1993], Gooding et al.
 [1992], Spergel et al.  [1991]  and references therein).

 My objective in this text is to fully develop gauge invariant linear
 perturbation theory to treat models with seeds.
 I then want to show in some detail, how all the linear perturbation theory
 aspects of a scenario of large scale structure can be investigated  with
 these tools. As an example, we discuss the texture scenario.
 I choose this scenario not because I think it is the solution of the
 problem of cosmological structure formation.  But it is the simplest
 worked out scenario, where initial fluctuations are induced by topological
 defects of a symmetry breaking phase transition, and  I believe
 that this class of models deserves thorough investigation as an
 alternative to models with initial perturbations from inflation.

 In the first chapter we set a frame for this review with a non--technical
 overview of the problem of
 cosmological structure formation. We also very briefly discuss some
 of the  presently  considered scenarios. A reader already familiar
 with the problem of structure formation may skip it.
 In the second chapter gauge--invariant cosmological perturbation theory
 \cite{H1,Ba,KS,DS,MFB}
 is presented  in a (hopefully)  pedagogical fashion. Although, some
 familiarity with general relativity and basic concepts of differential
 geometry are needed to follow the derivations in this chapter.
 We extensively discuss scalar, vector and tensor perturbations for
 fluids and for collisionless matter. Furthermore,
 general formulae for the  deflection of light in
 a perturbed Friedmann universe are  derived. This part is new.
 Finally,  seeds as initial perturbations are introduced \cite{d90}. Here it is
 assumed that
  non--gravitational interactions of the seeds with the surrounding
 matter are negligible. For topological defects, this is certainly a good
 approximation soon after the phase transition.

 In Chapter~3, some simple but important applications of cosmological
 perturbation theory are described. We first discuss the ideal fluid
 \cite{Ba,KS}.
 Then, a gauge invariant form of Boltzmann's equation for Compton scattering
 is derived. We study it in the limit of many collisions to obtain an
 approximation
 for the damping of cosmic microwave background (CMB) fluctuations
 by photon diffusion in a reionized universe. We then list
  all  mechanisms proposed to induce anisotropies in the cosmic microwave
 background. Finally, two applications of light deflection are
   discussed: Lensing by global monopoles \cite{BV} and light deflection
 due to a passing gravitational wave. This  effect is new and
 might be an alternative
 way to detect gravitational waves of very far away sources.

 In Chapter~4 we review the concept of  texture defects and present
 the exact flat space solution found by  Turok and Spergel [1990]. We then
  calculate the gravitational potentials induced by this solution
 and apply the formalism developed in Chapter~2  to derive photon
 redshift, light deflection and  perturbations of baryons and dark matter.
 All results can be obtained analytically and provide a nice
 application of gauge invariant perturbation theory. They are, however,
  cosmologically relevant only for scales
 substantially beyond horizon scale. Most of the results of this chapter were
 originally derived by Durrer [1990] and Durrer et al. [1992a], but the
 calculation of
 light deflection is new and  the work on collisionless particles is
  presented in  a much simpler way and a physically more sensible limit is
 performed.

 We  conclude with a chapter on textures in expanding space. Here we present
 a method to calculate the induced cosmic microwave background fluctuations
 by statistically distributing individual textures which are modeled
 spherically symmetric. The detailed numerical results of these investigations
  will be published elsewhere \cite{DSH}.

In Appendix~A the basics of 3+1 formalism of general relativity
are outlined. In Appendix~B, we provide a glossary of the variables used in
this review.

\chapter{The Problem of Large Scale Structure Formation}
\section{The Standard Cosmological Model}
We begin with some facts on the standard model of cosmology.
Extensive treatments of this subject can be found in the books by
Tolman [1934], Bondi [1960], Sciama [1971], Peebles [1971], Weinberg [1971],
Zel'dovich and Novikov [1983], B\"orner [1988], Kolb and Turner [1990],
Peebles [1993].

 The assumption that the universe is
homogeneous and isotropic on very large scales, leads to a very special class
of solutions to the gravitational field equations, the
Friedmann--Lema\^{\i}tre universes. Accepting furthermore the
cosmological origin of quasars and/or the cosmic background radiation, one
can reject the possibility of a so called bouncing solution and show that
our universe necessarily started in a big bang \cite{ER}. Due to homogeneity
and isotropy, the energy momentum tensor is described by the
total energy density
$\rho$ and pressure $p$ which are functions of time only. The metric is
of the form
\[ ds^2 = a^2(-dt^2+\ga_{ij}dx^idx^j) ~, \]
where $\ga$ is a 3-d metric of constant curvature $k$. The simply connected
3--spaces corresponding to $\ga$ depend on the sign of the curvature:
For $k>0$, the three--space is a sphere, for $k<0$ a pseudo sphere and for
$k=0$
flat, Euclidean space. (Note, however, that the metric $\ga$ cannot
decide on the topological structure of 3--space. For $k=0$, e.g.,
it may well be, that three space is topologically equivalent to
${\bf R}^3/{\bf Z}^3$, i.e., a torus with finite volume. The often
stated phrase that for $k\le 0$ 3--space is infinite is thus wrong.)

The time dependent function $a$ is the scale factor
and the physical time (proper time) $\tau$ of an observer at rest
is given by $d\tau = adt$. The time coordinate $t$ is  called
conformal time.

Einstein's equations imply the Friedmann equation
which determines the scale factor as a
function of the density:
\be \left({\dot{a}\over a}\right)^2 + k = {8\pi\over 3}G\rho a^2 +\La a^2/3
 \label{1F}~.\ee
Here $G$ is Newton's gravitational constant, and $\La$ is the famous
cosmological constant which has been resurrected
several times in the past \cite{We89}.

Important quantities in Friedmann cosmology are the Hubble parameter, $H$
and the density parameter $\Om$, which are defined by
\[ H = \dot{a}/a^2 \mbox{ ~~~and~~~~} \Om = \rho/\rho_c~,
 \mbox{~~~with~~~} \rho_c = 3H^2/(8\pi G)~\].
Here $\rho_c$ is the critical density, the density of a universe with
$k=\La =0$.
The  present value of the Hubble parameter, $H_0$, is
usually parameterized in the form $H_0 = h\times 100km/s/Mpc$. Observations
limit the value of $H_0$ in the range
\[0.4\le h\le 1   \mbox{~~~ and}\]
\[ 0.05\le \Om_0 \le  2 ~.\]
  From estimates of the mass to light ratio one obtains a value for the
amount of luminous matter, $\Om_{lum} \approx 0.007$ and from velocity
measurements one estimates the amount of 'clustered matter' to be
$\Om_{dyn}\approx 0.1$. These values result from many difficult observations
and are correspondingly uncertain \cite{KoT}. Unfortunately, the
measurements which lead to them
 can not provide information about a dark, poorly clustered
matter contribution and therefore do not yield  an upper limit for the
density parameter. The upper bound comes from a comparison of lower
limits to the age of the universe with the Hubble parameter.

Using the energy ``conservation'' equation
\be  \dot
{\rho} = -3{\dot{a}\over a}(\rho + p) ~, \label{1c} \ee
one finds that for $p\ge 0$, $\rho$ grows at least like $a^{-3}$ for
 $a\ra 0$ (i.e. approaching the big bang). Therefore, at an early enough
epoch the density term in
(\ref{1F}) always dominates over the curvature term and the
cosmological constant, and $\Om$ becomes arbitrarily close to 1. In other
words, the evolution of a Friedmann universe  always ``starts'' very close
to  the unstable fixpoint $\Om =1$.

On the other hand, (\ref{1F}) shows, if the cosmological constant $\La$
 once dominates the expansion of  the universe, the curvature term $k$
and the energy density
$\rho$ become less and less important. The universe then expands
 approximately  exponentially in terms of  physical time $\tau$
\[ a(\tau) = a(\tau_i)\exp(\sqrt{\La/3}\int_{\tau_i}^\tau d\tau') ~~. \]
This rapid expansion can only be stopped if some or all of the vacuum
energy inherent in the cosmological constant is radiated into particles.
Such an intermediate phase of $\La$ dominated expansion is the basic idea of
most scenarios of inflation.

Today, the cosmological constant is  severely limited from observations:
\[ \Om_\La \equiv |{\La\over 8\pi G H_0^2}| \le {\cal O}(1) ~.\]
This is very much smaller than the values expected from particle physics,
but might still be enough to influence the expansion of the universe and
structure formation substantially \cite{Ho}. It can, e.g., lead to a 'loitering
period' during which the universe is nearly non--expanding \\ \cite{DK}, and
thus fluctuations grow exponentially. Since such a small value for $\La$
requires an enormous fine tuning, we  set $\La=0$ in the following.

For $\La=0$, one has $\Om>1$ in a closed universe (3--sphere) and
$\Om < 1$ in a negatively curved (open)
universe.

Later, we  also use the following consequences of (\ref{1F}) and
(\ref{1c}) (setting $w=p/\rho$ and $c_s^2 = \dot{p}/\dot{\rho}$):
\bea  \ddot{a}/a &=& -{1\over 2}[(3w-1)({\dot{a}\over a})^2 +(3w+1)k] \\
	\dot{w} &=& 3(\dot{a}/a)(w-c_s^2)(1+w) \label{1wdot} ~.\eea

Like the standard model of particle physics, the standard model of cosmology
has many impressive successes.
The most important ones are:
\begin{itemize}
\item Uniform Hubble expansion (Detected by Wirtz [1918], Slipher [1920]
	and Hubble [1927]).
\item The prediction of Gamov [1946] and detection  by Penzias and
  Wilson [1965]
  of the cosmic microwave background radiation, its 'perfect' black body
  spectrum \cite{Ma} and
  its extraordinary uniformity (see Figs.~1 and 2).
\item The abundance of the light elements ($H$, $~^2H$, $~^3H$, $~^3He$,
  $~^4He$, $~^7Li$)
 can be calculated, and the comparison with observational estimates predicts
 $0.02\le \Om_B\le 0.1$ which is consistent with direct determinations
  of $\Om_0$.  This calculations were originally performed by Alpher et al.
  [1948], Wagoner et al. [1967] and by many others later. A comprehensive
  review is Boesgaard and Steigman [1985]. For  more recent developments,
  see e.g. Kurki--Suonio et al. [1988] and Walker et al. [1991].
 \end{itemize}
However, many questions are left open. Most of them can be cast in terms of
 very improbable initial conditions:
\begin{itemize}
\item The question of the cosmological constant, $\La$: Why is it so small?
 i.e., so much smaller than typical vacuum energies arising from particle
physics  \cite{We89}.
\item The flatness / oldness / entropy problem: why is the universe
 so old, $t\gg t_{pl}$, and still ${\cal O}(\Om) = 1$? Here, $t_{pl}$
	is the Planck time,
	$t_{pl} =(\hbar G/c^5)^{1/2} = 5.39\times 10^{-44}$sec .
\item The horizon problem: how can  different patches of the universe
 have been at the same temperature long before they ever where in causal
 contact?
\end{itemize}

So far, the most successful approach to answer the second and  third
questions is based on the hypothesis of an
 inflationary phase during which expansion is dominated by a 'fluid' with
negative effective pressure, $p\le -\rho/3$ (e.g. a cosmological term,
for which  $p=-\rho$) and physical distances thus grow faster than the
size of the effective particle horizon, $(\dot{a}/a^2)^{-1}$. I do not
explain how inflation answers these questions, but
 refer the reader to some  important publications on the subject:
\cite{Gu,AS,L1,L2,L3,LS,Ol}.

These open problems  provide  some of the very
few and important hints to new fundamental physics beyond the standard
models of  cosmology and of particle physics.

\section{Structure Formation}
Besides the fundamental issues, related to the very early universe
and unknown basic physics, there is  the important problem of
structure formation which, to some extent, should be solvable within
the standard model. This is the problem I want to
address here:

How did  cosmological structures like  galaxies, quasars, clusters,
voids, sheets... form?

  From difficult and expensive determinations of the three--dimensional
distributions of optical and infrared galaxies on scales up to
$150h^{-1}$Mpc \cite{LGH6,LGH8,GH,BEKS,SFR,LEPM}, \\ \cite{FDSYH},
we know that  galaxies
are arranged
in sheets which surround seemingly empty voids of sizes  up to
$50h^{-1}Mpc$.
Observations also show that the distribution of clusters of galaxies
 \\   \cite{BaS,Bah,WB} is inhomogeneous with
\[ {\de n_c\over n_c} \ge 1  \mbox{ \hspace{0.6cm} on scales up to 25 Mpc.} \]

In addition, it is  important to observe that not all galaxies are young.
There are quasars with redshift up
to $z\approx 5$ and galaxies with $z\approx 3$. On the other hand, there
are indications that the galaxy luminosity function is still evolving
considerably in the recent past, i.e., that galaxy formation
still continues \cite{LPEM}.

  From a naive Newtonian point of view one might say: {\em ``Gravity is
an unstable interaction. Once the slightest perturbations (e.g. thermal)
are present they start growing and eventually form all these observed
structures. The details of this process might be complicated but there
seems to be no basic difficulty.''}

This is true in a static space where small density fluctuations on
large enough scales grow exponentially. But, as we shall see in Chapter~2,
in an expanding space, this growth is reduced to  a power law:
\[ {\de\rho\over \rho} \propto \left\{ \begin{array}{ll}
  a & \mbox{for pressureless matter, dust} \\
\mbox{const. }& \mbox{ for radiation, relativistic particles.}
\end{array} \right.  \]
In a radiation dominated universe, pressure prohibits any substantial
growth of density perturbations. Once the  universe  is  matter dominated
$\de\rho/\rho$ grows proportionally to the scale factor, but the gravitational
potential
\[ \Psi \propto \int(\de\rho/r)r^2dr = \int\de\rho rdr \]
remains constant. This led Lifshitz [1946], who first investigated the
general relativistic theory of cosmological density fluctuations,
to the statement: {\em ``We can apparently conclude that
gravitational instability is not the source of condensation of matter into
separate nebulae''}. Only 20 years later  Novikov [1965] pointed out that
Lifshitz  was
not quite right. But to some extent Lifshitz's point remains valid:
Gravity cannot be the whole story. Imagine an $\Om = 1$ universe which
 was matter dominated after $z\approx 2\times 10^4$. Since
density perturbations in cold matter evolve like $1/(z+1)$, they have
grown  by less than $10^5$. But today there are galaxies
with masses of $M\approx 10^{12}M_\odot \approx 10^{69}m_N$~. Statistical
inhomogeneities of these $N$ nucleons have an amplitude of
$1/\sqrt{N} \approx 10^{-34}$, which would have grown to less than $ 10^{-29}
\ll 1$ today. Therefore, there must have been some non--thermal {\em initial
fluctuations} present with an amplitude on the order of $10^{-5}-10^{-4}$.
We also know that the amplitude of these initial perturbations did not
exceed  this amount because of the high degree of isotropy of the
cosmic microwave background radiation (CMB). In fact, only recently CMB
anisotropies have been
found in the COBE experiment \cite{Co1,Co2}. They are on the level
\[{\De T\over T} \approx 10^{-5} ~~~\mbox{ or }~~~
	\de\rho_{(rad)}/\rho_{(rad)} \approx 4\times 10^{-5}\]
on all angular scales larger than  $10^o$ and
compatible with a scale invariant Harrison--Zel'dovich spectrum. On
 smaller angular scales other experiments led to limits, $\De T/ T \le$
a few$\times 10^{-5}$  (see Fig.~2).
We disregard the dipole anisotropy which is due to our motion with about
600km/s with respect to the microwave background radiation.

There are scenarios of cosmological structure formation which are
not based on the gravitational instability picture. The most recent of them
is based on explosions (of supernovae or super conducting cosmic strings)
\cite{Os} which are supposed to sweep away baryons, producing shock
fronts in the form of sheets surrounding roughly spherical voids.
Usually such scenarios face enormous
difficulties in producing the energy required to account for the large scale
observations and to satisfy at the same time  the limits for perturbations
of the CMB on small angular scales. None of them has thus been worked out
in detail. In fact, the explosion scenario also needs gravitational
instability to amplify perturbations by a significant factor
\cite{MKO}. In this paper, we concentrate on scenarios which rely on
gravitational instability.
The fact, that the anisotropies measured by COBE just coincide
 with the amount of growth necessary to form structures today is taken as
a hint that the gravitational instability picture may be correct.

\section{The General Strategy}
We now want to outline in some generality  the ingredients that go into a
model of structure formation which is based on the gravitational
instability picture.

\subsection{Initial fluctuations}
We saw in the last section that small density fluctuations in a
Friedmann universe may have grown (by gravitational instability) by
about a factor of $2\times 10^4$ during the era of matter domination.
Therefore, a complete scenario of structure formation must lead to
 initial matter density fluctuations with amplitudes on the
order of $10^{-4}$. Two  possibilities to obtain these initial
fluctuations are primarily  investigated. \vspace{0.2cm}\\
{\bf A) ~Initial perturbations produced during inflation:~~} Here it is
assumed that  density fluctuations are generated
during an inflationary phase \cite{Gu,AS,L0,L1,L2,LS,L3} from initial
quantum fluctuations of
scalar fields  \cite{BD,H2,Sta},\\ \cite{GP,FRS}. Due to the nature of
quantum fluctuations, the distribution of the
amplitudes of these initial perturbations is usually   Gaussian.

There are several different more or less convincing models of inflation.
One  divides  them into: Standard (or old) inflation \cite{Gu}, new
inflation \cite{AS,L0}, chaotic inflation \cite{L1} and the most recently
proposed possibilities of extended and hyper--extended inflation \cite{LS}.
Reviews on the subject are found in Linde [1984], Linde [1990],
Olive [1990] and Steinhardt [1993]. All these
models of inflation  differ  substantially, but the mechanism
to produce initial fluctuations is basically the same:\\
The scales $l$ of interest ($l\le l_H(t_o)\approx 3000h^{-1}Mpc$) are
smaller than
 the effective particle horizon at the beginning of the inflationary
phase. The scalar field
that drives inflation therefore experiences quantum fluctuations on all
these scales. Their amplitude  for a minimally
coupled scalar field $\phi$ in  de Sitter universe\footnote{Except for
extended and hyper--extended inflation, the universe  during the
inflationary phase is a de Sitter universe, expansion is driven by a
cosmological constant.}
 within a volume $V$ can be calculated \cite{BD}:
\[ (\De\phi)^2_k = {Vk^3\over 2\pi^2}|\de\phi_k|^2 = (H/2\pi)^2~,\]
with
\[ \de\phi_k = V^{-1}\int \phi(\bm{x})\exp(i\bm{k\cd x})d^3x  ~. \]
In the course of inflation, the interesting scales inflate outside the horizon
and  quantum fluctuations 'freeze in' as classical fluctuations of the
scalar field. This leads to energy density perturbations according to
\[ \de\rho_{\phi} = \de\phi{d{\cal V}\over d\phi}~, \]
where $\cal V$ denotes the  potential of the scalar field. By
causality, on  scales larger than the effective particle horizon, these
perturbations cannot grow or decay by any physical mechanism.
The spurious growth of super--horizon size density perturbations in certain
gauges like, e.g., synchronous gauge, is a pure coordinate effect!

 This mechanism yields density perturbations $D_{\la}$ of a given
size $\la$ which have constant amplitude at the time they re--enter
the horizon after inflation is completed:
\[ D_{\la}(t=\la) = A ~.\]
A natural consequence is thus the well known scale invariant
Harrison--Zel'dovich spectrum \\ \cite{Ha,Ze2}.
Under certain conditions on, e.g., the potential of the scalar field (which
may require fine tuning of coupling constants), one
can achieve that these fluctuations have the required amplitude of
$A \approx 10^{-4}$.

Recently, the failure to account for the largest scale structure of the
otherwise so successful cold dark matter (CDM) model has caused some effort
to find inflationary models with spectra that differ substantially
  from Harrison--Zel'dovich on large scales \cite{PS}. Although of
principle interest, this work may turn out to be
unimportant after the new COBE results have shown such a striking
consistency with a scale invariant spectrum.\vspace{0.2cm}\\
{\bf B) ~~Seeds as initial perturbations, topological defects:} Initial
fluctuations  might have been triggered by seeds, i.e., by an
 inhomogeneously distributed matter
component which contributes only a small fraction to the total energy
density of the universe. Examples of seeds are a first generation of stars,
primordial
black holes, bosonic stars, cosmic strings, global monopoles and textures.
 We  restrict our discussion to  the latter three, the so
called topological defects,  which  can arise  naturally
during phase transitions in the early universe \cite{K1}.

To understand how they form,  consider a symmetry group $G$ which
is broken by a scalar field $\phi$ to a subgroup $H$ at a temperature $T_c$.
The vacuum manifold of the cooler phase is then generally given by
${\cal M}_0=G/H$. Since the order parameter field $\phi$ (Higgs field) has a
finite correlation length $\xi\le l_H$ ($l_H \equiv $ horizon size)
which is limited from above by the size of the horizon, the field varies
in ${\cal M}_0$ if compared over distances
larger than $\xi$. If the topology of ${\cal M}_0$ is non--trivial, the
scalar field can vary  in such a way that there are points in spacetime
where, by continuity reasons, $\phi$ has to leave $\mm_0$ and assume
values of higher energy.  This is the Kibble mechanism \cite{K2}.

The set of points with higher energy  forms a
connected sub--manifold  without boundary in four dimensional spacetime. The
dimensionality, $d$, of this sub--manifold is determined by the order, $r$, of
the corresponding homotopy group  $\pi_r({\cal M}_0)$: ~ $d = 3-r$, for
$r\le 3$.  The points of higher energy
of $\phi$ are often just called 'singularities' or 'defects'.

 For illustration, let us look at the simplest example, where ${\cal M}_0$ is
not connected, $\pi_0({\cal M}_0)\neq 0$:
At different positions in space with distances larger
than the correlation length $\xi$, the field can then assume values which
belong to disconnected parts of ${\cal M}_0$ and thus, by continuity,
$\phi$ has to leave $\mm_0$ somewhere in between. The sub--manifold of
points of higher energy is  three dimensional (in spacetime)
and called a {\bf domain wall}. Domain walls are  disastrous for
cosmology except if they originate from  late time phase transitions.

A non--simply connected vacuum manifold, $\pi_1(\mm_0)\neq 0$,
 leads to the formation of two dimensional defects, {\bf cosmic strings}.
Domain walls and cosmic strings are either infinite or closed.

If $\mm_0$ admits topologically non--trivial (i.e. not shrinkable to a point)
continuous maps from the two sphere, $\phi : {\bf S}^2\ra \mm_0$ ,
then $\pi_2(\mm_0)\neq 0$, one dimensional defects, {\bf mono\-poles}, form.

Finally, continuous mappings from the three sphere determine $\pi_3$.
If $\pi_3(\mm_0)\neq 0$, zero dimensional {\bf textures} appear
which are events of non--zero potential energy. Using Derricks theorem
 \cite{De},   one can argue
that a scalar field configuration with non--trivial $\pi_3$
winding number (i.e. a texture knot) contracts
and eventually unwinds, producing a point of higher energy for one instant of
time. This type of defect is discussed in more detail in Chapters~4 and 5.

 Topological defects are  very well known in solid state
physics. Important examples are vortices in a super conductor or the
vorticity lines in a super fluid. All four types of defects discussed
above can also be found in liquid crystals
see e.g. Chuang et al. [1991] and references therein.

Depending on the nature of the broken symmetry, defects can  either be
{\bf local}, if the symmetry is gauged, or {\bf global}, from a
global symmetry like, e.g., in the Peccei--Quinn mechanism.
In the case of local defects, gradients in the scalar field can be compensated
by the gauge field and the energy density of the defect is confined to the
defect manifold which has a thickness  given roughly by the inverse symmetry
breaking scale. On the other hand, the energy density of global defects
is dominated by gradient energy, with a typical  scale given by  the horizon
size at defect formation. The extension of the induced energy density
perturbation is thus about the horizon size at its formation. This leads to a
Harrison--Zel'dovich initial spectrum. The difference to initial perturbations
arising from inflation is that  density fluctuations induced by topological
defects are not Gaussian  distributed.

Local monopoles would dominate the energy density of the universe  by far
($\Om_M\approx 10^{11}$ !!) and must be excluded.
Local textures are not energetic enough to seed large
scale structure. But global monopoles and global textures are quite promising
candidates. Cosmic strings, both global or local, are
interesting. One of the models which we introduce in the next section
is based on local cosmic strings.

We will find in Chapter~4,  that the
properties of the scalar field $\phi$ other than the induced  homotopy
groups, e.g., the specific form of the Higgs potential, are of no importance
for the defect dynamics. But the probability of defect formation via the
Kibble mechanism might well depend on $\mm_0$. The symmetry breaking
scale just determines the energy of the defects, i.e.,
the amplitude of the fluctuations. From simulations of
 large scale structure formation one finds that successful models requires
$T_c^2G \approx m_c^2/m_{pl}^2 \approx 10^{-6}$, which corresponds to
 a typical GUT scale of $\sim 10^{16}$GeV. This can also be understood
analytically: From the perturbation equations derived in Chapter~2, we shall
see that defect induced structure formation leads to initial perturbations
with amplitudes $A\sim 16\pi GT_c^2$.

\subsection{Linear perturbation theory}
Since  the amplitudes of  the initial  fluctuations are tiny,
at early times their evolution can be calculated within linear cosmological
perturbation theory.

 Chapter~2 is devoted to a thorough discussion of this subject, we
thus skip it here. The reader just interested in the results is referred
to Chapter~3 where some important consequences from cosmological
perturbation theory are discussed.

\subsection{N--body simulations}
On relatively small scales (up to about 20Mpc or so) perturbations become
non--linear at late times. Structure formation must then be followed by
non--linear numerical simulations, whose input is  the perturbation spectrum
resulting from linear perturbation theory. Since at the time when the
non--linear regime takes over, the corresponding
 scales are much smaller than the horizon size, and  since
the gravitational fields and velocities are small, Newtonian
gravity is sufficient for these simulations. For a realistic calculation of the
process of galaxy formation, the hydrodynamical evolution of baryons has to
be included. The different heating and cooling processes and
the production of radiation which might partially provide the X--ray background
and might reionize the intergalactic medium have to be accounted for.
 Furthermore, the onset of nuclear burning which produces the
light in  galaxies has to be modeled, in order to obtain a mass to
light ratio or the bias parameter (which may well be scale dependent).

The inclusion of the heating and cooling processes of baryons into N--body
simulations is a very complex
computational task and the  results published so far are still preliminary
\cite{CJLO,Ce1,Ce2,COST}. Only very recently, the first calculations taking
into account nuclear burning have been carried out \cite{Onew}. High quality
simulations which only contain collisionless particles have been
performed to very good accuracy \\ \cite{CM,Wh,CDM}, but
unfortunately they
leave open the question how  light traces mass and therefore,
how  these results are to be compared with observations of galaxy
clustering.

I shall not discuss this important and difficult part of structure
formation in this review and refer the reader to the references given above.

\subsection{Comparison with observations}
Of course the final goal of all the effort is to confront the results of
a given model with observations. The easiest and least uncertain part
 is to analyse the induced microwave background fluctuations.
They can be calculated reliably within linear
perturbation theory.
Observations of the microwave background are also a  measurement
of the initial spectrum and it is very remarkable that the new COBE
results are compatible with a Harrison--Zel'dovich spectrum \cite{Co1,Co2}.

On smaller scales ($\th\le 6^o$), the CMB anisotropies depend crucially on
the question whether the universe has been reionized at early times
($z\ge 100$) and therefore indirectly on the formation of  non--linearities
in the matter density perturbation (which would provide the ionizing
radiation).  The different possibilities how the process of structure formation
can induce anisotropies in the cosmic microwave background are
discussed in detail in Chapter~3.

 On  scales up to about 20Mpc, the up today most extensively used
tool to compare
models with observations is the galaxy galaxy correlation function,
$\xi_{GG}(r)$,
\[ \xi_{GG}(r) = {1\over \langle n\rangle^2}\int n(\vec{x})
	 n(\vec{x}+r\vec{e})d^3xd\Om_{\vec{e}}~,\]
where $n$ is the number density of galaxies.
Usually, the amplitudes of initial fluctuations in a given scenario are
normalized by $J_3(R=10Mpc)$:
\[ J_3(R) = \int \xi_{GG}(r)W_R(r)r^2dr ~,\]
with an observational value of  $J_3(10Mpc)/(10Mpc)^3\approx 0.27$.
Here $W_R(r)$ is a (Gaussian or top hat) window function windowing scales
smaller than $R$.
Since the new COBE results now  provide the amplitude of fluctuations on very
large scales which are not influenced by non--linearities,
future calculations will clearly be normalized on these scales.
Observations fix the amplitude and slope of the
galaxy galaxy and cluster cluster correlation functions to be
\bea \xi_{GG}(r) &\approx& ({r\over r_0})^{-1.8} ~, \\
  \xi_{CC}(r) &\approx& ({r\over r_0})^{-1.8} ~.\eea
For galaxies one finds $r_0\approx 5.4h^{-1}Mpc$ \cite{P88}, whereas for
clusters $r_0$ depends on the ``richness class'' of the clusters considered.
For rich clusters $ r_0\approx (20 - 25)h^{-1}Mpc$ \\ \cite{BaS}.
Recently, a smaller amplitude for the cluster cluster correlation function
has been found from the APM survey, $\xi_{CC} = 4\xi_{GG}$ and
$\xi_{CC} = 2\xi_{CG}$ \cite{DEMS}.

A disadvantage of the correlation function is its insensitivity to
lower dimensional structures like sheets and filaments\footnote{ Recently
the amplitudes of the correlation function have been criticized to depend
crucially on the sample size and thus to be physically meaningless. New
analyses \protect\cite{EKS,Pi,DMSdCY,CP} have shown that $r_0$ depends
on the sample size, hinting that the distribution of galaxies may be
fractal up to
the largest scales presently accessible in volume limited samples,
$R_{\max}\approx 30h^{-1}$Mpc. If this objection is justified, the
normalization procedure with the help of the correlation function is
useless! }.

There are  various other statistical tests which one can perform  and compare
 with the sparse observations. I just mention a few:\\
The Mach number, which gives the ratio between the average velocity and the
velocity dispersion on a given length scale, $M = <v>^2/\si^2$
\cite{OS}.\\
 The genus test, where one calculates the number of holes minus the
number of islands in an iso--density contour \cite{GMD}.\\
The 3--point or 4--point correlation functions, which determine the deviation
  from Gaussian statistics of the distribution of perturbations on a given
length scale \cite{Pe}. They can  be cast in the skewness,
$ <\de\rho^3>/<\de\rho^2>^{3/2}$ and the kurtosis
$ <\de\rho^4>/<\de\rho^2>^2$.

Additional more qualitative results are:
 - The earliest galaxies must have formed at $z\ge 5$
but that there still must be substantial galaxy formation going on at
$z\approx 2-1$.

- From  observations \cite{LGH6,LGH8}, \\ \cite{GH} we
can conclude that galaxies are arranged in sheetlike structures around
seemingly empty voids.

- Velocity observations have found large ($\approx 100h^{-1}Mpc$) coherent
velocity fields with \\ $\langle v\rangle \approx 500km/s$.

- A successful model should of course also obtain flat rotation curves
of galaxies which have been observed with increasing accuracy since the
seventies \cite{Ru}.

\section{Models}

 The most
simple picture, a universe with $\Om h^2 = \Om_B h^2 \approx 0.1$ and
adiabatic initial fluctuations is definitely ruled out by the limits of the
microwave background fluctuations \cite{GSS1,GSS2}.
This is a first, very important result in the discussion of  different
models which might account for the formation of large scale structure.

To give the reader a taste of the presently favored scenarios, we present
here five cases.  A sixth possibility, the ``texture scenario'', will be
discussed in Chapters~4 and 5.
It is clear that a mixture of the hot dark matter (HDM) and cold dark matter
(CDM) models presented here,
 as well as defects with HDM or any of the proposed scenarios with
addition of a cosmological constant might lead to models
that fit the observations better. The ones presented here are  partly chosen
by reasons of simplicity. The first attempt to treat a class of  models
systematically is given by Holtzman [1989], where 94 different  combinations
of ($\Om_\La,\Om_{CDM},\Om_{HDM},\Om_B$) are investigated.

In this section we  give a short description of the models chosen, and
 compare them with some observational phenomena. Our aim is only to
give a sketchy overview of these models; readers interested in more
details are referred to the literature given in the text.

\subsection{The isocurvature baryon model}
 As the most conservative alternative to the adiabatic baryon model,
Peebles [1987] pursued the question, whether it is possible to construct a
viable scenario of structure formation without the assumption of any exotic,
i.e. up to now unobserved, form of energy, a
universe with  $\Om_0 = \Om_B \approx 0.1$, which is still marginally
consistent with nucleosynthesis limits on $\Om_B$.

In order not to overproduce CMB fluctuations, one has to assume
isocurvature initial perturbations,
i.e. no perturbations in the geometry on cosmologically
relevant scales (see Section~3.1). From this one can conclude $\de\rho/\rho
\approx 0$ on scales which are larger than the size of the horizon.
Since  the universe is radiation dominated initially, $\rho_r \gg \rho_B$
this yields $|\de\rho_r/\rho_r| \ll |\de\rho_B/\rho_B|$, i.e. isothermal
fluctuations.

Isocurvature perturbations allow for  relatively high initial values of
$\de\rho_B$ and therefore lead to early structure formation. Galaxies
form at $z_g \ge 10$. The even earlier formation of small objects
reionizes the universe. Small scale anisotropies in the CMB can
then be damped by photon diffusion. For photon diffusion to be effective,
reionization must take place before the universe becomes optically thin.
This provides a lower limit to a 'useful' reionization redshift (see
Chapter~3):
\[ z_i\ge z_{dec} \approx 100({\Om_Bh\over .03})^{2/3}  ~.\]
Photon diffusion then damps fluctuations on all  scales smaller than
the horizon scale at decoupling, $l_H(z_{dec})$ and correspondingly
all angular scales with
\[  \th \le  \th_{dec} \approx 6^o\sqrt{\Om(100/z_{dec})}  ~. \]
The problems of this scenario are twofold:
First,  the quadrupole anisotropy of the CMB turns out to
be unacceptably large   \cite{GSS2}.  A way out of this problem is
a steep  initial perturbation spectrum, but then it is
 difficult  to reproduce the large amplitudes of the galaxy
correlations on large scales.(Although, for steep a enough spectrum,
even quadrupole fluctuations may be damped by photon diffusion, see
Section~3.2.3!)

Secondly, observations hint that galaxy formation might peak around
$z\approx 2$ or, at least, is still going on around $z\approx 1$, whereas
in the isocurvature baryon model the process of galaxy formation is most
probably over before $z=3$.

\subsection{Hot dark matter}
 Massive neutrinos
with $\sum m_i \approx 200eVh^2$ they can provide the dark matter of
the universe and dominate the total energy density  with
 $1= \Om_{tot}\approx \Om_\nu $  \cite{DZS,BES}.
Large scale structure  then develops as follows:

Initial fluctuations from inflation give rise to a scale invariant
spectrum of Gaussian
fluctuations. These initial fluctuations are constant until they 'reenter
the horizon' (i.e. their scale becomes smaller than the size of the
horizon, $l<l_H$). Thereafter they decay by free streaming if the
universe is still radiation dominated;  if the universe is matter
dominated ($z\le 2\times 10^4$) they grow in proportion to the scale
factor \cite{BS,d89}.
This leads to a short wavelength cutoff of the linear perturbation
spectrum at $l_{FS} \approx 40(m_\nu/30eV)^{-1}h^{-1}Mpc$. The
corresponding cutoff mass is $M_{FS}\approx 10^{15}(30eV/m_\nu)^2M_\odot$.
The linear spectrum for HDM is given in Fig.~3.

In this model, large objects with  mass $\approx M_{FS}$ (large clusters)
form first . They then fragment into galaxies. Gravitational
interaction of collisionless particles generates sheets, pancakes
\cite{Ze1}, and  galaxies are thought to lie
on the intersections of these sheets. This leads to a filamentary structure.
Simulations of HDM show \cite{CM,WFD} that in order to obtain the correct
galaxy  correlation function today, the large scale structure
becomes heavily overdeveloped (see Fig.~4). The other main problem is
that galaxy formation starts only very recently ($z\approx 1$). The model
has serious difficulties to account for quasars with redshifts $z\ge 3-4$.
In addition to these grave objections, CMB fluctuations turn out
to be too large in this model. Since  galaxy formation is only
a secondary process, initial fluctuations which determine
the amplitudes of CMB anisotropies must be rather large.

Because their thermal velocities are relatively high,
 massive neutrinos  can only marginally provide the dark matter of
 galaxies but they cannot be bound to dwarf galaxies. If
neutrinos are to constitute the dark matter of a virialized object with
velocity dispersion $\si$ and size $r$, their mass is limited from below
by the requirement \cite{TG}
\[ m_\nu \ge 30eV({200km/s\over \si})^{1/4}({r\over 10kpc})^{1/2} ~. \]
Since  also dwarf galaxies do contain substantial amounts
of dark matter \\  \cite{CF}, this is another serious constraint
for  the HDM model.

\subsection{Cold dark matter}
In this scenario one assumes the existence of a cold dark matter particle
which at present dominates the universe with $1=\Om_{tot} \approx
\Om_{CDM}$. Particle physics candidates for
such a matter component are the axion or the lightest super--symmetric
particle. A more extended list can be found in Kolb and Turner [1990].

Again it is assumed that an inflationary phase leads to a scale invariant
spectrum of  Gaussian initial fluctuations.
After the universe becomes matter dominated at
$z_{eq}\approx 2\times 10^4h^2$,
perturbations smaller than the horizon,
$l\le l_{eq} \approx 10(\Om h^2)^{-1}$.
 start growing. Damping due to free streaming is negligible. Because of the
logarithmic growth of matter fluctuations in the radiation dominated era
(see Section~3.1), the spectrum is slightly enhanced on smaller scales
$l <l_{eq}$.

According to this linear spectrum, sub--galactic objects form first. Once
the perturbations become non--linear, these objects virialize
and develop flat rotation curves. Only very recently, big
structures begin to form through tidal interactions and mergers \cite{DEFW}.

To obtain a mass distribution with $\Om_{dyn}\approx 0.1\; -\; 0.2 < \Om =1$,
it is necessary that most of the mass is in the form of a dark
background which is substantially less clustered than the luminous matter.
This  can be achieved with the idea of biasing: Luminous galaxies only
form at high  peaks of the density distribution. Since for a Gaussian
distribution high peaks are more strongly clustered than average, this
simple prescription has the desired effect \cite{Ka,BBKS}. Usually one
introduces a bias parameter $b$ and requires that galaxies form only
in peaks of height $b\si$, where $\si$ is the variance of the Gaussian
distribution of density peaks. The best results are obtained for a
bias parameter $b \approx 1.5 \; - \; 2$. Clearly, once  the correct
hydrodynamical treatment of baryons is included in the numerical simulations,
such a bias parameter (which otherwise is just an assumption of how
light traces mass) could be calculated. First preliminary results of such
 calculations \cite{CJLO,Onew} indicate that the above assumption
is probably quite reasonable. A problem of the biasing hypothesis is the
prediction of many dwarf galaxies in the voids which have not been found
despite extensive searches.

The large scale structure obtained in this scenario looks at first sight
rather realistic (see Fig.~5). It leads to the right galaxy galaxy
correlation function up to 10Mpc. But the recently detected huge structures
like the great wall (see Geller and Huchra [1989]) are very unlikely in
this model.
This situation has  been quantified by comparing the angular
correlation function from the deep IRAS survey with the one predicted by CDM
\cite{MESL}. There, a substantial excess of power (as compared to the CDM
model) on large angular scales, $\th>2^o$ is found (see Fig.~6).

Since galaxies form relatively late (at $z\approx 2$), it might also be
difficult to produce the very high redshift quasars ($z\approx 5$).
 But since their statistics
are still so low, and since very little is known on the ratio of normal
galaxies
to quasars at high redshift, this may not be a real problem.

\subsection{Cosmic strings}
Here, initial fluctuations are seeded by cosmic strings which form
via the Kibble mechanism (see Section~3) after a phase transition at
$T_c\approx 10^{15}GeV$ \cite{Vi80}.

Inter--commutation and gravitational radiation of cosmic string loops (see
Vachaspati and Vilenkin [1984], Durrer [1989b]) determine the
 evolution of a network of (non--super\-con\-duc\-ting) cosmic strings.
Numerical simulations support  analytical
arguments for a scaling law, $\rho_{string}\propto 1/(at)^2
\propto \rho$, for  the energy density of a cosmic string network.
In contrast to gauge monopoles or domain walls, strings do not
dominate the energy density of the universe and are cosmologically
allowed \cite{AT}.

It is well known that a static straight cosmic string does not accrete
matter, whereas a cosmic string loop  from far away acts like a point mass
\cite{Vi80}. High resolution simulations of cosmic string networks \cite{BB}
have shown that the loops that chop off the network are too small
($l_{loop}\le 0.01 l_H$) for efficient accretion. But moving
 long strings produce large accretion wakes behind them which might
provide the sheets and walls observed in the universe \cite{Be,Br}.
In this scenario galaxies form via fragmentation and/or accretion onto
loops.

A relatively new idea is that  chopping off small loops could lead
to wiggles on the long strings. Cosmic strings with such small scale
wiggles give rise to  strings with an effective tension which is smaller
than its effective energy density. In contrast to the original
cosmic strings, wiggly strings
 can accrete matter even if they are static \cite{Ca,Vi90,VV91}.
In order for the large scale structure not to be 'drowned', in these small
scale structures, the scenario works best if the dark matter is hot,
 massive neutrinos (HDM), so that the small scale fluctuations
are damped by free streaming.

The requirement for successful structure formation on one hand and the
difficulty not to overproduce microwave background anisotropies \cite{St}
on the other hand tightly constrain the possible value of the symmetry
breaking scale
\[ \mu = G\eta^2 \approx (1~ -~ 2)\times 10^{-6} ~. \]
Recent work using new cosmic string simulations in which these estimates were
 redone to compare the CMB anisotropies with the COBE measurements led to
 a similar value\\ \cite{BB2,Per}.

\subsection{Global monopoles}
Like for  cosmic strings also the energy density of global monopoles
produced by the Kibble mechanism obeys a scaling law. Therefore, they
are candidates for a model of structure formation. In contrast to local
monopoles,
the gradient energy of global monopoles introduces a long range interaction,
so that monopole anti--monopole pairs annihilate, leaving always only a few
per horizon \cite{BV,BR}.

One assumes, like in the CDM model, that the matter content of the universe
is dominated by  cold dark matter with $1=\Om\approx \Om_{CDM}$.
The large scale structure induced by global monopoles seems to look quite
similar to the texture scenario. Galaxy formation starts relatively early,
the galaxy correlation function and large scale velocity field are in
agreement with observations. The CMB fluctuations are similar
to those obtained for the texture scenario (see Chapter~5) \cite{BR92}.

Recent investigations \cite{PST} claim that all models with global defects
and CDM miss some large scale power on scales $l\ge 20$Mpc.

\chapter{Gauge Invariant Perturbation Theory}

For linear cosmological perturbation theory to apply, we must assume that the
spacetime manifold $\cal M$ with metric $g$ and energy momentum tensor $T$
of ``the real universe'' is somehow close to a Friedmann universe, i.e.,
the manifold $\cal M$ with a Robertson--Walker metric $\bar{g}$ and a
homogeneous and isotropic energy momentum tensor $\overline{T}$. It
is an interesting, non--trivial unsolved problem how to construct
$\bar{g}$ and $\overline{T}$ from the physical fields $g$ and $T$ in
practice.  There are two main
difficulties: Spatial averaging procedures depend on the choice of a
hyper--surface and do not commute with derivatives, so that the averaged
 fields $\bar{g}$ and $\overline{T}$
 will in general not satisfy Einstein's equations. Furthermore, averaging
is in practice impossible over super--horizon scales.

We now  assume that there exists an averaging procedure which
 leads to a Friedmann
universe with spatially averaged tensor fields $\overline{Q}$, such that
the deviations
$(T_{\mu\nu}-\overline{T}_{\mu\nu})/\max_{\{\al\b\}}\{|\overline{T}_{\al\b}|\}$
and
$(g_{\mu\nu}-\overline{g}_{\mu\nu})/\max_{\{\al\b\}}\{\overline{g}_{\al\b}\}$
  are small, and $\bar{g}$ and $\overline{T}$ satisfy Friedmann's equations.
Let us call such an averaging procedure 'admissible'.
There might also be another admissible averaging procedure (e.g. over a
different hyper--surface) leading to a slightly different Friedmann background
$(\bar{g}_2,\overline{T}_2)$.
In this case, the averaging procedures are isomorphic via an isomorphism
 $\phi$ on $\cal M$ which is close to unity:
\bean \bar{g}_2 &=& \phi_*\bar{g}     \\
 \overline{T}_2 &=& \phi_*\overline{T} ~, \eean
where $\phi_*(Q)$ denotes the pushforward of the tensor field $Q$ under $\phi$.
The isomorphism $\phi$ can be represented
as the infinitesimal flow of a vector field $X$,
$\phi = \phi_\ep^X$. Remember the definition of the flow: For the integral
curve $\ga_x(s)$ of $X$ with starting point $x$, i.e., $\ga_x(s=0)=x$
we have $\phi_s^X(x) = \ga_x(s)$. In terms of the vector field $X$, the
relation of the two averaging procedures is given by
\bea \bar{g}_2 &=& \phi_*\bar{g} = \bar{g} -\ep L_{\!X}\bar{g}
   + {\cal O}(\ep^2)  \label{2ggt} \\
 \overline{T}_2 &=& \phi_*\overline{T} = \overline{T} -\ep L_{\!X}
\overline{T}   + {\cal O}(\ep^2) ~. \eea
In the context of cosmological perturbation theory, the isomorphism
$\phi$ is called a gauge transformation. And the choice of a background
$(\bar{g},\overline{T})$ corresponds to a choice of gauge.
The above relation is of course true for all averaged tensor fields
$\overline{Q}$ and $\overline{Q}_2$.
Separating $Q$ into a background component and a small perturbation,
$ Q = \overline{Q} + \ep Q^{(1)}=  \overline{Q}_2 + \ep Q^{(2)}$ , we
obtain the following relation:
\be Q^{(2)} = Q^{(1)} +L_{\!X}\overline{Q}   \;. \label{2gt} \ee
Since each vector field $X$ generates a gauge
transformation $\phi = \phi_\ep^X$, we can conclude that only perturbations
 of tensor fields with  $ L_X\overline{Q}=0$ for all vector fields $X$,
i.e., with vanishing (or constant) 'background contribution' are gauge
invariant. This simple result is sometimes referred to as the
'Stewart Lemma' \cite{StW}.

The gauge dependence of perturbations has caused many controversies in the
literature, since it is often difficult to extract the physical
meaning of gauge dependent perturbations, especially on super--horizon
scales. This has led to the development of gauge invariant
perturbation theory
which we are going to present in this chapter. The advantage of the
gauge--invariant fromalism is that the variables used have simple geometric
and physical meanings and are not plagued by gauge modes.
Although the derivation requires somewhat more work, the final system
of perturbation variables is usually simple and well suited for numerical
treatment. We shall also see, that on subhorizon scales, the
gauge invariant matter perturbations variables approach the usual, gauge
dependent ones, and one of the geometrical variables approaches the
Newtonian potential, so that the Newtonian limit can very
easily be performed.

There are two review articles
on the subject \cite{KS,MFB}. Our treatment will be in some way
complementary. Collisionless particles, photon propagation and seeds
(Sections~2.3, 2.4, 2.5) are not discussed in these reviews. On the
other hand, we do not investigate perturbation theory of scalar
fields which is presented
extensively in the other publications, since it is needed to treat
fluctuations induced by inflation. We want to discuss perturbations
induced by topological defects. Here, the scalar field itself is a small
perturbation on a matter or radiation dominated background. This issue
will become more clear later.

First we note that since all  relativistic
equations are covariant (i.e. can be written in the form $Q=0$ for some
tensor field $Q$), it is always possible to express the corresponding
perturbation equations in terms of gauge invariant variables.

\section{Gauge Invariant Perturbation Variables}
In this section we introduce  gauge invariant variables which
describe the perturbations of the metric and the energy momentum tensor
in a Friedmann background.

There are two main approaches to find gauge invariant quantities:
 One possibility is to make full use of the above statement that tensor
fields with vanishing background contribution are gauge invariant and
use them  to define gauge invariant perturbation variables. Examples
are the Weyl tensor, the acceleration of the energy velocity field,
anisotropic stresses, shear and vorticity of the energy velocity field.
 This covariant approach was originally proposed by Hawking
[1966] and later extended by Ellis and Bruni [1989], Ellis et al. [1989],
Hwang and Vishniac [1989], Bruni et al. [1992b], Dunsby [1992] and others.

The other possibility is to arbitrarily parametrize the perturbations of
the metric, the energy momentum tensor, the distribution function, a
scalar field... and
discuss the transformation properties of these gauge dependent variables
under gauge transformations. One can then combine them into
gauge invariant quantities. This way was initiated by Gerlach and Sengupta
[1978] and Bardeen [1980]
and later continued by Kodama and Sasaki [1984],Kasai and Tomita [1986],
Durrer [1988], Durrer and Straumann [1988], Durrer [1989a], Durrer [1990],
Mukhanov et al. [1992] and others.
In this approach one usually performs a harmonic analysis and the
gauge invariant perturbation variables found in this way may be acausal
(e.g., they may require inverse Laplacians over spatial hyper--surfaces).

As in the second approach we  divide the perturbations into scalar,
vector and tensor contributions, but we do not perform the
harmonic analysis. Our perturbation variables are thus
space (and time) dependent functions and not just amplitudes of harmonics.
In order to obtain unique solutions, we require (for non--positive spatial
curvature) all the perturbation variables to vanish at infinity.
I  partly relate the two approaches by originally
performing the second, but in many cases identify gauge invariant variables
with tensor fields which vanish in an unperturbed Friedmann universe.
It could not be avoided to use a somewhat extensive vocubulary for all
the variables used in this text. To help the reader not to get lost,
I have included a glossary in Appendix~B.

Like in Chapter~1, the unperturbed line element is given by
\[ ds^2 = a^2(-dt^2 + \ga_{ij}dx^idx^j)~,\]
where $\bm{\ga}$ is the metric of a three space with constant curvature
$ k = \pm 1$ or 0 and overdot denotes derivatives w.r.t conformal time $t$.
 We first define scalar perturbations of the lapse function $\al$,
the shift vector $\bm{\beta}$ and the 3-metric $\bm{g}$  of the slices
of constant time\footnote{A short description of the 3+1 formalism of
general relativity is given in Appendix~A} by
\bea
\al &=& a(1+A)  \label{2alpha} \\
\bm{\beta} &=& \al B^{|i}\dd_i   \label{2beta}\\
\bm{g}     &=& a^2[(1+2H_L-(2/3)l^2\lap H_T)\ga_{ij} +
                  2l^2H_{T|ij}]dx^idx^j \;\; . \label{2g}  \eea
We denote 3--dimensional vector and tensor fields  by bold face letters;
$|$ and $\lap$ denote the covariant derivative and Laplacian
with respect to the metric $\bm{\ga}$. The variables
$A,\;B,\;H_L,$ and $H_T$ are arbitrary functions of space and time. To keep
them dimensionless, we have introduced a length $l$ which, in applications,
will be chosen to be the typical scale of the perturbations so that, e.g.,
${\cal O}(B) \approx {\cal O}(lB_{|i})$.

By choosing the metric perturbations in the form  (\ref{2alpha}--\ref{2g}),
we restrict ourselves to { scalar} type perturbations, but
we do not perform a harmonic analysis.
Vector perturbations of the geometry are of the form
\bea \beta &=& aB^i\dd_i  \label{2betav}\\
\bm{g} &=& a^2[\bm{\ga}_{ij} + lH_{i|j} + lH_{j|i}]dx^idx^j ~,
  \label{2gv}  \eea
where $B^i\dd_i$ and $H^i\dd_i$ are divergencefree vector fields which
vanish at infinity.

Tensor perturbations are given by
\be \bm{g} = a^2[\bm{\ga}_{ij} + 2H_{ij}]dx^idx^j ~. \ee
Here $H_{ij}$ is a traceless, divergencefree, symmetric tensor field on the
slices of constant time.

Writing the 4--dimensional metric in the form
\be g_{\mu\nu} = \bar{g}_{\mu\nu} + a^2h_{\mu\nu} \; , \ee
the above definitions of the perturbation variables for scalar perturbations
yield
\be h^{(S)} = -2A(dt)^2 + 2lB,_idtdx^i + 2(H_L - {l^2\over 3}\triangle H_T)
\ga_{ij}dx^idx^j
         +2 l^2H_{T|ij}dx^idx^j  ~ .  \label{2h} \ee
For vector and tensor perturbations one  obtains correspondingly
\bea h^{(V)} &=& 2B_idx^idt +   l(H_{i|j}+H_{j|i})dx^idx^j~~~ \mbox{ and}
  \label{2hv} \\
     h^{(T)} &=& 2H_{ij}dx^idx^j ~. \label{2ht}  \eea
  From  (\ref{2alpha}), (\ref{2beta}), (\ref{2g}), (\ref{2betav}) and
(\ref{2gv}), one can calculate the
3--dimensional Riemann scalar and the extrinsic curvature given by
$K_{ij}= -n_{i;j}$.
Palatini's identity (see Straumann [1985]) yields the general formula
\[ \de R(g+\de g)= (\de g^j_i)^{|i}_{|j} -\lap\de g^i_i ~.\]
This leads to the following result for scalar perturbations
\be \de\bm{R} =-4 a^{-2}(\lap+3k){\cal R} \mbox{ , \hspace{1cm} } {\cal R} =
     H_L - {l^2\over 3}\lap H_T  \label{2R} \; ,  \ee
\be  K^{(aniso)} = -[2\dot{a}l^2(H_{T|ij}-
  {1\over 3}\ga_{ij}\lap H_T)+
  a(l\si_{|ij} - {l\over 3}\ga_{ij}\lap\si)]dx^idx^j \mbox{ , \hspace{0.2cm}  }
    l\si = l^2\dot{H_T} - lB  ~ . \label{2sigma} \ee
For vector perturbations one obtains
\bea \de {\bf R}^V &=& 0 \\
 K^{V(aniso)} &=& -[\dot{a}(H_{i|j}+H_{j|i}) +
          {a\over 2}(\si_{i|j}+\si_{j|i})]dx^idx^j~~\mbox{ with}~~
  \si_i = l\dot{H}_i-B_i \label{2siv}~.\eea
$K^{(aniso)}$ is the traceless contribution to the extrinsic curvature
of the slices of constant time or, what amounts to the same thing, the shear
of the normal to the slices.

We now investigate the gauge transformation properties of the variables
defined above. We introduce the vector field
describing the gauge transformation by
\[  X =T\dd_t + L^i\dd_i   ~.\]
Using simple identities like $L_X(df) = d(L_Xf)$  and $(\bm{L_X\ga})_{ij} =
 \bm{X}_{i|j}+ \bm{X}_{j|i}$, we obtain the Lie derivative
of the unperturbed metric
\[ L_X\bar{g} = a^2[-2({\dot{a}\over a}T +\dot{T})dt^2 + 2(\dot{L}_i -
  T_{|i})dx^idt +(2{\dot{a}\over a}T\ga_{ij} +L_{i|j}+L_{j|i})dx^idx^j]~.\]
For scalar type gauge transformations, $L^i$ is of the form $L^i=lL^{|i}$
for some scalar function $L$. Inserting this above and comparing with the
parametrization of the perturbed metric $h^{(S)}$, eq. (\ref{2h}), we find
the following transformation laws:
\bea A &\ra& A +{\dot{a}\over a}T +\dot{T} \\
     lB &\ra& lB -T +l\dot{L} \\
     H_L &\ra& H_L +{\dot{a}\over a}T +(l/3)\lap L \\
     l^2H_T &\ra& l^2H_T +lL ~.\\   \eea
This yields the transformation properties for $\cal R$ and $\si$
\be {\cal R} \ra {\cal R} +{\dot{a}\over a}T~;~~~~~
  \si \ra \si+T ~.\label{2gtRsi} \ee
The following  scalar perturbation variables, the so called Bardeen potentials,
are thus gauge invariant (see \cite{Ba,KS} and \cite{DS}):
\be \Phi = {\cal R} - (\dot{a}/a)l\si   \label{2Phi} \ee
\be \Psi =  A  - (\dot{a}/a)l\si -l\dot{\si} \; . \label{2Psi} \ee
We shall later see that $\Psi$ is a relativistic analog of the
Newtonian potential.

For vector type gauge transformations, where $T=0$ and $L^i$ is a
divergence free vector field on the hyper--surfaces of constant time, one
immediately sees that $\si_i$ is gauge-invariant.

Tensor perturbations are of course always gauge invariant (there are no
tensor type  gauge transformations).

Gauge transformations remove two scalar and one vector
degrees of freedom, so that geometrical perturbations are fully characterized
by  six degrees of freedom which we parametrize in the gauge invariant
variables $\Psi,\Phi,\si_i$ and $H_{ij}$.

Some combinations of these quantities have  simple
geometric interpretations. From the Stewart Lemma, we know that the Weyl
tensor, which vanishes in an unperturbed Friedmann universe, is gauge
invariant.
A somewhat lengthy calculation  shows that the electric and
magnetic components of the Weyl tensor, are given by
\bea
  E_{ij} &\equiv& {1\over a^2}C_{i0j0} \nonumber \\
  &=& {1\over 2}[(\Phi-\Psi)_{|ij}-{1\over3}\lap(\Phi-\Psi)\ga_{ij}
   + \dot{\si}_{ij}]     \label{weylE} \\
  B_{ij} &\equiv& {-1\over 2a^2}\ep_{ilm}C_{lmj0} \nonumber \\
     &=& \ep_{(ilm}[\si_{j)m|l}-{1\over 3}\ga_{j)l}\si^s_{m|s}]
\label{weylB} \\
  \mbox{ where} && \si_{lm} = \si^{(V)}_{lm} + \dot{H}_{lm} ~~
	\mbox{and (i..j) denotes symmetrization in i and j}\nonumber
\eea
is the sum of vector and tensor contributions to the extrinsic curvature
(this result is obtained in Bruni et al. [1992b]).

We now  discuss perturbations of the energy momentum tensor.
We define the perturbed energy density $\rho$ and energy velocity field $u$
as the timelike eigenvalue and eigenvector of the energy momentum tensor
(note that, apart from symmetry, we do not make any assumptions on the
nature of $T_{\mu}^{\;\;\nu}$):
\[ T_{\mu}^{\;\;\nu}u^{\mu} = -\rho u^{\nu} \;\;,\;\; u^2 = -1 \; .\]
We then define the perturbations in the energy density and energy
velocity field by
\be \rho = \overline{\rho}(1+\de)  \;\;, \label{2de} \ee
\be u    = u^0\dd_t +u^i\dd_i ~;     \label{2v} \ee
$u^0$ is fixed by the normalization condition, $u^0=a^{-1}(1-A)$.
In the 3--space orthogonal to $u$ we define the stress tensor by
\be \tau^{\mu\nu} \equiv P^{\mu}_{~~\al}P^{\nu}_{~~\beta}T^{\al\beta}~,\ee
where $P= u\otimes u + g$ is the projection onto the sub--space of
$T_q{\cal M}$ normal to $u(q)$. One obtains
\[ \tau^0_0 = \tau^0_i =\tau^i_0 = 0 ~ .\]
The perturbations of pressure and anisotropic stresses can be parametrized by
\be \tau_i^{\;j} = \bar{p}[(1+\pi_L  )\de_i^{\;j} + \pi_i^{\;\;j} ]
  ~~\mbox{ , with }~ \pi^i_i = 0 \;. \label{2pi} \ee
Again, we decompose these perturbations into scalar, vector and tensor
contributions. For scalar perturbations one can set
\[    u^0 = (1-A) \;,\; {u^i\over u^0} = -lv^{,i} \; ~\mbox{ and } \pi^i_{~j} =
   l^2(\pi^{|i}_{~|j}-{1\over 3}\lap\pi\de^i_{~j})~.\]

As before, the Lie derivatives of the unperturbed quantities $\bar{\rho}~,
  ~~\bar{u} = a^{-1}\dd_t~, ~~~\bar{\tau}= \bar{p}\dd_i\otimes dx^i$ determine
the transformation laws of the perturbation variables. One finds
\bea L_X\bar{\rho} &=& T\dot{\bar{\rho}} = -3(1+w){\dot{a}\over a}T\bar{\rho}
	\label{2tr}\\
   L_X\bar{u} &=& [X,\bar{u}] = -a^{-1}[({\dot{a}\over a}T +
	\dot{T})\dd_t - L^i\dd_i] \\
\L_X\bar{\tau} &=& L_X\bar{p}\dd_i\otimes dx^i =
    T\dot{\bar{p}}\dd_i\otimes dx^i
  -3{c_s^2\over w}(1+w){\dot{a}\over a}T\bar{p}\dd_i\otimes dx^i ~,
	\label{2tt}\eea
where we have used the background energy equation,
\[ \dot{\bar{\rho}} = -3(1+w){\dot{a}\over a}\bar{\rho}~, \]
and the definitions
\[ w = \bar{p}/\bar{\rho}~, ~~~ c_s^2 = \dot{\bar{p}}/\dot{\bar{\rho}}~.\]
  From eqs. (\ref{2tr}) -- (\ref{2tt}) we obtain the transformation laws
\bean \de &\ra & \de -3(1+w){\dot{a}\over a}T \\
   lv &\ra & lv+ l\dot{L} \\
\pi_L &\ra & \pi_L -3{c_s^2\over w}(1+w){\dot{a}\over a}T \\
\pi  &\ra& \pi  ~.  \eean
A first gauge invariant variable is therefore
\[ \Pi = \pi ~,\]
the scalar potential for anisotropic stresses. The other gauge invariant
combination which can be constructed from matter variables alone is
\[   \Ga = \pi_L - (c_s^2/w)\de  ~.\]
 Defining an entropy flux $S^\mu$ of the perturbations in the sense of
small deviations from thermal equilibrium  \cite[ Appendix~B ]{StB},  one
finds for the entropy production rate of the perturbation \cite{DS}
\[ S^{\mu};_\mu = 3(\dot{a}/a^2T)\Ga ~, \]
where $T$ denotes the temperature of the system. The variable $\Ga$ thus
measures the entropy production rate.

We now split the covariant derivative of the velocity field in the usual way
into acceleration, expansion, shear and vorticity:
\[ u_{\mu;\nu} = a_\mu\otimes u_\nu + P_{\mu\nu}\th +\si_{\mu\nu} +
	\om_{\mu\nu} ~.\]
Here
\[ a = \nabla_uu ~~\mbox{ is the acceleration}, \]
\[ P_{\mu\nu} = g_{\mu\nu} +u_\mu u_\nu \]
 denotes the projection onto the sub--space of tangent space normal to $u$,
$\th =u^\mu_{;\mu}$ is the expansion,
\[ \si_{\mu\nu} = {1\over 2}P_\mu^\la P_\nu^\rho(u_{\la;\rho}+u_{\rho;\la})
	-P_{\mu\nu}\th \]
is the shear of the vector field $u$ and
\[ \om_{\mu\nu} = {1\over 2}P_\mu^\la P_\nu^\rho(u_{\la;\rho} - u_{\rho;\la})
\]
is the vorticity.
A short calculation shows $\om_{\mu\nu} = 0$, $\si_{0\mu}=0$ and
\be \si_{ij} = {a\over l}V_{|ij} ~~~\mbox{ with }~~~ lV = lv -l^2\dot{H}_T
	~ .   \label{2V}\ee
We choose $V$ as gauge invariant scalar velocity variable.
For the acceleration one obtains $a^0 = 0$ and
\[ a_i\equiv {\cal A},_i = \Psi,_i +\dot{V}_i + {\dot{a}\over a}V_i ~, \]
 which shows for the first time the analogy of $\Psi$ with the Newtonian
potential.

There are several different useful choices of gauge invariant density
perturbation variables:
\bea
  D_s &=& \de +3(1+w)(\dot{a}/a)l\si \\
  D_g &=& \de +3(1+w){\cal R} = D_s +3(1+w)\Phi\\
  D   &=& D_s +3(1+w)(\dot{a}/a)lV  \; .  \eea
For a physical interpretation of these variables
note that
\bea D,_i &=& P_i^\mu\rho,_\mu  \\
 (D_s)_{(ij)} +3(1+w)\Psi_{(ij)}&=& P_i^\mu P_j^\nu\rho_{;\mu\nu} \\
 (D_g)_{(ij)}+3(1+w)(\Psi-\Phi)_{(ij)} &=& P_i^\mu P_j^\nu\rho_{;\mu\nu} ~.\eea
Here we have set $ S_{(ij)} \equiv S_{|ij} -(1/3)\lap S\ga_{ij}$ for an
arbitrary scalar quantity $S$.\\
Therefore, $D$ and $D_s+3(1+w)\Psi$ are potentials for the first and second
``spatial derivatives'' of the energy density.

For vector perturbations only
\[ u^i = {1\over a}v^i  ~~ \mbox{ and }
   \pi^i_{~j}= {l\over 2}(\pi^i_{|j}+\pi_j^{~|i})  \]
survive.  Vector type gauge transformations yield the
transformation laws
\bean  v^i &\ra & v^i + \dot{L}^i \\
   \pi^i &\ra & \pi^i ~. \eean
In addition to the anisotropic stress potential
 $\pi^i\equiv \Pi^i$ , two  interesting
gauge invariant quantities are the shear and vorticity of the vector field
$u$:
\bea u_{i;j}+u_{j;i} = a(V_{i|j}+V_{j|i}) ~,~~ &\mbox{ with }&
   V^i = v^i-l\dot{H}^i ~,  \\
 u_{i;j}-u_{j;i} = a(\Om_{i|j}-\Om_{j|i}) ~, ~~ &\mbox{ with }&
   \Om^i = v^i-B^i ~. \eea
For tensor perturbations the only variable $\pi^i_{~j}\equiv\Pi^i_{~j}$
is of course gauge invariant.

 We now  show that for perturbations which are small
compared to the horizon distance, $l_H$, in a generic gauge the gauge invariant
combinations $V$ and $D_{(.)}$ approach the original $v$ and $\de$.
Let us 	choose our free length scale $l$ to be the typical
size of a given perturbation. From the above equation it is then
clear that for $l\ll l_H = t$, $D \approx D_s$.

Noting that perturbations of the Einstein tensor are given by second
derivatives of the metric perturbations (Palatini's identity, see e.g.
Straumann [1985]), we obtain the following order of magnitude equation:
\be {\cal O}(\frac{\de T}{T}){\cal O}(8\pi GT_{\mu\nu}) =
   {\cal O}(t^{-2}\frac{\de g}{g} +
   (lt)^{-1}\frac{\de g}{g} + l^{-2}\frac{\de g}{g}) \label{2O} \; .\ee
Using Friedmann's equation
\[ {\cal O}(8\pi GT_{\mu\nu}) = {\cal O}(\dot{a}/a)^2 = {\cal O}(1/t^2)
	\equiv  {\cal O}(1/l_H^2)  \]
this yields
\be {\cal O}(\frac{\de T}{T}) =
   {\cal O}(\frac{\de g}{g} + (l_H/l)\frac{\de g}{g} +
   (l_H/l)^2\frac{\de g}{g}) \; . \label{2ord}\ee
On sub--horizon scales, $l_H\gg l$ the metric perturbations are thus
generically much smaller than the matter perturbations and the difference
between the gauge invariant quantities $V$, $D_{(.)}$, $V^i$, $|Om^i$  and
$v$, $\de$, $v^i$ becomes  negligible.

\section{The Basic Equations}
 In this section we write down
the perturbation equations resulting from  Einstein's equation,
and energy momentum ``conservation''  in a form
which will be convenient later. All these equations are
most easily derived using the 3+1 formalism of gravity  (see  Appendix~A)
as we shall demonstrate for a few examples.

The perturbations of Einstein's equations and energy momentum conservation
can be expressed in terms of the gauge invariant variables defined above.
(A simple derivation of the equations for scalar perturbations is given in
Durrer and Straumann [1988].) To simplify the notation we now drop  the
bar over background density and pressure. \vspace{12pt} \\
{\bf A) Constraint equations}
\bea
  4\pi Ga^2\rho D &=& -(\lap +3k)\Phi  \label{2C1}  \\
  4\pi Ga^2(\rho +p)lV &=& (\dot{a}/a)\Psi -\dot{\Phi}  \label{2C2} \;
  .\\
  && \nonumber  \eea
{\bf B) Dynamical equations}
\bea  -8\pi Ga^2pl^2\lap\Pi &=& \lap(\Phi+\Psi)  \label{2D1}  \\
       8\pi Ga^2p(\Ga + (c_s^2/w)D_g -(2/3)l^2\lap\Pi) &=&
  (\dot{a}/a)\{\dot{\Psi}-[(1/a)(\frac{a^2\Phi}{\dot{a}})^{\cdot}]^
{\cdot}\} +
  \nonumber \\
 \{2a(\dot{a}/a^2)^.+3(\dot{a}/a^2)^2\}[\Psi-1/a(\frac{a^2\Phi}{\dot{a}})^
{\cdot}]
\label{2D2} \;\; . \eea

Since vector and tensor type perturbations are not treated in
Durrer and Straumann [1988] and Durrer [1990], we present an explicit
derivation of the vector perturbation equations, making use of the 3+1
formalism of general relativity (see Appendix~A and Durrer and
Straumann [1988]). For vector perturbations, the unit normal to
the equal time slices is given by
\[ n = \al^{-1}(\dd_t -\bm{\beta}) = a^{-1}(\dd_t -B^i\dd_i)~.\]
We now decompose the energy momentum tensor in the form
\[ T = \ep n\otimes n + n\otimes \bm{S} +\bm{S}\otimes n +\bm{T} \]
where, as before, bold type vector and tensor fields, $\bm S$ and $\bf T$,
are tangent to the equal
time hyper--surfaces. Using the Gauss--Codazzi--Mainardi formulas to express
the four dimensional curvature in terms of the three dimensional and the
second fundamental form, one can derive the following 3+1 split of Einsteins
equations (Appendix~A4):
\bea \bm{R} + (tr\bm{K})^2 - tr(\bm{K}^2) &=& 16\pi G\ep \\
 \bm{\nabla}\cd\bm{K} - \bm{\nabla}tr(\bm{K}) &=& 8\pi G\bm{S} \label{2c}\\
\dd_t\bm{K} -\bm{L_\beta K} +{\bf Hess}(\al) - && \nonumber \\
 -\al[ {\bf Ricci} -2\bm{K\cd K}
  +(tr\bm{K})\bm{K}] &=& -4\pi G\al [2\bm{T} +\bm{g}(\ep -tr\bm{T})]
  ~~, \label{2dy}  \eea
 where $ \bm{K} = {1\over 2\al}(\bm{L_\beta g}-\dd_t\bm{g})$
	 is the second fundamental form.

For vector perturbations only the second constraint equation and the
traceless part of the dynamical equation contribute. From our definiton
of vector perturbations of the energy momentum tensor, one finds
\[\bm{S} = a(\rho+p)(v^i-B^i)\dd_i ~. \]
In what follows we use the notation $X^{(V)}_{ij}$ for the symmetric,
traceless tensor constructed  from  the vector field $X$, i.e.,
$X^{(V)}_{ij} = {1\over 2}(X_{i|j}+X_{j|i})$. We then have
\[ \bm{g} = a^2(\ga_{ij} + 2lH^{(V)}_{ij})dx^idx^j \]
and with (\ref{2siv}) the second fundamental form is given by
\[ \bm{K} = -a[{\dot{a}\over a}(\ga_{ij}+2lH^{(V)}_{ij})+\si^{(V)}_{ij}
   ]dx^idx^j~.\]
This leads to
\[ \bm{\nabla\cd K} = -{1\over 2a}(\lap\si^i +\si_j^{~|ij})\dd_i = -{1\over 2a}
    (\lap+2k)\si^i\dd_i \]
The constraint equation (\ref{2c}) thus results in
\be -{1\over 2}(\lap +2k)\si^i = 8\pi G(\rho+p)a^2\Om^i ~.\label{2Om}\ee

Let us now proceed to the dynamical equation. Up to first order we obtain
the following expressions for the terms in (\ref{2dy}):
\bean  \bm{L_\beta K} & =& -2\dot{a}B^{(V)}_{ij}dx^idx^j \\
\bm{K}^2 &=& [({\dot{a}\over a})^2(\ga_{ij}+2lH^{(V)}_{ij}) +
   2{\dot{a}\over a}\si^{(V)}_{ij}]dx^idx^j  \\
\bm{Ricci(g)} &=& 2k(\ga_{ij}+2lH^{(V)}_{ij})dx^idx^j  \\
8\pi G\bm{T} &=& 8\pi Ga^2p(\ga_{ij} +2lH_{ij}^{(V)} +l\pi^{(V)}_{ij})
   dx^idx^j\\
 4\pi G \bm{g}(\ep-tr\bm{T}) &=& 4\pi G(\rho-3p)a^2
   (\ga_{ij}+2lH^{(V)}_{ij})dx^idx^j  \\
 \dd_t\bm{K} &=& -[\ddot{a}(\ga_{ij}+2lH^{(V)}_{ij}) +\dot{a}
  (\si_{ij}^{(V)}+2l\dot{H}^{(V)}_{ij})+a\dot{\si}^{(V)}_{ij}]dx^idx^j  \eean
The result for $\bm{Ricci(g)}$ is again easily derived using Palatini's
identity \cite{StB}.

With the help of the background relation
\[ 4\pi Ga^2(\rho-p) = {\ddot{a}\over a} +({\dot{a}\over a})^2 + 2k~,\]
equation (\ref{2dy}) then yields
\[ \dot{\si}^{(V)}_{ij} + 2({\dot{a}\over a})\si^{(V)}_{ij} =
   8\pi Ga^2pl\Pi^{(V)}_{ij}~.\]
Since we require $\lim_{r\ra\infty}\si_i = \lim_{r\ra\infty}\Pi_i = 0$,
 the  tensor fields, $\si^{(V)}_{ij}$
and  $\pi^{(V)}_{ij}$ uniquely determine the corresponding vector fields:
\bea \dot{\si}_{i} + 2({\dot{a}\over a})\si_{i} &=&
   8\pi Ga^2pl\Pi_{i}~, \label{2dyv}   \\
   \mbox{or }~~(a^2\si_{i})^.  &=&
   8\pi Ga^4pl\Pi_{i}~.
\eea
In the absence of anisotropic stresses, vector anisotropies in the extrinsic
curvature thus decay like $1/a^2$.

In a similar way one finds the tensor perturbation equation
\be \ddot{H}_{ij} + 2{\dot{a}\over a}\dot{H}_{ij} + (2k-\lap)H_{ij} =
    8\pi Ga^2p\Pi_{ij}~. \label{2dyt} \ee
This is  a wave equation with source term. It describes the
creation,  propagation and damping  of gravitational waves in a Friedmann
background \vspace{12pt}.\\
{\bf C) Conservation equations}

The energy and the momentum conservation equations of each independent
type of matter (i.e. each matter component which does not interact other
than gravitationally with the rest)  yields the
following  equations of motion for
the scalar perturbation variables $D_\al$ and $V_\al$, where the index $ _\al$
denotes the different matter components (e.g. radiation, dark matter,
baryons ...):
\bea
  \dot{D_{\al}} -3w_{\al}(\dot{a}/a)D_{\al}  &=&
 (\lap +3k)[(1+w_{\al})lV_{\al} +2(\dot{a}/a)w_{\al}l^2\Pi_{\al}]
  \nonumber   \\ &&+
  3(1+w_{\al})4\pi Ga^2(\rho +p)(lV-lV_{\al})
  \; , \label{2con1}\\
l\dot{V}_{\al} +(\dot{a}/a)lV_{\al} &=& \frac{c^2_{\al}}{1+w_{\al}}D_{\al} +
\frac{w_{\al}}{1+w_{\al}}\Ga_{\al} + \Psi + 2/3(\lap + 3k)
\frac{w_{\al}}{1+w_{\al}}l^2\Pi_{\al}  \; . \label{2con2}  \eea

The total perturbations are defined as the sums:
\be \rho D = \sum_{\al}\rho_{\al}D_{\al}  \;\;, \;\;\;
 (\rho +p)V = \sum_{\al}
  (\rho_{\al} + p_{\al})V_{\al} ~. \ee
 The adiabatic sound speed, $c_{\al}$ and
enthalpy, $w_{\al}$ are
\[ c_{\al}^2 = \dot{p_{\al}}/\dot{\rho_{\al}} \;\;\;,\;\;\;\;\;
  w_{\al}= p_{\al}/\rho_{\al} \; \mbox{ and }\;\;
  w = \frac{\sum_{\al}p_{\al}}{\sum_{\al}\rho_{\al}}\equiv p/\rho  \;\; ,
\;\;\ c_s^2 = \dot{p}/\dot{\rho} . \]
The corresponding equations for interacting matter components are derived
in \\ \cite{KS}.

We shall later also use equations (\ref{2con1}) and (\ref{2con2}) for a one
component fluid in terms of the density perturbation variable $D_g$ and $V$.
One easily finds
\be
\dot{D}_g +3(c_s^2-w)(\dot{a}/a)D_g -(1+w)l\lap V +3w(\dot{a}/a)\Ga = 0
	\label{2cog1}\ee
\be l\dot{V} +(\dot{a}/a)(1-3c_s^2)lV = (\Psi-3c_s^2\Phi)
 	+{c_s^2\over 1+w}D_g +{w\over 1+w}[\Ga +
  	{2\over 3}(\lap+3k)l^2\Pi]
		\label{2cog2}\ee

\vspace{10pt}

Again we derive the conservation equation for for vector perturbations in
some detail. We start with the spatial part of the 3+1 split of \\
 $T^{\mu\nu}_{~~;\nu} = 0$ (see Appendix~A.3):
\be {1\over\al}(\dd_t-\bm{L_\beta})\bm{S} = -\bm{\nabla}(\ln\al)\ep + 2\bm{KS}
  +(tr\bm{K})\bm{S} -{1\over\al}\bm{\nabla}\cd(\al\bm{T})~. \label{2cv}\ee
All the terms in this equation are readily calculated. One obtains for
each matter component (suppressing the index $\al$)
\bean
\bm{S} &=& (\rho+p)a^{-1}(v^i -B^i)\dd_i \\
 {1\over\al}\bm{\nabla}(\al\bm{T}) &=& \bm{\nabla\cd T} =
   {1\over 2a^2}p(\lap l\Pi^i +2kl\Pi^i)\dd_i \\
2\bm{KS} &=& -2{\dot{a}\over a^3}(\rho+p)(v^i-B^i)\dd_i  \\
(\tr\bm{K})\bm{S} &=& -3{\dot{a}\over a^3}(\rho+p)(v^i-B^i)\dd_i \\
\bm{L_\beta S} &=& 0~~~ \mbox{ (in first order) }\\
\dd_t\bm{S} &=& [-\dot{a}a^{-2}(\rho+p)(v^i-B^i) + a^{-1}(\dot{\rho}+\dot{p})
   (v^i-B^i)
  +a^{-1}(\rho+p)(\dot{v}^i-\dot{B}^i)]\dd^i \\
 &=&  a^{-1}(\rho+p)[ \dot{v}_i-\dot{B}_i-(4+3c_s^2){\dot{a}\over a}(v_i-B_i)]
   \dd_i ~. \eean
 For the last equality sign, we made use of the background equation
$ \dot{\rho}+\dot{p} = -3(\dot{a}/a)(1+c_s^2)(\rho+p)$. Inserting all these
results, eq. (\ref{2cv}) becomes
\be  \dot{\Om}^i + (1-3c_s^2){\dot{a}\over a} \Om^i = -{p\over 2(\rho+p)}
  (\lap+2k)l\Pi^i   \label{2dom} ~.\ee
 If there are no sources present, $\Pi^i = 0$ , and if $c_s^2=w=$ constant,
the  amplitude of the vorticity is proportional to  $a^{3c_s^2-1}$ .
In comparison to the expansion velocity, $\dot{a}/a$, the vorticity behaves
like
\be |\Om|/{\dot{a}\over a} \propto a^{0.5(9w-1)} \label{2Omt}\ee
(as long as curvature is negligible, i.e. for $ k\ll 1/t^2$).
Especially, an initial vorticity in a radiation dominated universe $(w=1/3)$
grows relative to the expansion velocity in the course of expansion.

In addition to Einstein's field equation and the conservation equations
which are of course a consequence of them, we have to add matter equations
to fully describe the system. If the fluid description is justified, these
can be given in the form
\[ \Ga = \Ga(D,V)~, ~~~~ \Pi = \Pi(D,V)~. \]
In Section~3.1 we discuss, for illustration,  the simplest possibility,
adiabatic perturbations of an ideal fluid, where we just set $\Pi=\Ga=0$

There are however situations where the description of matter as a fluid is not
sufficient.
One then has to resort to the matter equations of more fundamental
quantities, e.g.  scalar fields and/or  gauge fields.

\vspace{12pt}

\section{Collisonless Matter}

In this section we discuss another approximate description of matter
 which can be used for  weakly interacting particles like photons in
 a recombined universe or massive neutrinos which might constitute the dark
matter. Here, the basic quantity is the one--particle distribution function
 $f$ which lives on the mass bundle,
  \[P_m = \{(p,x)\in T\mm|~g(\!x\!)(p,p) = -m^2\}~ . \]
When collisions can be neglected, the matter equation is the one particle
Liouville equation (for a thorough treatment of the kinetic approach in
general relativity see Stewart [1971] and references therein). Choosing
coordinates
$(x^\mu,p^i)$ on $P_m$  (where $p=p^i\dd_i + p^0(x^\mu,p^i)\dd_0$),
Liouville's equation reads
\[ (p^\mu\dd_\mu -\Ga^i_{\al\b}p^\al p^\b{\dd\over \dd p^i})f =0 ~.\]
In an unperturbed Friedmann universe, this equation  is equivalent to
the requirement that the distribution function, $\bar{f}$ is a function of
the redshift corrected momentum, $ v := a\sqrt{\bar{g}_{ij}p^ip^j}$ alone.

\subsection{A gauge invariant variable for perturbations of the
	distribution function}
We want to split the distribution function $f$ in a perturbed Friedmann
universe
into a background component and a perturbation. We cannot do this directly
since the background  distribution function, $\bar{f}$ is not defined on
$P_m$ but on the
background mass bundle $\bar{P}_m = \{(p,x)|~\bar{g}(\!x\!)(p,p) = -m^2\}$ .
Therefore, to split $f$, we first have to choose an isomorphism
$\iota : P_m \ra \bar{P}_m : (x,p) \mapsto (x,\bar{p})$. Then we can define
the perturbation according to
\[ f = (\bar{f} + F)\circ\iota ~.\]
It is clear that the perturbation $F$ in general depends on the isomorphism
$\iota$ which deviates in first order from identity. Choosing two basis
$(e_\mu)$ and $(\bar{e}_\mu)$ which are tetrads with respect to $g$ and
$\bar{g}$, we may set for $p=p^\mu\dd_\mu = \pi^\mu e_\mu$
\[ \iota(x,\pi^\mu e_\mu) = (x,\pi^\mu\bar{e}_\mu) ~.\]
On the other hand, every isomorphism $\iota$ is of this form
(i.e. $\bar{p}(\pi^\mu e_\mu) = \pi^\mu\bar{e}_\mu$) for suitably chosen
tetrads $(e_\mu)$ and $(\bar{e}_\mu)$. The tetrad $e_\mu$ can be
defined by
\be e_\mu = \bar{e}_\mu + R_\mu^{~\nu}\bar{e}_\nu  ~, \label{2tet}\ee
where the symmetrical part of $R$ is determined by the orthogonality
condition:
\[ R_{\mu\nu} + R_{\nu\mu} = -a^2 h(\bar{e}_\mu,\bar{e}_\nu) ~ ~~
    (g= \bar{g}+a^2h )~.\]

To determine how $F$ transforms under gauge transformations, we
consider a vector field $X$ which defines a gauge transformation.
The flow of $X$ on $\cal M$, $\phi_s^X$, induces the flow $T\phi_s^X$ on
$T{\cal M}$. The natural lift, $TX$, of $X$ to $T{\cal M}$ is defined by
$T\phi_s^X = \phi_s^{TX}$. A short calculation shows that for $X=X^\mu\dd_\mu$
\[ TX = (X^\mu\dd_\mu, X^\mu,_\nu p^\nu{\dd\over\dd p^\mu}) \]
for a coordinate basis $(x^\mu, p=p^\mu\dd_\mu)$.
For the full distribution function this leads to the transformation law
\be [f:P_m\ra {\bf R}]~ \longrightarrow [(T\phi)_*f : T\phi(P_m)\ra {\bf R}]
	~.\ee
In linearized form this yields
\[ f \ra f + L_{TX}f ~.\]

We now want to use our split of $f$ to obtain a transformation law for
$F$.
The first problem to note here is that
$\bar{f}$ is not defined on all of $T\mm$ but only on $\bar{P}_m$,
 and since $TX$ is in general not tangent to $\bar{P}_m$, $L_{TX}\bar{f}$
is not  well defined. But of course it is possible to extend the
definition of $\bar{f}$ to an open sub--set of $T\mm$ containing $\bar{P}_m$.
We thus do not have to bother about this technical point. We just keep
in mind that the gauge transformation properties of $F$ should not
depend on such an extension.

More important  is that $F$  also depends on the
choice of the isomorphism $\iota$, the transformation properties of which
have also to be taken into account. We
 now choose two admissible splittings  of $f$  given by
\[ f = \bar{f}_1\circ\io_1 +F_1 = \bar{f}_2\circ \io_2 + F_2 ~. \]
Then, there exists a gauge
transformation given by a vector field $X$, such that
$ \bar{f}_2 = \bar{f}_1 - L_{TX}\bar{f}_1$, and therefore
\[ \bar{f}_1\circ \io_1 + F_1 = \bar{f}_1\circ \io_2 + F_2 -
                                                 L_{TX}\bar{f}_1 ~.\]

The change of $F$ under a gauge transformation is thus given by
\[ F_2-F_1 = \bar{f}\circ\io_1 -\bar{f}\circ\io_2 + L_{TX}\bar{f} \equiv
	L_{(TX)_\|} F_{1}~, \]
where we have dropped the index $1$ on $\bar{f}$.

If $\io_{1,2}$ are specified by the tetrads $e_{(1,2)\mu}$ which are related
to the background tetrad $\bar{e}_\mu$  according
to eq. (\ref{2tet}) with matrices $R^{(1,2)\nu}_\mu$, we obtain
\be \bar{f}\circ\io_1 - \bar{f}\circ\io_2 = {\dd\bar{f}\over\dd \pi^\nu}
  \pi^\mu (R^{(2)\nu}_\mu - R^{(1)\nu}_\mu) \equiv - (TX)_\perp\bar{f}~,
 \label{2TXperp} \ee
\[ \mbox{ where we have introduced}~~~ (TX)_\perp =
 (R^{(2)\nu}_\mu - R^{(1)\nu}_\mu)\pi^\mu{\dd\over \dd\pi^\nu}~.\]

For  $ p =p^\mu\dd_\mu = \pi^\mu e_\mu $ we find
\[ (TX)_\| = TX - (TX)_\perp = X^\mu\dd_\mu + X^\mu,_\nu p^\nu
	{\dd\over\dd p^\mu} - (R^{(2)\nu}_\mu - R^{(1)\nu}_\mu)
					\pi^\mu{\dd\over \dd\pi^\nu}~.\]

It is easy to see that $(TX)_\|(\bar{g}(p,p)) =0$ which shows that $(TX)_\|$ is
parallel to $\bar{P}_m$, i.e., $(TX)_\|\bar{f}$ does not depend on the
extension of $\bar{f}$ which is necessary to make $(TX)\bar{f}$ and
$(TX)_\perp\bar{f}$ well defined.

We now explicitly calculate $(TX)_\|$ for $X=T\dd_t + L^i\dd_i$ and a
tetrad $(e_\mu)$ which is adapted to the splitting of spacetime into
$\{t=\mbox{const.}\}$ hyper--surfaces , $\Si_t$:
\[ e_0 = n ~, ~ \bar{e}_0 =\bar{n} = a^{-1}\dd_t~,~~ g(\bm{e}_i,n)=0 ~.\]
The triad $(\bm{e}_i)$ is an orthogonal basis on the slices $\Si_t$. We
furthermore
use the fact that there exist coordinates such that the metric of a space
of constant curvature $k$   is given by
\[ \ga_{ij} = (1+{k\over 4}r^2)^{-2}\de_{ij} \equiv \la^2\de_{ij} ~.\]
In this coordinates we can choose $\bar{e}_i = (a\la)^{-1}\dd_i$.
According to our definition of the perturbed lapse function, shift vector
and three metric, we then have
\bean
 e_0 &=& \al^{-1}(\dd_t -\bm{\beta}) = (1-A)\bar{e}_0 -B^i\la\bar{e}_i\\
 \bm{e}_i &=& (1-H_L)\bar{e}_i - H_{i}^{~~j}\bar{e}_j ~,\eean
where the indices of the perturbation variables are, as usual raised and
lowered with the metric $\bm{\ga}$ and $H_i^j$ is traceless, but may contain
scalar, vector and tensor contributions. Using the gauge transformation
properties of these variables and the definition of $(TX)_\perp$ we obtain
\bean (TX)_\perp &=& [(\dot{a}/a)T +\dot{T}]\pi^0{\dd\over\dd\pi^0}
  -\la(T^{|i} - \dot{L}^i)\pi^0{\dd\over\dd\pi^i} \\
 && + [(\dot{a}/a)
  -T\de_i^{~j} + 1/2(L_i^{~|j} +L^j_{~|i})]\pi^i{\dd\over\dd\pi^j} ~. \eean
Inserting the expression for $TX$ we find
\bean (TX)_\| &=& T\dd_t + L^i\dd_i +T,_ip^i{\dd\over\dd p^0} + L^i,_jp^j
   {\dd\over\dd p^i} + \dot{L}^ip^0{\dd\over\dd p^i} -
  (\dot{a}/a)T\pi^0{\dd\over\dd\pi^0} \\ &&+\la T'^i\pi^0
   {\dd\over\dd\pi^i} - \la\dot{L}^i\pi^o{\dd\over\dd\pi^i}
	+ 1/2(L_i^{~|j} +L^j_{~|i})]\pi^i{\dd\over\dd\pi^j}
   + ({\dot{a}\over a}T\pi^i{\dd\over \dd\pi^i}~. \eean
We now use that $\bar{f}$ is only a function of
\[ v = a\sqrt{\bm{g}(\bm{p},\bm{p})} = a\sqrt{\sum_i\pi_i\pi_i} =
	a^2\la\sqrt{\sum_ip_i p_i} ~.\]
 Denoting the direction cosines of the momentum by $\ep^i$ we obtain
($ \ep^i = p^i/\sqrt{\sum_ip_ip_i} = \pi^i/\sqrt{\sum_i\pi_i\pi_i}$),
\bean {\dd\bar{f}\over\dd p^i} &=& a^2\la\ep^i{d\bar{f}\over dv} \\
 {\dd\bar{f}\over\dd \pi^i} &=& a\ep^i{d\bar{f}\over dv} \\
 \left({\dd\bar{f}\over\dd t}\right)_{p^i} &=&
				 2(\dot{a}/a)v{d\bar{f}\over dv} \\
 \left({\dd\bar{f}\over\dd x^j}\right)_{p^i} &=&
			(\la,_j/\la)v{d\bar{f}\over dv} \\
  p^j{\dd\bar{f}\over\dd p^i} &=& \pi^j{\dd\bar{f}\over\dd\pi^i} =
	\ep^j\ep^iv{d\bar{f}\over dv} \\
  {\dd\bar{f}\over\dd p^0} &=&   {\dd\bar{f}\over\dd \pi^0} = 0 ~. \eean
Furthermore,
\[ 1/2((L_i^{~|j} +L^j_{~|i})\ep^i\ep^jv{d\bar{f}\over dv} =
	(L^j,_i\ep^i\ep^j + (\la,_j/\la)L^j)v{d\bar{f}\over dv}~. \]
 Thus, the terms containing $L^i$ in $(TX)_\|\bar{f}$ cancel.
Introducing $q=\la\sqrt{v^2+m^2a^2}$ we finally obtain
\[ (TX)_\|\bar{f} = [v(\dot{a}/a)T  + q\ep^iT,_i]{d\bar{f}\over dv}~.\]
This  leads to the transformation law
\be F \ra F {d\bar{f}\over dv}[v(\dot{a}/a) + q\ep^i\dd_i]T
						~. \label{2traF}\ee
We first note that $F$ is invariant under vector type gauge transformations.
Comparing (\ref{2traF}) with the transformation properties of $\cal R$ and
$\si$ given in (\ref{2gtRsi}) we find the following gauge invariant
combination:

\be \FF = F -[v{\cal R} +ql\ep^i\dd_i\si]{d\bar{f}\over dv}~.\label{2FF}\ee

\subsection{The perturbation of Liouville's equation}

Choosing an arbitrary basis of vector fields $(e_\mu)$ and corresponding
momentum coordinates, $p = \pi^\mu e_\mu$ on $T\mm$, Liouville's equation is
given by
\be L_{X_g}f = \pi^\mu e_\mu(f) - \om^i_{~\mu}(p)\pi^\mu{\dd f\over\dd\pi^i}
      =0  ~ . \label{2Li1}	\ee
If we
select, as above, a tetrad adapted to the slicing of $\mm$ into slices $\Si_t$
of constant time,
\[ e_0 = n = \al^{-1}(\dd_t - \bm{\beta})~~,~~~~  \bm{e}_i \in T\Si_t ~, \]
we find (see Appendix A5)
\bean \lefteqn{X_g = {\pi^0\over\al}(\dd_t -\bm{\beta}) +\bm{p} } \\
 && -[\bm{\om}^i_{~j}
	(\bm{p}-{\pi^0\over \al}\bm{\beta})\pi^j +(\pi^0)^2{\al^{|i}\over\al}
	-\al^{-1}(\beta_j^{|i}-c^i_{~j})\pi^0\pi^j]{\dd \over\dd\pi^i} ~,\eean
 where $\bm{p}=\pi^i\bm{e}_i$ is the component of $p$ tangent to the slices
and $c^i_{~j}$ is defined by
\[ \dd_t\bm{\vth}^i = c^i_{~j}\bm{\vth}^j ~. \]
The $(\bm{\vth}^i)$ are the basis of one forms dual to the vector fields
$(\bm{e}^i)$ on $\Si_t$. More details are found in Appendix~A.
 We now rewrite $X_gf$ in terms of the  variables \\
$t,~\bm{x}$, $v = a\sqrt{\pi_i\pi^i}$, $\ep^i = a\pi^i/v = \ep_i$  and
$ q=a\la\pi^0$. For this we use the following easily established identities:
\bean
\bm{\beta} &=& B^i\dd_i = a\la B^i\bm{e}_i \\
\al &=& a(1+A)~,~~ \al^{-1} =a^{-1}(1-A) \\
c^i_{~j} &=& (\dot{a}/a)\de^i_j +\dot{H}_L\de^i_j +\dot{H}_{Tj}^i \\
\beta^j_{|i} &=&\beta^{|i}_j = \bm{e}_i(\beta^j) + \bar{\bm{\om}}^i_{~l}
   (\bm{e}_j)\beta^l = B^j,_i +{k\la\over 2}
	(B^ix^j-B^jx^i -B^lx^l\de^j_i)~.\eean
Since the background distribution function only depends on $v$, i.e.,
$ {\dd\over\dd\pi^i}\bar{f}= a\ep^i{d\bar{f}\over dv}$,  only
the symmetrical part of $\beta_j^{~|j}$ contributes to the Liouville equation
in first order and the term $\bm{\om}^i_{~j}(\bm{\beta})$ gives no
contribution in this
approximation. With the help of the splitting $f = (\bar{f}+F)\circ\io$ and
the background Liouville equation for $\bar{f}$ we finally obtain
\[ q\dd_tF + v\ep^i\dd_iF -vk\la/2(x^i-x^j\ep^j\ep^i){\dd F\over\dd\ep^i} =
  [q^2\ep^iA,_i -qv\ep^i\ep_j(B^j_{|i} -\dot{H}_{i}^j) +v^2H_L]{df\over dv}
	~.\]
For the gauge invariant tensor and vector contributions to $F$ this yields,
setting $F^{(T)}\equiv \FF^{(T)}$ and  $F^{(V)}\equiv \FF^{(V)}$
\bea q\dd_t\FF^{(T)} +v\ep^i\dd_i\FF^{(T)} -v\Ga^i_{jk}\ep^j\ep^k
   {\dd\FF^{(T)}\over \dd\ep^i} &=& qv\ep^i\ep_j\dot{H}^i_j{d\bar{f}\over dv}
	\label{2liouT} \\
 q\dd_t\FF^{(V)} +v\ep^i\dd_i\FF^{(V)} -v\Ga^i_{jk}\ep^j\ep^k
 {\dd\FF^{(V)}\over \dd\ep^i} &=& qv\ep^i\ep_j\si^{(V)i}_{|j}{d\bar{f}\over dv}
	\label{2liouV}	~.\eea

 For scalar perturbations the situation is somewhat more complicated. Since
 the scalar contribution $F^{(S)}$ to $F$ is not gauge invariant  we
want to express the Liouville equation in terms of the gauge invariant
combination
\[ \FF^{(S)} = F^{(S)} -[v{\cal R} + ql\ep^i\si,_i]{d\bar{f}\over dv} ~.\]
After carefully calculating $\dd_t[v{\cal R} + ql\ep^i\si,_i]$,
$\dd_i[v{\cal R} + ql\ep^i\si,_i]$ and
${\dd\over \dd\ep^i}[v{\cal R} + ql\ep^i\si,_i]$, we finally obtain the
Liouville equation for scalar perturbations in a Friedmann universe \cite{d90}:

\be q\dd_t\FF^{(S)} + v\ep^i\dd_i\FF^{(S)} -v\bm{\Ga}^k_{ij}\ep^i\ep^j
	{\dd\FF^{(S)}\over\dd\ep^k}=
 (q^2\dd_i\Psi -v^2\dd_i\Phi)\ep^i{d\bar{f}\over dv} ~.\label{2liouS} \ee

\subsection{ Momentum Integrals}

To connect this equation of motion to Einstein's field equations, we
calculate the energy momentum tensor from $f$, which is given by
\be T^{\mu\nu} = \int_{P_m(x)}p^\mu p^\nu f\mu(p) ~, \ee
where $\mu$ is an invariant measure on $P_m(x)$ (for a general definition see
Stewart [1971]). With respect to a tetrad $p=\pi^\nu e_\nu$,  $\mu$ looks
like in special relativity
\be \mu(p) = {d\pi^1\wedge d\pi^2\wedge d\pi^3\over|\pi^0|} =
	{\la v^2\over a^2q}dvd\Om ~,\ee
where we have used the definition of $v$ and $q$, and $d\Om$ is the usual
surface element on the 2-sphere integrating over the momentum directions
$\bm{\ep}$. Let us, as an example, consider $T^0_0$:
\[ T^0_0 = \bar{\rho}(1+\de) = {\la\over a^2}\int {p^0p_0v^2\over q}
	(\bar{f}(v) +F)dvd\Om ~.\]
Expressing $p^0$ and $p_0$ in terms of $v$ and $q$ and separating into a
background and  first order contribution yields
\[ \bar{\rho} = {4\pi\over\la a^4}\int v^2q\bar{f}dv~~,~~~
   \rho\de =  {1\over\la a^4}\int v^2qFdvd\Om   ~. \]
Using this we obtain
\[ {1\over\rho\la a^4}\int v^2q\FF^{(S)}dvd\Om = \de -
	{{\cal R}\over\rho\la a^4}
	\int v^3q{d\bar{f}\over dv}dvd\Om   ~. \]
After a partial integration, inserting ${dq\over dv}= \la^2v/q$ and
\[ \bar{p} = {4\pi\la\over 3a^4}\int(v^4/q)\bar{f}dv \]
we end up with
\be  {1\over\rho\la a^4}\int v^2q\FF^{(S)}dvd\Om = \de + 3(1+w){\cal R} = D_g
   \label{2Dint}	~.\ee
For the last equality sign, the definition of the gauge invariant density
perturbation variable $D_g$ is inserted.

In a similar way all the other momentum integrals are obtained:
\bea
l\lap V &=& \frac{-1}{a^4(\rho+p)}\int v^3\ep^i\dd_i\FF^{(S)} dvd\Om
          \label{2Vint}   \\
l^2(\Pi_{|ij}-1/3\ga_{ij}\lap\Pi) &=&{\la\over a^4p}\int{v^4\over q}
	(\ep^i\ep^j - (1/3)\de_{ij})  \FF^{(S)} dvd\Om    \label{2Pint} \\
\Ga &=& \frac{1}{a^4p}\int ({v^4\la\over 3q} - {c_s^2v^2q\over \la})
	\FF^{(S)} dvd\Om \; .  \eea
For the vector and tensor perturbations one finds
\bea
V^{(V)i} &=& \frac{-1}{a^4(\rho+p)}\int v^3\ep^i\FF^{(V)} dvd\Om
          \label{2VintV}   \\
(l/2)(\Pi^{(V)}_{i|j}- \Pi^{(V)}_{j|i}) &=&{\la\over a^4p}\int{v^4\over q}
	(\ep^i\ep^j - (1/3)\de_{ij})  \FF^{(V)} dvd\Om    \label{2PintV} \\
\Pi^{(T)}_{ij} &=&{\la\over a^4p}\int{v^4\over q}
	(\ep^i\ep^j - (1/3)\de_{ij})  \FF^{(T)} dvd\Om   ~.  \label{2PintT}
\eea
 These matter variables inserted in Einstein's equations (\ref{2C1})
(\ref{2D1}), (\ref{2Om}) and (\ref{2dyt}) yield the geometrical perturbations
$\Psi$,  $\Phi$, $\si^i$ and $H^{(T)}_{ij}$ which
enter in (\ref{2liouS},\ref{2liouV},\ref{2liouT}). In Section~5, we
discuss how this
 closed system is altered in the presence of  seeds.

\subsection{The ultrarelativistic limit}

Here we briefly investigate the special case of extremely relativistic
particles for which we can set $m=0$. Since curvature only may play a role
in the late, matter dominated stages of the universe, we neglect it here,
$k=0, \la=1$, so that $q=v$. (The generalization to $k\neq 0$
is straight forward.) In the extremely relativistic  case all the integrals
above contain the energy integral $\cd\int v^3{\cal F}dvd\Om$. Therefore,
 it makes sense to introduce the perturbation of the energy integrated
distribution function, the brightness:
\be \mm \equiv {4\pi\over\bar{\rho}a^4}\int_0^\infty{\cal F}v^3dv ~,
  \label{2M} \ee
which is  a function of the momentum directions $\ep^i$ only.
Defining
\[ \io = {4\pi\over\bar{\rho} a^4}\int_0^\infty Fv^3dv ~, \]
one finds, using (\ref{2FF}) and the gauge invariance of $F^{(V)}$ and
$F^{(T)}$
\bea  \mm^{(S)} &=& \io^{(S)} +4{\cal R} +4l\ep^i\dd_i\si  \label{2Mio}\\
	 \mm^{(V)} &=& \io^{(V)} \\
	 \mm^{(T)} &=& \io^{(T)} ~.\eea
In the case where \mm~ describes thermal radiation, we may
interpret the perturbation in the distribution function as a perturbation
of the temperature:
\be f = \bar{f}\left({\bm{g}(\bm{p,p})^{1/2}\over T(x^\mu,\bm{\ep})}\right) =
      \bar{f}\left({v\over aT(x^\mu,\bm{\ep})}\right)~~  \mbox{ with }
  ~~  T(x^\mu,\bm{\ep}) = \bar{T}(t) + \de T(x^\mu,\bm{\ep})  ~.\ee
Inserting this form of $f$ one  obtains $ F  = -v{d\bar{f}\over dv}\cd
{\de T\over T}$. The integral (\ref{2M}) then yields
\bean {1\over 4}\mm^{(S)} &=& {\de T^{(S)}\over\overline{T}} + {\cal R} +
	l\ep^i\si_i \\
{1\over 4}\mm^{(V)} &=& {\de T^{(V)}\over\overline{T}} \\
{1\over 4}\mm^{(T)} &=& {\de T^{(T)}\over\overline{T}} ~.\eean
Therefore, $(1/4)\mm$ can be interpreted as  a gauge invariant variable
for the temperature perturbation.\footnote{Note that even though $\FF$, $\mm$
and $\de T$ are scalar functions they do in general contain vector and tensor
perturbations, since they depend not only on position but also on momentum
or momentum direction. They may contain terms of the form $\al^i\ep_i$ or
$\tau^{ij}\ep_i\ep_j$ where $\al$ is a divergence free vector field and
$\tau$ is a traceless, divergencefree tensor field. These are the type of
contributions which we indicate with $^{(V)}$ and $^{(T)}$.}

In terms of \mm~ the perturbation equations (\ref{2liouS},\ref{2liouV},
\ref{2liouT})
become (for $k = 0$)
\bea  \dot{\mm}^{(S)} + \ep^i\dd_i\mm^{(S)} = 4\ep^i\dd_i(\Phi-\Psi)
   \label{2li0S}  \\
\dot{\mm}^{(V)} + \ep^i\dd_i\mm^{(V)} = -4\ep^i\ep_j\si^{(V)i}_{|j}
   \label{2li0V}  \\
\dot{\mm}^{(T)} + \ep^i\dd_i\mm^{(T)} = -4\ep^i\ep^j\dot{H}_{ij}
  ~.  \label{2li0T}  \eea
 The evolution of the distribution of massless particles
only depends on the Weyl part of the curvature. This is geometrically
very reasonable since null geodesics are conformally invariant.

By similar calculations like in the preceding paragraph, one finds the
perturbations of the energy
momentum tensor for extremely relativistic particles
\bea
D_g  &=& \frac{1}{4\pi}\int \mm d\Om  \label{2Dint0} \\
l\lap V &=& \frac{-3}{16\pi}\int \ep^i\dd_i\mm d\Om
          \label{2Vint0}   \\
l^2(\Pi_{|ij}-1/3\ep_{ij}\lap\Pi)  &=& \frac{3}{4\pi}\int(\ep^i\ep^j -
    {1\over 3}\de_{ij})      \mm d\Om  \label{2Pint0}     \\
\Ga &=& 0 \\
V^{(V)i} &=& \frac{1}{4\pi}\int \ep^i\mm^{(V)}d\Om
          \label{2V0V}   \\
(l/2)(\Pi^{(V)}_{i|j}- \Pi^{(V)}_{j|i}) &=&{3\over 4\pi}\int
	(\ep^i\ep^j - (1/3)\de_{ij})  \mm^{(V)} d\Om    \label{2P0V} \\
\Pi^{(T)}_{ij} &=&{3\over 4\pi}\int(\ep^i\ep^j - (1/3)\de_{ij})
	\mm^{(T)} dvd\Om    \label{2P0T} ~.\eea
We  use  Liouville's equation for massless particles for the
numerical calculation of perturbations of the cosmic microwave background
in Chapter~5. In Section~3.2 we use (\ref{2li0S}) to derive the
Boltzmann equation for photons in an electron proton plasma.

\section{The Propagation of Photons in a Perturbed Friedmann Universe}

On their way from the last scattering surface into our antennas,
the microwave photons travel through a perturbed Friedmann geometry.
Thus, even if the photon temperature was completely uniform at the
last scattering surface, we would receive it slightly perturbed  \cite{SW}.
In addition, a photon
traveling through a perturbed universe is in general deflected. In
this section we  calculate both these effects
in first order perturbation theory. I  present the calculation
rather explicitly, since I haven't found a complete gauge invariant
treatment of this problem anywhere in the literature.
For sake of simplicity we  restrict ourselves to $k=0$.

As already mentioned, two metrics  which are conformally equivalent,
\[ d\tilde{s}^2 = a^2ds^2 \; , \]
have the same lightlike geodesics, only the corresponding affine parameters
are different. We may thus discuss the propagation of light in a
 perturbed Minkowski geometry.  This simplifies things greatly. We denote
the affine parameters by $\tilde{\la}$ and $\la$
respectively and the tangent vectors to the  geodesic by
\[ n = \frac{dx}{d\la} \; \mbox{ and } \tilde{n} =
\frac{dx}{d\tilde{\la}} \;\;,
\;\;\; n^2 = \tilde{n}^2 = 0 \;, \]
with unperturbed values $  n^0 =1$ and $\bm{n}^2 =1$.
If the tangent vector of the perturbed geodesic is $ (1,\bm{n})  +\de n$,
the geodesic equation for the metric
\[ ds^2 = (\eta_{\al\nu}+h_{\al\nu})dx^{\al}dx^{\nu}  \]
yields to first order
\be \de n^\mu |_i^f = -\eta^{\mu\nu}[h_{\nu 0} + h_{\nu i}n^i]_i^f
+{\eta^{\mu\si}\over 2}
   \int_i^fh_{\rho\nu,\si}n^{\rho}n^{\nu}d\la  \; , \label{2deltan} \ee
where the integral is along the unperturbed photon trajectory and the
unperturbed values for $n^\mu$ can be inserted.
Starting from this general relation, let us
first discuss the photon redshift.
The ratio of the energy of a photon measured by some observer at $t_f$
to the energy emitted at $t_i$ is given by
\be E_f/E_i = \frac{(\tilde{n}\cdot u)_f}{(\tilde{n}\cdot u)_i} = (T_f/T_i)
     \frac{(n\cdot u)_f}{(n\cdot u)_i}  \; , \label{Ef/Ei} \ee
where $u_f$ and $u_i$ are the four velocities of the observer and the emitter
respectively and the factor $T_f/T_i$ is the usual redshift which relates
 $n$ and $\tilde{n}$. We write $T_f/T_i$ and not $a_f/a_i$ here, since also
this redshift is slightly perturbed in general, and we want $a$ to
denote the unperturbed background expansion factor.

Since this is a physical, intrinsically defined quantity it is
independent of coordinates. It must thus be possible to write it
in terms of gauge invariant variables. We now calculate the gauge
invariant expression for $E_f/E_i$.
The observer and emitter are comoving with the cosmic fluid.
We have
\[ u = (1-A)\dd_t -lv^{,i}\dd_i \; . \]
Furthermore, since the photon density $\rho^{(r)}\propto T^4$ may itself
be perturbed
\[ T_f/T_i = (a_i/a_f)(1+\frac{\de T_f}{T_f} - \frac{\de T_i}{T_i}) =
    a_i/a_f(1 + (1/4)\de^{(r)}|_i^f) \; ,  \]
where $\de^{(r)}$ is the intrinsic density perturbation in the radiation.
This term was neglected in the original analysis of Sachs and Wolfe,
but since it is gauge dependent, doing so violates gauge invariance.
We therefore have to include $\de^{(r)}$ to obtain a gauge invariant
expression.
Inserting all this  and (\ref{2deltan}) into (\ref{Ef/Ei})  yields
\be E_f/E_i = (a_i/a_f)[1+ n^jv,_j|_i^f + A|_i^f +(1/4)\de^{(r)}|_i^f
    - 1/2\int_i^f\dot{h}_{\mu\nu}n^{\mu}n^{\mu}d\la]  \;. \label{2z} \ee
 With the help of equation (\ref{2h}) for the definition of $h_{\mu\nu}$
one finds for {\bf scalar} perturbations after several integrations by part
\bea (E_f/E_i)^{(S)} &=& (a_i/a_f)\{1+[(1/4)D^{(r)}_s + l V_{|j}^{(m)}n^j +
   \Psi]|_i^f
   - \int_i^f(\dot{\Phi}-\dot{\Psi})d\la\}   \label{2deltaE}  \\
&=& (a_i/a_f)\{1+[(1/4)D^{(r)}_g + l V_{|j}^{(m)}n^j +
   \Psi-\Phi]_i^f
   - \int_i^f(\dot{\Phi}-\dot{\Psi})d\la\} \label{2dE} \; . \eea
Here $D_s^{(r)},~D_g^{(r)}$ denote the gauge invariant density perturbation
in the radiation field and
$V^{(m)}$ is the peculiar velocity of the matter component (the emitter
and observer of radiation).  From the second of these equations one sees
explicitly that the geometrical part of the perturbation of the photon
redshift depends  on the Weyl curvature  only (specialize eq.
(\ref{weylE}) to purely scalar perturbations), i.e., is conformally
invariant.

For a discussion of the Sachs--Wolfe effect alone we neglect the intrinsic
density perturbation of the radiation, i.e., we set $D_g^{(r)} = 0$,
which now is a gauge invariant statement (but a bad approximation in
many circumstances like, e.g. for adiabatic CDM perturbations). $V^{(m)}$
is a Doppler term due to the relative motion of  emitter and
receiver. The $\Psi-\Phi$ -- term accounts for the redshift due to the
difference of the gravitational field at the place of the emitter and
receiver and the integral is a path dependent contribution to
the redshift.

For {\bf vector} perturbations $\de^{(r)}$ and $A$ vanish and eq. (\ref{2z})
leads to
\be (E_f/E_i)^{(V)} = (a_i/a_f)[1 - V_j^{(m)}n^j|_i^f +
  \int_i^f\dot{\si}_jn^jd\la]   ~. \label{2dev} \ee
Again we obtain a Doppler term and a gravitational contribution.
For {\bf tensor} perturbations, i.e. gravitational waves,  only the
gravitational part remains:
\be (E_f/E_i)^{(T)} = (a_i/a_f)[1 -   \int_i^f\dot{H}_{lj}n^ln^jd\la]
   ~. \label{2det} \ee
Equations (\ref{2deltaE}), (\ref{2dev}) and (\ref{2det}) are the
manifestly gauge invariant results for the Sachs--Wolfe effect for
scalar vector and tensor perturbations.

 In addition to  redshift, photons in a perturbed Friedmann universe
also experiences  deflection. We now calculate this effect.

The direction of the light ray with respect to a comoving observer is given by
the direction of the spacelike vector
\[ n_{(3)} =  n + (un)u     \]
which lives on the sub--space of tangent space normal to $u$. Let us
also define the vector field $n^{||}_{(3)}$, which coincides with
$n_{(3)}$ initially (i.e. at $t_i$) and is  Fermi transported
along $u$, i.e.
\be \nabla_un^{||}_{(3)} =(n^{||}_{(3)}\nabla_uu)u ~.\label{2Fer} \ee
Note, that we have to require Fermi transport and not parallel
transport since $u$ is in general not a geodesic and therefore
$(un^{||}_{(3)} )=0$ is not conserved under parallel transport!
(For an explanation of Fermi transport, see e.g. Straumann [1985]).
$n^{||}_{(3)} = (0,\bm{n}) + \de n^{||}_{(3)}$,  where
$ \de n^{||}_{(3)}$ is determined by the Fermi transport equation
(\ref{2Fer}).  Since the observer Fermi transports her frame of
reference with respect to which angles are measured, she would
consider the light ray as not being deflected if
 $n^{||}_{(3)}(t_f)$ is parallel to $ n_{(3)}(t_f)$. The difference between
the direction of these two vectors is thus the light deflection:
\be \varphi e = \left[ n_{(3)} - {(n^{||}_{(3)}\cd n_{(3)})\over
    (n^{||}_{(3)}\cd n^{||}_{(3)})}n^{||}_{(3)}\right](t_f)
   ~. \label{2gendef}\ee
Here $e$ is a spacelike unit vector normal to $u$ and normal to
$n^{||}_{(3)}$ which determines the direction of the deflection
and $\varphi$ is the deflection angle. (Note that (\ref{2gendef}) is the
general formula for light deflection in an arbitrary gravitational field.
Up to this point we did not make any assumptions about the strength of
the field.) For a spherically symmetric
problem, as we shall encounter when discussing the collapsing
texture, $e$ is uniquely determined by the above conditions since
the path of a light ray is confined to the plane normal to the angular
momentum. In the general case, when angular momentum is not conserved
$e$ still has one degree of freedom. We now calculate $\varphi e$
perturbatively.
Let us recall and define the perturbed quantities:
\bean
 n &=& (1, \bm{n}) + \de n  ~~\mbox{ with }~~ \bm{n}^2 = 1  \\
 u     &=& (1,0) + \de u ~=~ (1+{1\over 2}h_{00}, \bm{v}) \\
 n_{(3)} &=& (0,\bm{n}) + \de n_{(3)}   ~~~\mbox{ and}\\
 n^{||}_{(3)} &=& (0,\bm{n}) + \de n^{||}_{(3)} ~.\eean
The perturbation $\de n$ is given in (\ref{2deltan}). Furthermore, we obtain
\be \de n_{(3)} = \ep u + \de n -\de u ~, \label{2den3} \ee
with
\[ \ep = [n^iv^i -\de n^0 +{1\over 2}h_{00}+n^ih_{i0}] \]
The Fermi transport equation leads to
\bea \de (n^{||}_{(3)})^0 &=& n^i(h_{i0}+v_i) \\
  \de (n^{||}_{(3)})^j &=&  -{1\over 2}[h_{lj}n^l )|_i^f+
    \int_i^fdt(h_{j0,l} - h_{l0,j})n^l \label{2denp}~.\eea
So that $(n^{||}_{(3)})^2 = 1$ and
\[ \varphi e = \de n_{(3)} - (n^{||}_{(3)}\cd n_{(3)})n^{||}_{(3)}=
   \ep u+ \de n -\de u -\de n^{||}_{(3)} -[h_{ji}n^in^j+n^i
    (\de (n^{||}_{(3)})^i+\de n^i -v^i)]\bm{n}~.\]
Inserting (\ref{2den3}--\ref{2denp}) into (\ref{2gendef}) we find
\bea  \varphi e^0 &=& 0 \\
  \varphi e^i &=& \de^i-(\bm{\de\cd n})n^i~, ~~\mbox{ with} \\
    \de_j &=& [\de n_j -v_j +{1\over 2} h_{jk}n^k]|_i^ f+{1\over 2}
   \int_i^f(h_{0j,k}-h_{0k,j})n^kdt ~. \label{2dej} \eea
This quantitiy is observable and thus gauge invariant. For scalar
 perturbations one finds (after integrations by parts)
\be
(\de_j)^{(S)} = V_{,j}|_i^f +\int_i^f(\Phi-\Psi)_{,j}d\la
 \label{2defs} ~.\ee
For spherically symmetric perturbations, where $e$ is uniquely defined,
we can write this result in the form
\be \varphi = V_{,i}e^i|_i^f +\int_i^f(\Phi-\Psi)_{,i}e^id\la ~.
 \label{2ldefl} \ee
The first term here denotes the special relativistic spherical aberration.
 The second term represents the gravitational light deflection. Here again
one sees that gravitational light deflection, which of course is
conformally invariant, only depends on the Weyl part of the curvature.
As an  easy test we insert the Schwarzschild weak field approximation:
$\Psi = -\Phi = -{GM\over r}$. The unperturbed geodesic is given by
$ x = (\la, \bm{n}\la + \bm{e}d)$, where $d$ denotes the impact parameter
of the photon. Inserting this into (\ref{2ldefl}) yields
Einstein's well known result
\[ \varphi = {4GM\over d} ~.\]

For vector perturbations we obtain from (\ref{2dej})
\be (\de_j)^{(V)} = \Om_{j}|_i^f -{1\over 2}[ \int_i^f
   (\si_{j,k}-\si_{k,j})n^kdt +\int_i^f\si_{k,j}n^kd\la]  ~.
\label{2defv} \ee
This result can be expressed in three dimensional notation  as follows:
\bean \lefteqn{\varphi\bm{ e}= \bm{\de}-(\bm{\de\cd n})\bm{n} =}  \\
 && -(\bm{\Om\wedge n})\wedge \bm{n}|_i^f + {1\over 2}\int_i^f
  (\bm{\nabla}\wedge\bm{\si})\wedge\bm{n}dt - \int_i^f
 (\bm{\nabla}(\bm{\si\cd n})\wedge\bm{n})\wedge\bm{n}d\la ~. \eean
The first term is again a special relativistic  "frame dragging" effect.
The second term is the change of frame due to the gravitational
field along the path of the observer and the third term gives the
gravitational light deflection. This formula
could be used to obtain in first order the light deflection in the vicinity of
a rotating neutron star or a Kerr black hole.  The special relativistic Thomas
precession is not recovered with this formula since it is of order $v^2$.

For tensor perturbations we find
\[ (\de_j)^{(T)} = -H_{jk}n^k|_i^f +\int_i^fH_{lk,j}n^ln^kd\la
  ~,\]
or, after an integration by parts,
\be (\varphi e_j)^{(T)} = -H_{jk}n^k|_i^f +\int_i^f
  (H_{lk,j} + \dot{H}_{kl}n_j)n^ln^kd\la ~.\label{2deft} \ee
 Only the gravitational effects remain. The first
 contribution comes from the difference of the metric before and
after the passage of the gravitational wave. Usually this term is
negligible. The second term accumulates along the path of the photon.
\vspace{1cm}

\section{Gauge Invariant Perturbation Theory in the Presence of Seeds}
In this section we add an inhomogeneous term to the perturbation equations.
Perturbations can then be generated even starting from an initially
unperturbed spacetime. Seeds produce  this inhomogeneous term in a
natural way. By seeds we mean density
perturbations originating from an inhomogeneously
distributed form of energy whose mean density is much smaller than the
density of the Friedmann background. We assume that, once they are
produced, these
seeds do not interact with the rest of the matter other than gravitationally.

\subsection{The energy momentum tensor}

Since the energy momentum tensor of the seeds, $T^{\mu\nu}_{(s)}$, has no
homogeneous background contribution, it is gauge invariant by itself
according to (\ref{2gt}).

$T^{\mu\nu}_{(s)}$  can be calculated by solving the matter equations for
the seeds in the Friedmann {\em background} geometry. (Since $T^{\mu\nu}_{(s)}$
has no background component it satisfies the unperturbed matter and
``conservation'' equations.) We again decompose
$T^{\mu\nu}_{(s)}$ into scalar, vector and tensor contributions.
They decouple within linear perturbation theory and it is thus  possible
to write  the equations for each of these contributions separately.
However, this decomposition is acausal. It
requires $T_{\mu\nu}(t,\bm{x})$ at a given time $t$ to be known for all
positions \bm{x} and not only within a causally connected region.
We  just ignore this problem now and, nevertheless, work with this
decomposition. We parametrize the scalar $(S)$ vector $(V)$ and tensor
$(T)$ contributions to $T^{\mu\nu}_{(s)}$ in the form
\bea  T_{00}^{(sS)} &=& a^2\rho^{(s)} = (M^2/l^2)f_{\rho}
     \label{3seed00} \\
     T_{i0}^{(sS)}  &=& a^2lv_{|i}^{(s)} = (M^2/l)f_{v\; |i}
     \label{3seed0i} \\
   T_{ij}^{(sS)}    &=& a^2[(p^{(s)} - (l^2/3)\lap\Pi^{(s)}) \ga_{ij}
                       +l^2\Pi^{(s)}_{|ij}]   \nonumber \\
    &=& M^2[(f_p/l^2 -(1/3)\lap f_{\pi})\ga_{ij} + f_{\pi\; |ij}]
  \label{3seedij}\\
     T_{i0}^{(sV)}  &=& a^2v_i^{(s)} = (M^2/l^2)w^{(v)}_i       \\
   T_{ij}^{(sV)}    &=& {a^2l\over 2}[\Pi^{(s)}_{i|j}+\Pi^{(s)}_{j|i}]
    = (M^2/2l)(w^{(\pi)}_{i|j}+w^{(\pi)}_{j|i}) \\
   T_{ij}^{(sT)}  &=&  a^2\Pi^{(s)}_{ij}
                = (M^2/l^2)\tau^{(\pi)}_{ij}~.
  \eea
As before, $l$ is introduced merely to keep the functions $f.$, the
vector fields
 $w^.$ and the tensor field $\tau^\pi$  dimensionless.
It may be chosen as a typical size of the seeds. $M$ denotes a
typical mass of the seeds. (It is of course possible to choose $l=M^{-1}$.)

If we are given the full energy momentum tensor $T^{(s)}_{\mu\nu}$ which may
 contain scalar, vector and tensor contributions, the scalar parts
$f_v$ and $f_{\pi}$ are in general determined by the non--local identities
\[ T^{(s)\; |j}_{0j} = (M^2/l)\lap f_v  ~, \]
\[ (T^{(s)}_{ij} - 1/3\ga_{ij}\ga^{kl}T_{kl})^{|ij} =
   \frac{2}{3}M^2(\lap + 3k)\lap f_{\pi}   \; . \]
On the other hand $\lap f_v$ and $\lap(\lap +3k)f_{\pi}$ are also
determined in terms of $f_{\rho}$ and $f_p$ by the ``conservation'' equations:

\be \dot{f}_{\rho} - l\lap f_v + (\dot{a}/a)(f_{\rho} +3f_p) = 0 ~, \ee
\be -l\dot{f}_v -2 (\dot{a}/a)lf_v + f_p +(2/3)l^2(\lap+3k)f_{\pi} = 0 ~.
 \label{3f}  \ee
Once $f_v$ is known it is  easy to get $w^v_i = l^2/M^2(
 T_{0i}) - lf_{v,i}$. To obtain $w^\pi_i$ we use
\[ T_{ij}^{(s)|j}- T_{ij}^{(sS)|j} = {M^2\over l}(\lap+2k)w^{(\pi)}_i~.\]
Again  $w^{(\pi)}_i$ can also be obtained in terms of  $w^{(v)}_i$ with
the help of the ``conservation'' euation:
\be \dot{w^{(v)}_i} +2({\dot{a}\over a})w^{(v)}_i -
	{l\over 2}(\lap+2k)w^{(\pi)}_i = 0 ~.\ee

\subsection{Perturbation equations}
 The energy momentum tensor of the seeds is determined by the unperturbed
equations of motion. The gravitational interaction with the perturbations
of  other components does not contribute to first order.
We assume non--gravitational interactions with other components can be
neglected. This is certainly a good approximation soon after the phase
transition and thus can only affect the initial conditions. (Situations
where non--gravitational interactions must not be neglected are discussed
by Magueijo [1992].) The geometrical
perturbations can then be separated into a part induced by the seeds
and a part caused by the perturbations in the remaining matter
components:
\bean \Psi &=& \Psi_s + \Psi_m \;\;\mbox{ , }\;\; \Phi = \Phi_s + \Phi_m
 \\
 \si_i &=& \si_i^{(s)}  + \si_i^{(m)}  \\
  H_{ij} &=& H_{ij}^{(s)}   + H_{ij}^{(m)} ~. \eean
Using Einstein's equations, we can  calculate the geometry perturbations
induced by the seeds:
\bea  -(\lap +3k)\Phi_s &=& \ep( f_{\rho}/l^2 + 3(\dot{a}/a)f_v/l)
        \label{2S1} \\
      (\dot{a}/a)\Psi_s -\dot{\Phi}_s &=& \ep f_v/l  \label{2S2} \\
  \lap(\Phi_s +\Psi_s) &=& -2\ep \lap f_{\pi}  \label{2S3}  \\
  (\dot{a}/a)\{\dot{\Psi}_s-[(1/a)(\frac{a^2\Phi_s}
 {\dot{a}})^{\cdot}]^{\cdot}\} + &&
  \nonumber  \\
  \{2a(\dot{a}/a^2)^{\cdot} +3(\dot{a}/a^2)^2\}[\Psi_s - 1/a(\frac{a^2\Phi_s}
  {\dot{a}})^{\cdot}] &=& 2\ep(f_p/l^2 - (2/3)\lap f_{\pi})
    \label{2S4} \\
 (\lap +2k)\si^{(s)}_i &=& 2\ep w^{(v)}_i/l^2  \\
 2{\dot{a}\over a}\si^{(s)}_i+\dot{\si}^{(s)}_i &=& -2\ep w^{(\pi)}_i/l  \\
\ddot{H}^{(s)}_{ij} +{\dot{a}\over a}\dot{H}^{(s)}_{ij} +
  (2k-\lap)H^{(s)}_{ij} &=&
  2\ep\tau^{(\pi)}_{ij}/l^2 ~. \eea
We  assume that $\ep \equiv 4\pi GM^2$ is  much smaller than 1, so that
 linear perturbation analysis is justified. \\
As before, the part of the geometrical perturbations induced by the matter
are determined by equations (\ref{2C1}) to (\ref{2D1}). But
in the conservation equations and in any matter equations the full
 perturbations, $\Psi$, $\Phi$, $\si_i$ and $H_{ij}$ have
to be inserted.

We now discuss the example of a single fluid with only scalar
perturbations where $\Pi$ and $\Ga$ are given
in terms of $D$ and $V$.
We assume that in addition to the seeds we have one perturbed
matter component which we indicate by a subscript $ m$. Other components
which contribute to the background, but whose perturbations can be
neglected, may also be present. The conservation equation (\ref{2con1})
then reads
\be
  \dot{D_{m}} -3w_{m}(\dot{a}/a)D_{m}  =
 (\lap +3k)[(1+w_{m})lV_{m} +2(\dot{a}/a)w_{m}l^2\Pi_{m}]
-  3(1+w_{m})\ep f_v/l \; . \label{3con1}\ee
The last term describes the influence of the seeds.

Solving this equation for $(\lap +3k)lV_{m}$ and inserting the result
and its time derivative into (\ref{2con2}) yields a second order equation
for $D_m$. Using
\[ (\lap + 3k)\Psi = 4\pi G\rho_m a^2(D_{m} -2w_{m}l^2(\lap +3k)\Pi_{m})
  +\ep(f_{\rho}/l^2 +3{\dot{a}\over a}f_v/l- 2(\lap +3k)f_{\pi}) \]
and the conservation equation (\ref{3f}) we find

\bea   &&
\ddot{D} -(\lap +3k)c_s^2D + (1+3c_s^2 -6w)(\dot{a}/a)\dot{D} -
\{3w(\ddot{a}/a)
 -9(\dot{a}/a)^2(c_s^2-w) + \nonumber  \\
&& + (1+w)4\pi G\rho a^2\}D \;\; =  \nonumber \\
&& (\lap +3k)w\Ga + 2(\dot{a}/a)wl^2(\lap+3k)\dot{\Pi}  \nonumber \\
&& + \{ 2(\ddot{a}/a)w
 -6(\dot{a}/a)^2(c_s^2-w)+ (1+w)8\pi Ga^2p +2/3(\lap+3k)w\}l^2(\lap+3k)\Pi
  \nonumber \\
&& + (1+w)\ep(f_{\rho} +3f_p)/l^2    \; , \label{2DD}  \eea
where we have dropped the subscript $m$. \\
This equation describes the behavior of density perturbations in the presence
of seeds in an arbitrary Friedmann background.  We have not used
Friedmann's equations to express $\ddot{a}/a$ in terms of $w$ and $\dot{a}/a$,
or $\rho$ in terms of $(\dot{a}/a)^2$ so that (\ref{2DD}) is valid also if
 there are unperturbed components which contribute to the expansion
 but not to
the perturbation. Note that within this gauge invariant treatment the
source term is, up to a factor $(1+w)$, just the naively expected term
$4\pi Ga^2(\rho^{(s)} +3p^{(s)})$ for all types of fluids.

We now simplify equation (\ref{2DD}) in the case where $\Pi = \Ga=0$
(adiabatic perturbations and no anisotropic stresses) and $k =0$:
If one chooses a realistic density parameter $ 0.2\le \Om_0 \le 2$,
the curvature term  can always be neglected
at early times, e.g., for redshifts $z \ge 5$. It is of the order
$(\max(l,l_H)/l_k)^2$
as compared to the other contributions. (Here $l$ and $l_k = k^{-1/2}$
denote the typical
size of the perturbation and the radius of curvature, respectively.)
Under these assumptions, eq. (\ref{2DD}) becomes

\bea \ddot{D} - c_s^2\lap D + (1+3c_s^2-6w)(\dot{a}/a)\dot{D} - &&\nonumber \\
 3[w(\ddot{a}/a) -3(\dot{a}/a)^2(c_s^2-w) + (1+w)(4\pi/3)G\rho a^2]D
&=& S \; ,\label{2D}\eea
where $ S = (1+w)\ep(f_{\rho}+3f_p)/l^2$ .

 We Fourier transform (\ref{2D}) (and denote the Fourier transform of $D$
with the same letter):
\bea \ddot{D} +k^2 c_s^2D + (1+3c_s^2-6w)(\dot{a}/a)\dot{D}&&\nonumber \\
 - 3[w(\ddot{a}/a) -3(\dot{a}/a)^2(c_s^2-w) + (1+w)(4\pi/3)G\rho a^2]D &
  =& \tilde{S} \; .\label{2DF}\eea
$\tilde{S} = (1+w)\ep(\tilde{f_{\rho}}+ 3\tilde{f_p})/l^2$ is the
Fourier transform of $S$.

  From the homogeneous solutions $D_1$ and $D_2$ of (\ref{2DF}),
we can find the perturbation induced by $S$ with the Wronskian method:
\be D = c_1D_1 + c_2D_2  ~;\ee
\be c_1 = -\int(\tilde{S}D_2/W)dt \;\;\mbox{ , }\;\;
    c_2 =  \int(\tilde{S}D_1/W)dt  \;\; , \ee
where $W =D_1\dot{D_2} - \dot{D_1}D_2$ is the Wronskian determinant of
the homogeneous solution.\newline
This leads to the following general behavior: If the time
dependence of $D_1$, $D_2$ and
$\tilde{S}$ can be approximated by power laws, $D$
 behaves like $D \propto \tilde{S}$ as long as $\tilde{S}\ne 0$.
If $D_1$ and $D_2$
are waves with approximately constant amplitude and frequency $\om$,
$D$ can be approximated by a wave with amplitude proportional to
 $\om^{-1}\int e^{i\om t}\tilde{S}dt$. Thus,
only typical frequencies of the source finally survive.

As a second example, we consider collisionless particles (again only scalar
perturbations and $k=0$ are considered). The source term on the
r.h.s. of Liouville's equation (\ref{2liouS}) can be separated
as above into a part
due to the collisionless component and a part induced by the seeds.
Equation (\ref{2liouS}) then becomes
\be  ( q\dd_t + v^k\dd_k )
 {\cal F} = \frac{d\bar{f}}{dv}[(q/v)v^k\dd_k\Psi_m - (v/q)v^k\dd_k\Phi_m]
   + {\cal S}\; , \label{3liou} \ee
with
 \be {\cal S} = \frac{d\bar{f}}{dv}[(q/v)v^k\dd_k\Psi_s - (v/q)v^k\dd_k\Phi_s]
 \; . \label{3source} \ee
In the same way, one obtains for massless particles
\be \dd_t\mm +\ga^i\dd_i\mm = 4\ga^i\dd_i(\Phi_m-\Psi_m)
     + {\cal S} ~, \label{3lioum}\ee
with corresponding source term
\[ {\cal S} = 4\ga^i\dd_i(\Phi_s-\Psi_s) ~. \]
With the integrals for the fluid variables $D_g$, $V$, $\Ga$ and $\Pi$
 given in Section~3 and Einstein's equations (\ref{2C1}) to (\ref{2D1})
for the geometrical perturbations $\Psi_m$ and $\Phi_m$ induced by the
collisionless component, this forms a closed system.
\vspace{1cm}

\chapter{Some Applications of Cosmological Perturbation Theory}

\section{Fluctuations of a Perfect Fluid}

A perfect fluid is free of anisotropic stresses, i.e. $\Pi=0~,~~
\Pi_i =0~,~~\Pi_{ij}=0$ for scalar, vector and tensor perturbations,
respectively. For scalar perturbations the dynamical equation (\ref{2D1})
then relates the Bardeen potentials:
\[ \Psi = -\Phi ~. \]
For vector perturbations we obtain from (\ref{2dyv})
\[ a^2\si_i = \mbox{const. , i.e. ~} \si_i = (a_*/a)^2\si^*_i ~.\]
 The vector contribution to the shear of the equal time hyper--sufaces of
perfect fluid perturbations thus decays like $1/a^2$. For tensor
perturbations eq. (\ref{2dyt}) yields
\[ \ddot{H}_{ij} +2(\dot{a}/a)\dot{H}_{ij} +(2k-\lap)H_{ij} = 0~,\]
which describes a damped gravitational wave with damping scale
$\dot{a}/a=t^{-1}$.

The often used notion of {\em isocurvature fluctuations} is defined by
$ \Psi=\Phi=0$ ( for scalar perturbations), $\si_i =0$ (for vector
perturbations) and $H_{ij}=0$ (for tensor perturbations) on super--horizon
scales.
{\em Adiabatic fluctuations} require $\Ga = 0$. In a coupled baryon/photon
universe this reduces to $D_g^{(B)} = (4/3)D_g^{(r)}$.
Note that fluctuations from topological defects are always isocurvature,
since they emerge in a causal way from an initially homogeneous and
isotropic Friedmann universe.

Let us now solve the  perturbation equations for adiabatic perturbations of
 a one component perfect fluid with $w=c_s^2=$const. and negligible spatial
 curvature. This simplification is a good approximation during some periods
of time (e.g. in the radiation dominated epoch). The very simple behavior
of vector perturbations  in this case is given by equation (\ref{2Omt}).
For scalar perturbations equations (\ref{2con1}--\ref{2con2}) reduce to
\bea \dot{D} -3w(\dot{a}/a)D &=& (1+w)l\lap V  \label{3c1} \\
 l\dot{V} + (\dot{a}/a)lV &=& {c_s^2\over w+1}D +\Psi \label{3c2} ~.\eea
Taking the Laplacian of (\ref{3c2})  (using (\ref{2C1}) and
$\Phi = -\Psi$) and inserting eq. (\ref{3c1}) we obtain a second
order equation for $D$
\be \ddot{D} + (\dot{a}/a)(1-3w)\dot{D} -[3w(\ddot{a}/a) +
	(1+w)4\pi G\rho a^2 -c_s^2\lap]D = 0 ~. \label{3DD1}\ee
For a spatially flat universe with $-1/3< w=c_s^2 =$const., the scale
factor obeys a power law
\[ a = (\nu\sqrt{C}t)^\nu~,~~\mbox{ with }~~~ \nu = {2\over 3w+1} ~~
	\mbox{ and }~~ C = (8\pi G/3)\rho a^{2(\nu+1)/\nu}=\mbox{ const.} \]
 We now Fourier transform $D$, so that the Laplacian is replaced by a factor
 $-k^2$. Equation (\ref{3DD1}) can then be expressed as
 ordinary differential equation in the dimensionless variable $\eta=kt$.
The regime $\eta\ll 1$ describes perturbations with wavelength substantially
larger than the size of the horizon and the  regime $\eta\gg 1$ describes
perturbations with wavelength much smaller than the size of the horizon.
Denoting the Fourier transform of $D$ again with $D$ we find
\[ D'' +{2(\nu-1)\over\eta}D' -{(2-\nu)(\nu-1)-\nu(\nu+1)\over\eta^2}D
		+wD = 0 ~.\]
In terms of $f = D\eta^{\nu-2} \propto \rho a^3D$, this equation becomes
\be f'' +{2\over\eta}f' +[w +{\nu(\nu+1)\over\eta^2}]f = 0 ~.\ee
For $w=c_s^2\neq 0$ this is the well known Bessel differential equation
 whose general solution is
\be f = Aj_\nu(c_s\eta) + B n_\nu(c_s\eta) \equiv Z_\nu(c_s\eta) \ee
(see Abramowitz and Stegun [1970]).
For our perturbation variables $D, \Psi$ and $V$ this yields
\bea D &=& \eta^{2-\nu}Z_\nu(c_s\eta) \\
  \Psi &=& -{2\over 3}\nu^2\eta^{-\nu}Z_\nu(c_s\eta) \\
     V &=&  {2\over 3}\nu[\eta^{1-\nu}Z_\nu(c_s\eta) -
	  {c_s\over 1-\nu}\eta^{2-\nu}Z_{\nu-1}(c_s\eta)] ~, \eea
where we have used $ Z'_\nu = c_sZ_{\nu-1} - (\nu+1/\eta)Z_\nu$ \cite{ASt}.
This solution was originally obtained by Bardeen [1980]. The asymptotic
behavior of Bessel functions yields
\bean
Z_\nu = & C\eta^\nu + E\eta^{-(\nu+1)}~,~~~~~~ & \mbox{ for } c_s\eta\ll 1 \\
 \mbox{and}&&\\
Z_\nu = & {A\over\eta}\cos(c_s\eta-\al_\nu) + {B\over\eta}\sin(c_s\eta-\al_\nu)
 	~,& \mbox{ for } c_s\eta\gg 1 \eean
with $\al_\nu = \pi(\nu+1)/2$.
Therefore, the ``growing'' mode behaves like
\be \left. \begin{array}{lll}
 \Psi &=& \Psi_0 \\
 D    &=& D_0\eta^2 \\
 V   &=&  V_0\eta   \end{array} \right\} \mbox{ for } c_s\eta\ll 1\ee
and
\be \left. \begin{array}{lll}
 \Psi &=& \Psi_0 \eta^{-(\nu+1)}\cos(c_s\eta-\al_\nu) \\
 D    &=& D_0\eta^{1-\nu}\cos(c_s\eta-\al_\nu) \\
 V   &=&  V_0\eta^{1-\nu}\cos(c_s\eta-\al_{\nu-1})
\end{array} \right\} \mbox{ for } c_s\eta\gg 1~. \label{3fs}\ee
On scales below the sound horizon, $c_s\eta\gg1$, density and
velocity perturbations grow only if $\nu<1$, i.e. for $w>1/3$ (or $w<-1/2$)
and they decay for $-1/2 <w<1/3$. Radiation ($w=1/3$) represents the limiting
case where the
amplitude of density perturbations remains constant. Perturbations in the
gravitational field always decay for $\nu\ge -1$, i.e. for $ w\ge -2/3$.
On scales
substantially larger than the sound horizon ($c_s\eta\ll 1$) it might seem
at first sight that density and velocity perturbations are growing. But one
easily
establishes that, e.g., the ``growing'' mode of the alternative gauge invariant
density variable $D_g = D -3(1+w)\Psi -3(1+w)(\nu/\eta)V$ is constant.
Therefore, in a coordinate system where $D_g$ represents the density
fluctuation (a slicing with ${\cal R} = 0$), density
perturbations do not grow.
This shows that the behavior of density perturbations on these scales
crucially depends on the coordinates  and can not be inferred from the
growth of the  particular gauge invariant variable $D$. This leads us to
define the perturbation amplitude $\cal A$ on  super--horizon scales as the
amplitude of the largest perturbation variable in a gauge where this
quantity is a minimum. For scalar perturbations thus
\[{\cal A} = \min_{\{\mbox{gauges}\}}(\max\{A,B,H_l,H_T,\de,v,\pi_L,
	\pi_T\})~.\]
It is clear that this quantity is of the same order of magnitude as
the largest gauge invariant variable. In our case therefore
\[ {\cal A} \approx |\Psi| ~,\]
i.e., {\em super--horizon perturbations do not grow in amplitude}, as one
would also expect for causality reasons.

Fortunately, this analysis,  which shows that no perturbations with
$0<w\le 1/3$ grow, does not hold for dust ($w=c_s^2=0$). In this case, equation
(\ref{3DD1}) reduces to
\[ \ddot{D} +(2/\eta)\dot{D} -(6/\eta^2)D = 0 ~,  \]
with the general solution $D= At^2+Bt^{-3}$, yielding the growing mode solution
\bean
\Psi &=& \Psi_0 \\
D    &=& D_0\eta^2~~~~~~~~~~ \eta =kt \\
 V   &=& V_0\eta ~. \eean
Again on super--horizon scales, $\eta<1$ , the perturbation amplitude is given
by $\Psi_0$ and is thus constant, whereas on sub--horizon scales, $\eta\gg 1$,
 the largest gauge invariant variable is $D = D_0\eta^2$, i.e., density
perturbations do grow on sub--horizon scales.

An additional important case is given by dust perturbations in a radiation
dominated universe with $\rho_r\gg \rho_m$ at times $t\ll t_{eq}$, where
$t_{eq}$ denotes the time when $\rho_r=\rho_m$, which  happens at some
time, because $\rho_r$ decays faster ($\propto a^{-4}$) than
$\rho_m\propto a^{-3}$. At $t\ll t_{eq}$  the scale factor grows according to
$a\propto t$ but
$4\pi G\rho_ma^2 \approx 4\pi G\rho_ra^2(t/t_{eq}) = (3/2){1\over tt_{eq}}$.
Equation (\ref{3DD1}) then yields
\[ \ddot{D} +(1/t)\dot{D} -{3\over 2tt_{eq}}D =0 ~, \]
which is approximately solved by
\be D = D_0\log(t/t_{eq})~, ~~~~ t\ll t_{eq} ~.\ee
This fact, that even dust perturbations cannot grow substantially in a
radiation dominated universe, is called the M\'ez\'aros effect \cite{Me'}.

We can now draw the following conclusions: For adiabatic
perturbations of a perfect fluid, in a universe where spatial curvature is
negligible, the time evolution of $D$, $\Phi$ and $V$ is given by
\be D \propto \left\{ \begin{array}{lll}
 	a ~,& & \mbox{in a matter dominated background,} \\
  \log(t/t_*)~,& \mbox{for}~kt\gg 1, & \mbox{in a radiation dominated
	background,}
  \end{array}  \right. \ee
\be \Phi = -\Psi = \mbox{const.}  ~.\ee
For pure radiation perturbations ($p = (1/3)\rho,~ \Ga = 0,~\Pi=0$)
one obtains
\bea
  D &\propto & \left\{ \begin{array}{ll}
  \mbox{const.} ~, & \mbox{for}~kt\ll 1 \\
 \exp(ik(x-\sqrt{1/3}t))~, & \mbox{for}~kt\gg 1~,  \end{array}\right.\\
\Phi = -\Psi &\propto &  \left\{ \begin{array}{ll}
  \mbox{const.} ~, & \mbox{for}~kt\ll 1  \\
 {1\over a^2} \exp(ik(x-\sqrt{1/3}t))~, & \mbox{for}~kt\gg 1~. \end{array}
 \right.  \eea
On super--horizon scales, $kt\ll 1$ the total perturbation amplitude is
given by $|\Psi|=|\Phi|$. Thus, even if one special variable like, e.g.,
$D$ is growing the perturbation amplitude $\cal A$ as defined above remains
constant.

We note the important result: The only substantial growth of linear
perturbation is that of sub--horizon sized, pressureless matter density
perturbations in a
matter dominated universe. Then $D\propto a$ and the gravitational potential
is constant.

\section{The Perturbation of Boltzmann's Equation for Compton Scattering}
In this section, we restrict ourselves to scalar type perturbations, i.e.,
 vector and tensor fields can be derived from
scalar potentials. Furthermore, we set $k=0$.

Following Section 2.3, the perturbed photon distribution function is
denoted  by
$f=(\bar{f}+F)\circ\io $ and lives on the one particle zero mass
phase space,
${\cal P}_0 = \{(x,p)\in T{\cal M}: g(x)(p,p)=0\}$.
 Choosing coordinates $(x^{\mu},p^i)$ on ${\cal P}_0$, Boltzmann's
equation for $f$ reads
\[ p^{\mu}\dd_{\mu}f-\Ga^i_{\mu\nu}p^\mu p^\nu{\dd f \over \dd p^i} =
  C[f] ~ , \]
where $C[f]$ is the collision integral depending on the cross section
and angular dependence of the interactions considered.

In terms of the gauge invariant  perturbation variable $\mm$
(see section 2.3.4) the
collisionless Boltzmann equation is (\ref{2li0S})
\be \dot{\cal M} +\ep^i\dd_i\mm = 4\ep^i\dd_i(\Phi-\Psi) \; ,\label{3li0} \ee
with (\ref{2M})
\bean \mm &\equiv& {4\pi\over\bar{\rho}a^4}\int_0^\infty{\cal F}v^3dv \\
	&=&\io +4{\cal R} +4l\ep^i\dd_i\si ~.\eean

To this we have to add the collision term which is given by
\[ C[\mm] = {4\pi\over\rho a^4}\int v^3dvC[f] = {4\pi\over\rho a^4}\int v^3dv
    {df_+\over dt}-{df_-\over dt} \equiv {d\io(\bm{\ep})_+\over dt}-
	{d\io(\bm{\ep})_-\over dt} \; , \]
where $f_+$ and $f_-$ denote the distribution of photons scattered into
respectively out of the beam due to Compton scattering.

In the matter  (baryon/electron) rest frame,
which we indicate by  a prime, we know
\[ {df'_+\over dt'}(p,\bm{\ep})= {\si_Tn_e\over 4\pi}\int
      f'(p',\bm{\ep}')\om(\bm{\ep,\ep}')d\Om'  \; , \]
where $n_e$ denotes the electron number density, $\si_T$ is the Thomson
cross section,  and $\om$ is the normalized
angular dependence of the Thomson cross section:
\[ \om(\bm{\ep,\ep}') = 3/4[1 + (\bm{\ep\cd\ep}')^2] =
  1 + {3\over 4}\ep_{ij}\ep'_{ij} ~~~\mbox{ with }~~
  \ep_{ij} = \ep_i\ep_j - {1\over 3}\de_{ij}  \; . \]
In the baryon rest frame which moves with four velocity $u$, the photon
energy is given by
\[ p' = p_\mu u^\mu \; . \]
We denote by $p$ the photon energy with respect to a tetrad adapted to
the slicing of spacetime into $\{ t=$ constant$\}$  hyper--surfaces:
\[ p =  p_\mu n^\mu \; ,~~~\mbox{ with }~~
  n = a^{-1}[(1-A)\dd_t +\beta^i\dd_i] ~,~ \mbox{ see Chap. 2 .}\]
The lapse function and  the shift vector of the slicing are given by
  $\al= a(1+A)$ and  $\bm{\beta}=
	-B^{,i}\dd_i$ .  In first order,
\[ p_0 = ap(1+A) - ap\ep_i\beta^i~~,\]
and in zeroth order, clearly,
\[ p_i = ap\ep_i ~ .\]
Furthermore, the baryon four velocity has the form
$ u^0 = a^{-1}(1-A)~~,~~~ u^i = u^0v^i $ up to first order.
This yields
\[ p' = p_\mu u^\mu = p(1+\ep_i(v^i-\beta^i)) \; . \]
Using this identity and performing the integration over photon energies,
we obtain
\[ \rho_r{d\io_+(\ep)\over dt'} = \rho_r\si_Tn_e[1+4\ep_i(v^i-\beta^i) +
 {1\over 4\pi}\int\io(\ep')\om(\ep,\ep')d\Om'] ~ .\]
The distribution of photons scattered out of the beam, has the well known
form \\ (see e.g. Lifshitz and Pitajewski [1983])
\[ {df_-\over dt'} = \si_Tn_ef'(p',\bm{\ep}) ~, \]
so that we finally obtain
\[ C' = {4\pi\over\rho_r a^4} \int dp({df_+\over dt'} -{df_-\over dt'})p^3 =
\si_Tn_e[
   \de_r -\io + 4\ep_i(v^i-\beta^i) + {3\over 16\pi}\ep_{ij}\int
  \io(\ep')\ep'_{ij}d\Om'] ~ , \]
where $\de_r = (1/4\pi)\int\io(\ep)d\Om$ is the photon energy density
perturbation.\\
Using the definitions of the gauge--invariant variables \mm ~and $V$, we
can write $C'$ in gauge--invariant form.
\be C'  = \si_Tn_e[D_g^{(r)} -\mm + 4\ep_il\dd_iV +
              {1\over 2}\ep_{ij}M^{ij}]~,\label{BC} \ee
with $D_g^{(r)} = (1/4\pi)\int\mm d\Om = \de_r + 4{\cal R}$ and  \[ M^{ij} =
  {3\over 8\pi}\int\mm(\ep')\ep'_{ij}d\Om']~. \]
Since the term in square brackets of (\ref{BC}) is already first order we have
$C = {dt'\over dt}C' = aC'$.
So that the Boltzmann equation becomes
\be \dot{\cal M} +\ep^i\dd_i\mm = 4\ep^i\dd_i(\Phi-\Psi) +
    a\si_Tn_e[D_g^{(r)}-\mm -4\ep^il\dd_iV + {1\over 2}\ep_{ij}M^{ij}]
	 \; . \label{3B} \ee
In the next two subsections we  rewrite this equation for two special cases.

Note that  perturbations of the electron density, $n_e = \bar{n}_e +\de n_e$
do not contribute in first order for a homogeneous and isotropic background
distribution of photons. The only first order term which accounts for
the perturbations
of baryons and therefore acts as a source term for the photon perturbations
is the Doppler term $4\ep^i\dd_iV$ which is due to the relative motion of
baryons and photons. A comparison of the numerical integration of equation
(\ref{3B}) with and without collision term (with a gravitational potential
$\Psi-\Phi$ originating from a collapsing texture) is shown in Fig.~7.
There one sees that the collision term has two effects: \\
 \hspace*{0.2cm} 1) Damping of the perturbations by several orders of
    magnitude.\\
  \hspace*{0.2cm} 2) Broadening of the signal to about $7^o$ (FWHM) which
    corresponds to the
    horizon scale at the time when Compton scattering 'freezes out', i.e.
    $t_T\approx t$ and  the mean free path of photons becomes larger than the
    size of the horizon. In general, there will always be the question
whether the damping term, $D_g^{(r)}-\mm$, or the source term,
$4\ep^i\dd_iV$, in equation (\ref{3B}) wins. It seems that for standard
CDM the source term is so strong, that no damping due to photon diffusion
occurs [Efstathiou, private communication]. It is an interesting and
partially unsolved problem,
in which scenarios of structure formation, photon perturbations on small scales
($\th\le 7^o$) are effectively damped by reionization. It seems plausible to
me, that models with topological defects, where  perturbations are highly
correlated, are affected more strongly than models with Gaussian perturbations.

The collision term above also appears in the equation of motion of the
baryons as a drag. The Thomson drag force is given by
\be F_j = {a\si_Tn_e\rho_r\over 4\pi}\int C[\mm]\ep_jd\Om =
	 {a\si_Tn_e\rho_r\over 3}(M_j +4l\dd_iV) ~,\label{3drag}\ee
\[  \mbox{ with }~~~ M_j ={3\over 4\pi}\int\ep_j\mm d\Om ~.\]
This yields the following  baryon equation of motion   in an ionized plasma
\be l\dd_j\dot{V} + (\dot{a}/a)l\dd_iV = \dd_i\Psi -
	{a\si_Tn_e\rho_r\over 3\rho_b}(M_j +4l\dd_iV) ~, \label{3bar}\ee
where we have added the drag force to eq. (\ref{2con2}) with $w=c_s^2=0$.

We now want to discuss equations (\ref{3B},\ref{3bar}) in the limit of very
many collisions. Clearly the photon mean free path  is given by
$t_T =l_T = (a\si_Tn_e)^{-1}$. In lowest order $t_T/ t$ and $l_T/ l$ these
equations reduce to
\be D_g^{(r)} +{1\over 2}\ep_{ij}M^{ij} +4\ep^il\dd_iV =\mm  \label{3r1}\ee
and
\be 4l\lap V = \dd_iM_i = 3\dot{D}_g^{(r)} ~, \label{3r2} \ee
where we made use of (\ref{3r1}) and (\ref{A0}) below, for the last
equal sign. Eq. (\ref{3r2}) is equivalent to (\ref{2cog1}) for radiation.
Using also (\ref{2cog1}) for baryons, $w=0$, we obtain
\[ \dot{D}_g^{(r)} ={4\over 3}l\lap V = {4\over 3}\dot{D}_g^{(m)} .\]
This shows that entropy per baryon is conserved, $\Ga=0$.
Inserting (\ref{3r1}) in (\ref{3B}) we find up to first order in $t_T$
\bea \mm &=& D_g^{(r)} -4l\ep^i\dd_iV +{1\over 2}\ep_{ij}M^{ij}
	-t_T[\dot{D}_g^{(r)}
	-4l\ep^i\dd_i\dot{V} +{1\over 2}\ep{ij}\dot{M}^{ij} \nonumber\\
	&&  +\ep^j\dd_jD_g^{(r)}
	-4l\ep^i\ep^j\dd_i\dd_jV +{1\over 2}\ep^i\ep{kj}\dd_iM^{kj}
	-4\ep^j\dd_j(\Phi-\Psi)]~ .   \label{3mm1} \eea
Using (\ref{3mm1}) to calculate the drag force yields
\[ F_i = (\rho_r/3)[4l\dd_iV -\dd_iD_g^{(r)} + 4\dd_i(\Phi-\Psi)] ~.\]
Inserting $F_i$ in (\ref{3bar}), we obtain
\[ (\rho_b +(4/3)\rho_r)l\dd_i\dot{V} +\rho_b(\dot{a}/a)l\dd_iV =
   (\rho_r/3)\dd_iD_g^{(r)} + (\rho_b+(4/3)\rho_r)\dd_i\Psi -
	(4\rho_r/3)\dd_i\Phi ~.\]
This is equivalent to (\ref{2cog2}) for $\rho=\rho_b+\rho_r$,
$p=\rho_r/3$  and $\Ga=\Pi =0$, if we use
\[ D_g^{(r)}= (4/3)D_g^{(m)} ~~~\mbox{ and }~~~~ D_g =
	{\rho_rD_g^{(r)} +\rho_mD_g^{(m)}\over \rho_m+\rho_r}   ~ . \]
In this limit therefore, baryons and photons behave like a single fluid
with density $\rho =\rho_r+\rho_m$ and pressure $p=\rho_r/3$.

  From (\ref{2cog1}) and (\ref{2cog2}) we can derive a second order
equation for $D_g$. To discuss the coupled matter radiation fluid we
regard a plane wave $ D = D(t)\exp(i\bm{k\cd x})$. We then obtain
\[ \ddot{D} +c_s^2k^2D + (1+3c_s^2 - 6w)(\dot{a}/a)\dot{D} -
  3[w(\ddot{a}/a) - (\dot{a}/a)(3(c_s^2-w) -(1/2)(1+w))]D = 0 ~.\]
For small wavelengths (which are required for  the fluid approximation to be
 valid),  $1/t_T \gg c_sk\gg 1/t$,
we may drop the term in square brackets. The ansatz
$ D(t) = A(t)\exp(-i\int kc_sdt)$
then eliminates the terms of order $c_s^2k^2$. For the terms of order
$c_sk/t$ we obtain the equation
\be 2\dot{A}/A + (1-3c_s^2 -6w)(\dot{a}/a) +\dot{c_s}/c_s = 0 \label{3dA}~.\ee
For the case $c_s^2 =w=$const. , this equation is solved by
$ A\propto (kt)^{1-\nu}$ with $\nu =2/(3w+1)$, i.e., the short wave limit
 (\ref{3fs}). In our situation we have
\bean
 w &=& {\rho_r \over 3(\rho_r+\rho_m)}  \\
 c_s^2 &=& {\dot{\rho_r} \over 3(\dot{\rho_r}+\dot{\rho_m})} ~
    = ~ {(4/3)\rho_r \over 4\rho_r+3\rho_m}~~\mbox{ and} \\
 \dot{c}_s/c_s &=& -3/2(\dot{a}/a){\rho_m \over 4\rho_r+3\rho_m} ~.\eean
Using all this, one finds that
\[ A = \left({\rho_m+(4/3)\rho_r \over c_s(\rho_r+\rho_m)^2a^4}\right)^{1/2}
     =  \left({\rho+p \over c_s\rho^2a^4}\right)^{1/2} \]
solves (\ref{3dA}) exactly, so that we finally obtain the
approximate solution
for the, tightly coupled matter radiation fluid
\be D(t) \propto  \left({\rho+p \over c_s\rho^2a^4}\right)^{1/2}
	\exp(-ik\int c_sdt) ~.\label{3mrt} \ee
Note that this short wave approximation is only valid in the limit
$t\gg 1/(c_sk)$, thus the limit to the matter dominated regime, $c_s\ra 0$
cannot be performed. In the limit to the radiation dominated regime,
$c_s^2\ra 1/3$ and $\rho\propto a^{-4}$ we recover the acoustic waves
with constant amplitude which we have already found in the last subsection.
But also in this limit, we still need matter to ensure $t_T =1/(a\si_Tn_e)
\ll t$. In the oppotite case, $t_T\gg t$, radiation does not behave like
an ideal fluid but it becomes collisionless and has to be treated with
Liouville's equation (\ref{3li0}).

In the paragraph 3.2.3 we  evolve \mm~ up to order $t_T^2$ and
obtain the damping of \mm~ by photon diffusion.

\subsection{Spherically symmetric Boltzmann equation}
In the spherically symmetric case \mm ~depends on the momentum direction
\bm{\ep} only via the variable $\mu = (\bm{r\cd\ep})/r$. Integrating out the
independent angle $\varphi$ and choosing the coordinates in photon momentum
space such that the third axis is parallel to $\bm{r}$, one finds
\[ {1\over 2\pi}\int_0^{2\pi}\ep_{ij}d\varphi = \left\{ \begin{array}{ll}
  0~,& i\neq j \\
\mu^2-1/3~,& (i,j)= (3,3) \\
{1\over 2}(1/3-\mu^2)~,& (i,j)=(1,1) \mbox{ or } (2,2)~. \end{array} \right. \]
We thus obtain
\[ \int_0^{2\pi}d\varphi[\ep_{ij}M^{ij}]
     = {3\over 4}(\mu^2-1/3)\int_{-1}^1
   (\mu'^2-1/3)\mm d\mu' \equiv {3\over 2}(\mu^2-{1\over 3})M_2 ~~. \]
For $\mm(t,r,\mu)$ we further have
\[ \ep^i\dd_i\mm = \mu\dd_r\mm + {1-\mu^2\over r}\dd_\mu\mm \]
so that Boltzmann's equation becomes
\bea \lefteqn{\dot{\cal M} +  \mu\dd_r\mm + {1-\mu^2\over r}\dd_\mu\mm =
4\mu\dd_r(\Phi-\Psi) +}
  \nonumber  \\ &&
    a\si_Tn_e[M-\mm -4\mu\dd_rV + (\mu^2-1/3){3\over 2}M_2
  ] \; . \label{ABs} \eea
This is the equation which we have integrated numerically, (in somewhat
different coordinates, see Chap.~4) for a $\Phi-\Psi$ from a spherically
symmetric texture, to produce Fig.~7.

\subsection{Moment expansion}

As long as collisions are efficient, $t>1/(a\si_Tn_e)\equiv t_T$, it is
reasonable to truncate $\mm$ at second moments
\be \mm = D_g^{(r)} + \ep_iM^i + 5\ep_{ij}M^{ij} \; . \ee
With this ansatz, the integration over directions of the zeroth, first and
second moments of (\ref{3B}) yields
\be \dot{D}_g^{(r)} +{1\over 3}\dd_iM^i = 0 ~, \label{A0} \ee
\be \dot{M}^j + \dd_jD_g^{(r)} + 2\dd_iM^{ij} = 4\dd_j(\Phi-\Psi) -
 a\si_Tn_e[M^j + 4\dd_jlV]~, \label{A1}  \ee
\be \dot{M}^{ij} + {1\over 15}(\dd_iM^j + \dd_jM^i) =
      -{9\over 10}a\si_Tn_eM^{ij}~. \label{A2} \ee
Taking the total divergence of (\ref{A1}) and (\ref{A2}), we obtain with
 the help of (\ref{A0})
\be \ddot{D}_g^{(r)} - {1\over 3}\lap D_g^{(r)} -{2\over 3}\al =
	-{4\over 3}\lap(\Phi-\Psi)
    + a\si_Tn_e[\dot{D}_g^{(r)}-{4\over 3}\lap lV] \label{Ad1} \ee
and
\be \dot{\al} -{2\over 5}\lap\dot{D}_g^{(r)} = -{9\over 10}a\si_Tn_e\al ~,
	\label{Ad2} \ee
where we have set $\al = \dd_i\dd_jM^{ij}$ \vspace{2cm}.

\subsection{Damping by photon diffusion}
In this subsection we want to estimate the damping of CBR fluctuations
in an ionized plasma using our gauge invariant approach, as it was done by
Peebles [1980] within synchronous gauge. We again consider eqs. (\ref{3B}) and
(\ref{3bar}), but since we are mainly interested in collisions which
take place on time scales $t_T\ll t$, we neglect gravitational effects and
the time dependence of the coefficients. We can then look
for solutions of the form
\[ V \propto \mm \propto \exp(i(\bm{k\cd x}-\om t)) ~. \]
In (\ref{3B}) and (\ref{3bar}) this yields (neglecting also the
angular dependence of Compton scattering, i.e., the term $\ep_{ij}M^{ij}$)
\be \mm = {D_g^{(r)} -4i\bm{k\ep}lV\over 1 -it_T(\om -\bm{k\cd \ep})}
	\label{3md}  \ee
and
\be t_T\bm{k}\om lV = (\rho_r/3\rho_m)(4i\bm{k}lV +\bm{M}) ~, \label{3Vd}\ee
with $\bm{M} = (3/4\pi)\int\bm{\ep}\mm d\Om$. Integrating (\ref{3B}) over
angles, one obtains $\dot{D}_g^{(r)} +(1/3)\dd_iM^i = 0$. With our ansatz
therefore $\bm{k\cd M} =3\om D_g^{(r)}$. Using this after scalar
multiplication
of (\ref{3Vd}) with $\bm{k}$, we find, setting $R =3\rho_m/4\rho_r$,
\[ lV = {(3/4)\om  D_g^{(r)} \over t_Tk^2R\om -ik^2} ~. \]
Inserting this result for $V$ in (\ref{3md}) leads to
\[ \mm = D_g^{(r)}{1+{3\mu\om/k\over1-it_T\om R}\over 1-it_T(\om -k\mu)}~,\]
where we have set $\mu =\bm{k\cd \ep}/k$. This is exactly the same result
as in Peebles [1980], where this calculation is performed in synchronous
gauge. Like in there (\S 92), one obtains in lowest order $\om t_T$
the dispersion relation
\be \om =\om_0 -i\ga~ \mbox{ with }~~ \om_0 =k/[3(1+R)]^{1/2} ~~ \mbox{ and }
  ~~~~ \ga = (k^2t_T/6){R^2+{4\over 5}(R+1)\over (R+1)^2} ~ .
	\label{3disp}\ee
In the matter dominated regime, $R\gg 1$, therefore
\be \ga \approx k^2t_T/6 ~.\ee

We  now consider a texture which collapses in a matter dominated universe
 ($a\propto t^2$) at time $t_c = t_0/(1+z_c)^{1/2}$.
The total damping, $\exp(-f)$, which this perturbation experiences is
given by the integral
\be f \approx \int_{t_c}^{t_{end}} \ga(t)dt ~. \label{df}\ee
The end time $t_{end}$ is the time, when the mean free path,
$t_T$ equals $t_c$, the size of the perturbation.
We define $z_{dec}$ as the redshift when
photons and baryons decouple due to free streaming. This is about the
time when the mean free path has grown up to the size of the horizon:
 $t_T(z_{dec}) \approx t(z_{dec})$. To obtain exponential
damping ($t_c < t_{end}$), we thus
need $z_c>z_{dec}$. 	In this case damping is effective until
\[ 1+z_{end} = (1+z_{dec})^{3/4}(1+z_c)^{1/4}~,  \]
 and we obtain
\be f \approx  2\left({1+z_{dec}\over 1+z_c}\right)^{3/2}\left[\left(
      {1+z_c\over 1+z_{dec}}\right)^{15/8}-1\right]
      ~,  \label{damp}  \ee
where we have set $k^2 = (2\pi/t_c)^2$.
In terms of angles this yields
\[ f(\th) \approx 2(\th/\th_d)^3[(\th_d/\th)^{15/4} -1] ~,~~~\th<\th_d\]
where $\th_d = 1/\sqrt{1+z_{dec}}\approx 6^0$.

If the plasma ionizes at a redshift $z_i$,  $z_{dec}<z_i<z_c$, after
the texture has
already collapsed, damping is only effective after $z_i$, if
$z_{end}< z_i$. Instead of formula (\ref{damp}), we then obtain
\be f  \approx 2\left({1+z_{dec}\over 1+z_c}\right)^{3/2}\left[\left(
    {1+z_c\over 1+z_{dec}}\right)^{15/8}-\left(
    {1+z_c\over 1+z_i}\right)^{15/8}\right] ~. \label{damp2} \ee
This damping factor can again be converted in an angular damping
factor in the above, obvious way.
The CBR signal of  textures is thus exponentially damped only if
  \[z_i>z_{dec} \approx 100(0.05/h_{50}\Om_B)^{2/3}  ~,\]
and in this case only textures with
$ 1+z_{dec}< 1+z_c< (1+z_i)^2/(1+z_{dec})$ are  affected.
So, if $z_i \le 50$, there is little exponential damping, even if $h_{50}= 2$
and $\Om_B =0.1$, but if $z_i= 200$, all textures which collapse at
redshifts $z_i\ge z_c$ are damped by a factor 10 or more.

In this  approximation, we have neglected the the amount of damping
which still may occur after $z_{end}$ and the induced fluctuations due to the
source term $\propto lV$ in (\ref{3B}). A more accurate numerical treatment
where just equation (\ref{3B}) is solved is shown in Fig.~7.
\vspace{0.1cm}

Astonishingly,  even perturbations on large scales up to the quadrupole
 may be damped by photon diffusion if the spectrum is steep enough
(Peebles, private communication): We assume perturbations of a given size
$l$ are uncorrelated and have an average amplitude $A$. On a larger scale
$L=Nl$, they statistically induce perturbations with typical amplitude
 $AN^{-3/2}=A(l/L)^{3/2}$. (A cube of size $L^3$ contains $N^3$ cubes of size
$l$. The statistical residual of $N^3$ perturbations with amplitude $A$
is thus $AN^{-3/2}$.)

This simple argument has two interesting conclusions:
\begin{itemize}
\item[i)] The effective spectrum $|D(k)|^2$ of Gaussian distributed
	fluctuations cannot decrease faster than $|D(k)|^2\propto k^3$
	towards large scales.
\item[ii)] In the limiting case, $|D(k)|^2\propto k^3$, all the power
	in large scales is induced due to the statistical residuals of
	small scale perturbations. Therefore if small scale perturbations
	are damped so are large scale perturbations up to the size of the
	horizon.
\end{itemize}
It is clear that this mechanism crucially depends on the assumption of
uncorrelated perturbations. It is thus not effective for fluctuations
induced by topological defects. It also does not work for a
Harrison--Zel'dovich spectrum,  $|D(k)|^2\propto k$.
\vspace{1cm}

\section{Perturbations of the Microwave Background}
The redshift of photons propagating in a perturbed Friedmann
geometry is given by (for scalar perturbations) (\ref{2dE}):
\be {\de E\over E} = [{1\over 4}D_g^{(r)} -V_jn^j +\Psi-\Phi]|_i^f - \int_i^f
  (\dot{\Psi}-\dot{\Phi})d\tau  ~. \label{3dE} \ee
The first of these terms is due to intrinsic fluctuations on the
surface of last scattering, the second term is the usual, special
relativistic Doppler shift, the third and the last terms are
gravitational redshift contributions, the Sachs---Wolfe effect \cite{SW}:
The third term is the difference of the potential at the emitter and
receiver and
the last term is due to the time dependence of the  gravitational potential
along the path of the photon.

Since perturbations of the cosmic microwave background (CMB) are probably the
most reliable observational tool for investigating the initial
perturbation spectrum, and since they are calculable  within linear
perturbation analysis, we present them here in some detail. There are
seven different physical mechanisms which perturb the microwave background
on different scales. The first four of them are given by equation
(\ref{3dE}):
\begin{itemize}
\item Intrinsic inhomogeneities on the last scattering surface,
\[ {\De T\over T} = {1\over 4}D_r  ~,\]
\item relative motions of emitter and observer,
\[ {\De T\over T} = -\bm{V\cd n}|_i^f ~, \]
\item the difference of the gravitational potential at the position of
emitter and observer \\ (Sachs--Wolfe I),
 \[ {\De T\over T} = (\Psi-\Phi)|_i^f ~, \]
\item and the time dependence of the gravitational field along the path of
the photon \\ (Sachs--Wolfe II),
\[ {\De T\over T} = \int_i^f(\dot{\Psi}-\dot{\Phi})d\tau~. \]
\item In an intergalactic ionized plasma, fluctuations are damped by photon
diffusion (Section~3.2.3). As long as the mean free path of photons is
considerably
smaller than the size of the horizon, $l_T=1/(\si_Tn_e)\ll l_H$,
this damping increases exponentially with the damping rate
$\ga \propto l_T/l^2$ and is
thus very effective for inhomogeneities with sizes smaller or of the
order of the Thompson mean free path, $l\le l_T $
\[({\de T\over T})_f = ({\de T\over T})_i\exp(-\int_i^f\ga dt) ~.\]
Clearly, damping by photon diffusion can only be efficient as long
as $l_T<l_H$.
At $z\approx 100$, $l_T\approx l_H$ (for $\Om=1$ and $\Om_Bh=0.05$)
and angular scales which are larger than
the horizon scale at $z\approx 100$, which corresponds to about $6^o$, are
 usually not affected (for an exception of this rule, see the note at the end
of the last section).
\item On the other hand, the passage of photons through a cloud of ionized gas
at a temperature $T_e$ different from the photon temperature $T$
induces deviations in the   black body spectrum, which can be cast in the so
called Compton $y$ parameter. This is the Sunyaev--Zel'dovich
effect \cite{SZ}. In the Rayleigh--Jeans regime the change of the spectrum
corresponds to  a temperature shift according to
\[ {\De T\over T} \approx -2y ~\mbox{, with }~ y =
  \int_i^f({T_e-T\over m_e})n_e\si_Td\tau ~.\]
\item Since the Bardeen potentials of linear dust perturbations are
constant,
they do not give rise to a path dependent contribution to the photon
redshift. But once a dust perturbation has become non--linear and is
virialized so that its density $\rho D\approx $ const.~, the corresponding
gravitational potential
$\Psi$ grows like $a^2$. Usually, the gravitational potential of such a
dust cloud is still very small, so that the redshift of photons passing
through it can be calculated in linear perturbation theory.
Equation (\ref{3dE}) then yields
\[ {\De T\over T} = -2(\Phi_i/a^2_i)(a^2_f-a^2_i)  \approx
	4\Psi{\De z \over 1+z}~,\]
where $\De z$ denotes the redshift difference of the two ends of the structure
and $z$ is the average redshift of the perturbation. Typically not
only $\Psi$ but also $\De z$ is very small. This non--linear contribution to
$\De T/T$ is  called the Rees--Sciama effect \cite{RS}.

\end{itemize}
  From these seven effects, for a long time only the Doppler term, point
two, was detected. It leads to the famous dipole anisotropy with
amplitude $(\De T/T)_{dipole} \approx 2\times 10^{-3}$ which shows that
we are moving with a relative velocity of about $ 600km/s$ with respect
to the microwave background. Now, the COBE team confirmed that also
the spectrum of the dipole anisotropy is the derivative of a blackbody
spectrum with $T=2.731K$ to an accuracy better than  1\% \cite{MaP}.

 Recently, also the Sunyaev--Zel`dovich effect
has been confirmed  \\  \cite{McH,BHA}, and with the  COBE satellite
\cite{Co1,Co2}, the sum of intrinsic fluctuations and
the Sachs--Wolfe effect (points one, three and four) have been
measured. With these  recent observations, the cosmic microwave
background has started to become a  very successful tool for
investigating cosmological perturbations on the linear level.

\section{Light Deflection}
We now want to present some applications of the effect of
light deflection in a perturbed Friedmann universe. The general formulae
are derived in Section~2.4.

\subsection{Monopoles}
As a first example, we discuss light deflection and lensing in the field of
a global monopole, see also Barriola and Vilenkin [1989]. We  discuss the
simple, static
hedgehog solution of a three component scalar field with $\phi^2=\eta^2$,
i.e., a non--linear sigma model on ${\bf S}^2$:
\[ \phi^i = \eta x^i/r ~.\]
This is an infinite action and infinite energy solution and should thus not
be taken seriously at large distances. In a cosmological context, when
monopoles form via the Kibble mechanism during a symmetry breaking phase
transition, the hedgehog solution may be approximately valid on distances
small compared to the distance to the next monopole or antimonopole, which
is about  horizon distance. This is also the scale where the approximation
of a static, i.e., non--expanding background, which we  adopt here,
 breaks down.

The energy momentum tensor of the hedgehog solution is readily calculated:
\bean T_{00} &=& \eta^2/r^2  \\
      T_{ij} &=& -\eta^2x^ix^j/r^4 \eean
Setting $\eta^2 =M^2$ we obtain, using the definitions (\ref{3seed00}) to
(\ref{3seedij}),
\bea f_\rho &=& l^2/r^2  \\
     f_v    &=& 0  \\
     f_p    &=& -(1/3)f_\rho    \\
     f_\pi  &=& -(1/2)\log(r/l) ~.\eea
It is interesting to note that the quantity $f_\rho+3f_p$ which enters as
source term in the evolution equation for density perturbations (\ref{2DD})
vanishes, which shows that static global monopoles do not produce an
attractive gravitational force, much like cosmic strings.

Setting $\ep = 8\pi G\eta^2$, we find from (\ref{2S1}) and (\ref{2S3})
\be \Phi = -\ep\log(r/l)~~,~~~~~~ \Psi = 0 ~. \label{3Mgr} \ee
The second of these equations again shows that the analog to the Newtonian
potential vanishes for  global monopoles, like it does for cosmic strings
(exercise for the reader).
 The divergence of $\Phi$ at large distances reflects the infinite
energy of our solution
which  needs a physical cutoff (at most at the distance to the next
monopole).

We consider a photon passing the monopole at $t=0$, say with an
impact parameter $b$. Its unperturbed trajectory is then given by
$\bm{x} = \la\bm{n} +b\bm{e}$. According to (\ref{2ldefl}), neglecting
spherical aberration, we obtain
\be \varphi =+\int_i^f(\Phi-\Psi)_{,i}e^id\la = -\ep\int_i^f
	{b\over\la^2+b^2}d\la = -\ep\arctan(\la/b) \approx -\ep\pi=
	-8\pi^2G\eta^2  ~. \label{3Mdef}   \ee
This result was originally obtained (by completely different means) by
Barriola and Vilenkin [1989].

We want to investigate the situation of gravitational lensing.
First, we treat the
special case, where source ($S$), monopole ($M$) and observer ($O$) are
perfectly aligned at distances $\overline{SM}=s$ and $\overline{MO}=d$
  from each other.
If emitted at a small angle $\al = b/s$, a photon will reach the
observer at an angle $\beta = b/d$, if the deflection angle
$|\varphi|=\al+\beta =b(s+d)/sd$. This leads to an Einstein ring with
opening angle
\be \beta = b/d =|\varphi| s/(s+d) = 8\pi^2G\eta^2s/(s+d) ~.\ee
If observer and source are slightly misaligned by an angle less than
$|\varphi|$, the ring is reduced to two points with the same angular
separation.

For monopoles produced at a typical GUT scale, $\eta\approx 10^{16}GeV$,
the deflection angle is $|\varphi| \approx 10$ arcsec and thus observable.
Since the density of global monopoles is about one per horizon cite{BR},
roughly 10 monopoles present at a redshift $z=4$ would be visible for us
today. The probability that one of them is within  less than 10 arcsec
of a quasar with redshift $z>4$, so that the lensing event discussed
above could occur, is very small indeed.

\subsection{Light deflection due to gravitational waves}
We discuss formula (\ref{2deft}) for light deflection due to a passing
gravitational wave pulse for which the difference of the gravitational
field before and after the passage of the wave is negligible:
\be \varphi e_j =\int_i^f(H_{lk},_j+\dot{H}_{lk}n_j)n^kn^ld\la ~. \ee
We consider a plane wave,
\[ H_{kl} = \Re(\ep_{kl}\exp(i(\bm{k\cd x}-\om t)) ~~\mbox{ with } ~~~
	\ep_{kl}k^l=0 ~. \]
For a photon with unperturbed trajectory $\bm{x} = \bm{x}_o +\la\bm{n}~,
 t=\la$, we obtain
\[ \varphi e_j = \left\{\begin{array}{lll}
  \Re[ie^{(i\bm{k\cd x_o})}\ep_{lm}n^ln^m{k_j-\om n_j\over \bm{\om-k\cd n}}[
e^{(i(\bm{k\cd n}-\om)\la_f)}   - e^{(i(\bm{k\cd n}-\om)\la_i)}]] &
  \mbox{ for }& \bm{k} \neq \om\bm{n} \\
 0 & \mbox{ for }& \bm{k} = \om\bm{n}~. \end{array} \right. \]
Setting $\bm{n} = (p/\om)\bm{k} + q\bm{n}_\perp$ with $\bm{n}_\perp^2=1$
and $p^2+q^2 = 1$, we have $\ep_{lm}n^ln^m =q^2\ep_\perp$, where
 $\ep_\perp =\ep_{ij}n_\perp^in_\perp^j$.
Inserting this above, we find for the deflection angle an equation of the
form
\be \varphi = \ep_\perp {\sqrt{2}q^2\over\sqrt{1-p}}\cos(\al +\om(p-1)t)
	~. \label{3gw} \ee
Here $\ep_\perp$ is determined by the amplitude of the gravitational wave
and $q~,~p^2=1-q^2$ is determined by the intersection angle of the photon
with the gravitational wave as explained above.

This effect for a gravitational wave from two coalescing black holes would
be quite remarkable: Since for this (most prominent) event
$\ep_\perp$ can be as large as $\approx 0.1(R_s/r)$, the rays of sources
behind the black hole with impact parameters up to $b<10^4R_S$ would be
deflected by a measurable amount:
\[ \varphi\approx 2''(10^4R_S/b) ~. \]
Setting the  the  source at distance $d_{LS}$ from the coalescing
black holes and at distance $d_S$ from us, we observe a deviation angle
\[ \al = \varphi d_{LS}/d_S ~. \]
The best source candidates would thus be quasars for which $ d_{LS}/d_S$
 is of order unity for all  coalescing black holes with, say $z<1$.
In the vicinity of the black holes ($r\le 10R_S$), linear perturbation
theory is of course
not applicable and also the 'monopole' component of the gravitational field
is not negligible. But in the wide range $10^7R_S>b>10R_S$ (for radio sources)
and  $10^4R_S>b>10R_S$ (for optical sources) our calculation
is valid and leads to an effect that is in principle detectable.

A thorough investigation of the possibility of detecting  gravitational
waves of coalescing black holes out to cosmological distances by this effect
may be worth while.

\chapter{Textures in Flat Space}
\section{The $\si$--Model Approximation for Texture Dynamics}
Global texture occurs in every symmetry breaking phase transition
where a global symmetry group $G$ is broken to a sub--group $H$, such that
$\pi_3(G/H) \neq 1$.

In the cooler, broken symmetry phase the Higgs field remains most of the time
in the vacuum manifold $M \equiv G/H$, where the effective potential
assumes its minimum value. It only leaves M if the
gradients, i.e., the kinetic energy of the massless Goldstone modes,
become comparable to the symmetry breaking scale $\eta$.
In order to discuss the dynamics of the massless modes at temperatures
$T\ll T_c \approx \eta$, it is thus
sufficient to neglect the potential and, instead, fix the Higgs field in
{\cal M} with a Lagrange multiplier.

For illustration, and since we believe that the results  remains
valid at least qualitatively also for other symmetry groups, we
discuss the simplest version, an  $SU(2)$ symmetry which is completely broken
by a ${\bf C}^2$ valued scalar field $\phi$. In this case, clearly
$M \equiv SU(2) \equiv {\bf S}^3$.
Here ${\bf S}^n$ denotes the $n$--sphere and $\equiv$ means topological
equivalence.

The Higgs field $\phi$ is described by the zero temperature Lagrangian
\be L = {1\over 2}\phi_{,\mu}\phi^{,\mu} -\la(\phi\cd\phi - \eta^2)^2~,\ee
with corresponding field equation
\be  \Box\phi +4\la\phi(\phi\cd\phi - \eta^2) = 0 ~. \label{4tf} \ee
As argued above, on energy scales well below the symmetry breaking scale
$\eta$ we may fix $\phi^2=\eta^2$ by the Lagrangian multiplier
$\la(\phi\cd\phi - \eta^2)$. Instead of the usual equations of motion, we
then obtain the scale invariant equation
\be \Box\phi - \eta^{-2}(\phi\cd\Box\phi)\phi = 0 ~, \label{fsi}  \ee
 i.e., $\phi$ describes a harmonic map from spacetime to ${\bf S}^3$.
If $\phi$ is asymptotically parallel, $\phi(r)\ra_{r\ra\infty}\phi_0$,
 we can regard it (at fixed time) as a map from compactified three
dimensional space  ($\equiv  {\bf S}^3$) into  ${\bf S}^3$. The degree
(winding number) of this map is called the texture winding number. It is
given by the integral of the  closed three--form
\[ \om = {1\over 12\pi^2}\ep_{abcd}\phi^ad\phi^b\wedge d\phi^c\wedge d\phi^d
 ~~,   \]
($\om$ is nothing else than the pullback of the volume form on ${\bf S}^3$).
This integral over some region of space (e.g. a horizon volume) is of
course  well defined also if $\phi$ is not parallel at infinity and is often
referred to as fractional texture winding number. Numerical simulations show
that a texture starts collapsing as soon as the fractional winding inside the
horizon exceeds about 0.5 \cite{LeP,BCL}.

In our case even an exact spherically symmetric solution to
equation (\ref{fsi}) is known: A spherically symmetric ansatz for $\phi$ is
\be \phi({\bf r},t) = \eta(\hat{\bf r}\sin\chi,\cos\chi)~,\ee
where $\hat{\bf r}$ denotes the unit vector in direction of {\bf r}
and $\chi$ is an angular variable depending on $r$ and $t$ only.
In flat space the field equations for $\phi$ yield
\be -\dd_t^2\chi + \dd_r^2\chi + {2\over r}\dd_r\chi = {\sin2\chi\over r^2}
  ~ . \label{fchi}  \ee

We now look for solutions which depend only on  the self similarity variable
$y = (t-t_c)/r$, where $t_c$ is an
arbitrary time constant. In terms of $y$, equation (\ref{fchi}) becomes
\[ (y^2 -1) \chi'' = \sin 2\chi ~,  \]
with the exact solutions $ \chi = 2\arctan(\pm y) \pm n\pi$ which were found by
Turok and Spergel [1990].
To describe a collapsing texture which has winding number 1 for $t<t_c$
and winding number 0 for $t>t_c$, we patch together these solutions
in the following way:
\be \chi(y) = \left\{ \begin{array}{ll}
          2\arctan y + \pi ~,& -\infty \le y\le 1 \\
	  2\arctan(1/y) + \pi~,& 1\le y \le \infty~ . \\
  \end{array}  \right.\label{chiy}  \ee
It is straight forward to calculate the integral of the density $\om$ given
above for this example and one of course obtains
\[ \int_{{\bf R}^3}\om(t) =\left\{\begin{array}{ll} 1, & \mbox{ if }t<t_c \\
     0, & \mbox{ if } t> t_c ~.\end{array} \right.  \]
The kink  of $\chi$ at $y = 1$ reflects the singularity of the $\si$--model
approach at the unwinding event $t-t_c=r = 0$. There the gradient energy
of the solution (\ref{chiy})
diverges, i.e., becomes bigger than the symmetry breaking scale $\eta$.
Therefore, the Higgs field leaves the vacuum manifold, unwinds and the
kinetic  energy can  be radiated away in massless Goldstone modes. To remove
the singularity at $r=0,~t=t_c$, we would have to evolve the innermost region
of the texture ($r\le\eta^{-1}$) with the true field equation (\ref{4tf})
during the collapse ($t_c-\eta^{-1}\le t\le t_c+\eta^{-1}$).
It is well known
\cite[Derrick's theorem]{De}, that a $\si$--model like (\ref{fsi})
cannot have static
finite energy solutions, but it is not proven that there is no
static infinite energy solution.  It is easy to see that (\ref{chiy})
represents an infinite energy solution. We shall
see that the infinite energy  of the solution will require some
'renormalization' or, equivalently, a physical cutoff.

{\bf Remark:} It seems quite natural that this scale invariant problem
admits scale invariant solutions. But this is by no means guarantied. It is
well known, e.g., that Yang Mills theories in general do not admit
non--trivial spherically symmetric  solutions. But also other $\si$--type
models do not admit them. As an example we have studied $SO(3)$ broken to
$SO(2)$ by a vector field (the Heisenberg model). The vacuum manifold in
this example is ${\bf S}^2$ with $\pi_3({\bf S}^2) = {\bf Z}$ (the Hopf
fibration). A spherically symmetric ansatz for a texture configuration in
this model is the Hopf map:
\be {\cal H} : \overline{\bf R^3} \ra {\bf S}^2 : {\bf r} \ra {\bf n} =
    R(\th(r),\hat{\bf r}){\bf n}_o  ~, \label{ho}  \ee
where $R(\th,{\bf e})$ denotes a rotation around {\bf e} with angle $\th$,
$\hat{\bf r}$ is the unit vector in direction of {\bf r} and ${\bf n}_o$
is an arbitrary but fixed unit vector. For the mapping (\ref{ho}) to be
well defined, we must require the boundary conditions
\be \th(r=0) = 0~,~~~~ \th(r=\infty) = N2\pi~.  \label{bound} \ee

$N$ is the Hopf invariant, or the $\pi_3$ winding number of ${\cal H}$.
The $\si$--model equations  for {\bf n} are
\be \bf \Box n - (n\cd\Box n)n = 0   .\ee
With the ansatz (\ref{ho}) above, this gives rise to the following two
equations  for $\th$:
\bea
 \Box\th -{2\sin\th\over r^2} &=& 0    \\
 -(\dd_t\th)^2 + (\dd_r\th)^2 -{2(1-\cos\th)\over r^2} &=& 0 ~, \eea
which have no common non--trivial solutions.

\section{Gravitational Effects of Textures in Flat Space}

As long as $\ep = 16\pi G\eta^2$ is much smaller than 1, the gravitational
field of a texture is weak and we can calculate it in first order
perturbation theory. We apply the gauge invariant formalism developed
in Chapter~2.

The energy momentum tensor,
\be T_{\mu\nu} = \phi,_\mu\phi,_\nu -1/2g_{\mu\nu}\phi,_\la\phi^{,\la}~,\ee
of solution (\ref{chiy}) is readily calculated with the result
\bea T_{00} &=& {2\eta^2\over r^2}{1+3y^2\over (1+y^2)^2} \label{Tt00} \\
     T_{0i} &=& -{4\eta^2\over r^2}{y\over (1+y^2)^2}\hat{\bf r}_i
       \label{Tt0i}\\
     T_{ij} &=& {2\eta^2\over r^2}{1-y^2\over (1+y^2)^2}\de_{ij}.
       \label{Ttij}  \eea

Clearly (\ref{chiy}) represents a solution with infinite action
and infinite energy,
\[  E(R)  = 4\pi\int_0^R T_{00}r^2dr  \propto R~, ~~~\mbox{ for } R>> t-t_c~,
\]
diverges linearly.

Since this texture solution is spherically symmetric, it only gives rise
to scalar perturbations.
We now set $M^2 = 4\eta^2$ and $l=t_c$, the  'radius' of the texture
(for an expanding universe, $t_c$ is the horizon scale at texture collapse).
We then find
\bea f_{\rho}   &=& {l^2\over 2r^2}{1+3y^2\over (1+y^2)^2}  \label{3fr}\\
     f_v        &=& {l\over 2r}\frac{y}{1+y^2}  \\
     f_p        &=& {l^2\over 2r^2}\frac{1- y^2}{(1^2+y^2)^2}   \\
     f_{\pi}    &=&  0 ~.  \label{3fpi}\eea
  From (\ref{3fr}) to (\ref{3fpi}) and the equations for the gauge invariant
Bardeen potentials of scalar seed perturbations (\ref{2S1}, \ref{2S2},
\ref{2S3}), we obtain in flat space ($\dot{a} =0~,~~ a=1$)
\bean -\lap\Phi_s &=& {\ep\over 2}{1+3y^2\over r^2(1+y^2)^2} \\
   \dot{\Phi}_s &=& {-\ep\over 2}{y\over r(1+y^2)}  \\
  \lap(\Phi_s+\Psi_s) &=& 0 ~,\eean
with the solution
\bea \Phi_s  &=& -{\ep\over 4}\ln((1+y^2)r^2/t_c^2)  \label{3Phik}\\
     \Psi_s &=&  \frac{\ep}{ 4}\ln({1+y^2\over y^2})  ~. \eea
$\Psi_s$ is only determined up to a function of time, which we have chosen
to ensure $\Psi_s \ra 0$, for $t\ra \pm\infty$.
\be \Psi_s  = {1\over 4}\ep\ln(\frac{r^2+(t-t_c)^2}{(t-t_c)^2}) \; .
\label{3Psik} \ee
Of course, physical observables do not depend on this choice.

It might seem unphysical that the potentials $\Phi_s$ and $\Psi_s$ do not
vanish at infinity $r\ra \infty$, in contrary, they diverge. This reflects
again the infinite energy of our solution. Noting this divergence, one might
fear that linear perturbation theory breaks down at large distances from the
texture, but we find that the relevant
geometrical quantities, like e.g. the
 3--dimensional Riemann scalar on the surfaces of constant time,
\be \de^3R = 4a^{-2}\lap{\cal R} \approx 4a^{-2}\lap\Phi = -\frac{2\ep}{a^2}
     \frac{r^2+3(t-t_c)^2}{(r^2+(t-t_c)^2)^2}  \; , \ee
do vanish at infinity. So that, far away
  from the collapsing texture and at early and late times, the solution does
  approach flat space.
The validity of linear perturbation theory for textures is also confirmed in
Durrer et al. [1991], where we find an exact solution for a texture coupled to
gravity and show that, for $\ep<0.1$ say, it deviates only very little from
the flat space solution used here.

We now calculate the behavior of baryons and collisionless
particles (dark matter or photons) and photons in this geometry.

\vspace{1cm}

\subsection{  Baryons around a collapsing texture }

\noindent
Let us briefly discuss the behavior of cosmic dust (baryons) in the field of
a texture. Equation (\ref{2DD}) for  a flat dust universe
($c_s^2 = w = 0\;\; , \;\; \dot{a} =0$) yields
\be \frac{d^2D}{dy^2} + 4\pi G\bar{\rho}t_c^2D = S \; , \label{3D} \ee
with
\[ S = 2\ep\frac{1}{(1+y^2)^2}   \;\; .\]
The term $4\pi G\bar{\rho} t_c^2D$ is the coupling of the perturbation to
its own gravitational field. It leads to exponential growth
of perturbations which is a feature of the
non--expanding universe only. But our approximation, neglecting expansion,
means that all times involved are much smaller than Hubble time. This
coincides with $4\pi G\bar{\rho}t_c^2 \ll 1$. Within our approximation, it
is thus consistent to neglect the self gravitating term in
(\ref{3D}). Direct integration then yields the solution
\bea D &=& \ep[y\arctg(y) + c_1y + c_2]  \nonumber \\
       &=&  \ep(t/r)[\arctg(t/r) +\pi/2]  + \ep
  \;\; , \eea
where we have chosen the integration constants $c_1$ and $c_2$ such that
$D$ converges to   $0$ for large negative times, $D(t= -\infty) = 0$,
and $D$ converges to a constant  for large radii, $D(t,r= \infty) =\ep$.
 Since
$D$ is only a function of the self similarity variable $t/r$, we cannot
consistently choose both boundary conditions to be 0.
For late times, $t/r \gg 1$, $D$ grows linearly with time:
\[  D = \ep\pi(t/r)   \; , \]
Near the time of collapse, $ |t/r| \ll 1$, $D$ is
of the order of
$\ep$,
$ D(t=0,r) = \ep$. A given time $t_*$ after texture  collapse
which is small compared to
the Hubble time, $D$ has the following profile:
For large radii $D \approx\ep$ and
roughly at $r=t_*$ bends into
 $D \approx \ep\pi t_*/r$ and diverges for $r\ra 0$.
This divergence leads to early formation of non--linear structure on
small scales. At time $t_*$ perturbations on scales of the order of
$r \le r_{nl} = \ep\pi t_*$ have become non--linear.

The total mass accumulated around a texture diverges like
the mass of the texture itself (see [4]). But in the real, expanding universe
one has to cut it of at roughly the Hubble radius at the time when
the texture collapses, $l_H$.

In this simple approximation,  we end up with the following picture:
Due to  textures forming at a time $t$ in the universe, objects of
 mass $M\approx 2\ep M_H(t)$, form at separations on the order of
 $p^{-1/3}l_H(t)$.
Where $M_H$ denotes the horizon mass at the time when the texture
collapses and $p$ is the probability that a four component  vector field
which is distributed in a completely uncorrelated manner
over a 2-sphere winds around a 3-sphere (i.e. the probability of
texture formation at the horizon). In  Gooding et al. [1991],
\AA minneborg [1992] this probability has been found numerically to be
about $1/25$.

  From matter conservation, (\ref{3con1}) for $w=0$ we obtain
\[ \lap V = \ep/r[\arctg(t/r) +\pi/2] - (\ep/2)\frac{t}{r^2 +t^2}   \]
and therefore
\be v_i = - \dd_iV = -{\ep\over 2}{r_i\over r}[\arctg(t/r) +\pi/2]
.\label{3vel}  \ee
The total change in a particle's velocity as the texture collapses is thus
independent of the particles distance from it and is given by
\be \De v_j = v_j(\infty) - v_j(-\infty) = -(\ep\pi/2) {r_j\over r}
\label{3deltav} ~. \ee
A result which was found by Turok and Spergel [1990] and Durrer [1990].

A numerical calculation for the distribution of texture in an expanding
Friedmann universe, where the growth of density perturbations is given
according to (\ref{2D}) with $w=c_s^2=0$, is presented in Gooding et
al. [1991].

\vspace{1cm}

\subsection{Collisionless particles in the gravitational field of a texture}

\noindent
Let us now calculate the perturbations in the distribution function
of collisionless particles induced by
a collapsing texture. Like for dust, we neglect self gravity.
We start with the gauge invariant perturbation equation for
 Liouville's equation in a Friedmann universe with
$k=0$ (\ref{3liou}):

\be q\dd_t{\cal F} +v^k\dd_k{\cal F} = \frac{d\bar{f}}{dv}
   [(q^2/v)v^k\dd_k\Psi_s - vv^k\dd_k\Phi_s] \equiv  {\cal S}
    \; , \label{4L} \ee
where $\Psi_s$ and $\Phi_s$ are the metric perturbations due to the
texture.

Making use of spherical symmetry and
inserting the results (\ref{3Psik}) and (\ref{3Phik}) for $\Psi$ and
$\Phi$ yields
\be q\dd_t{\cal F} + v\mu\dd_r{\cal F} + {v(1-\mu^2)\over r}\dd_\mu{\cal F}
     = {\cal S} \label{fli} \ee
with
\[ {\cal S} = (\ep/2)\frac{d\bar{f}}{dv}\cd (q^2 +v^2)\frac{\mu}{r(1+y^2)}
  \; . \]

The  solution of (\ref{fli}) is easily found with the help of the
physical coordinates
\be \tau = t-(q/v)\mu r ~~,~~~~ b=r\sqrt{1-\mu^2}~~ \mbox{ and } ~~ t  ~. \ee
It is straight forward to see that $\tau$ is just the time of
closest encounter of
the given particle with the texture, the impact time, and $b$ is the
impact parameter.
In these coordinates eq. (\ref{fli}) reduces to
\[ q\dd_t{\cal F}(t,\tau,b) = {\cal S}(t,\tau,b)  ~~\mbox{ and thus} \]
\be {\cal F}(t,\tau,b) = {1\over q}\int_{t_o}^t{\cal S}(t,\tau,b)dt \ee

A short calculation gives
\bea {1\over q}\int_{t_1}^{t^2}{\cal S}dt &=&
 {\ep\over 2}\frac{d\bar{f}}{dv}\left\{ {v\over 2}
 \ln[q^2(b^2+t^2) +v^2(t-\tau)^2] - \right. \nonumber \\ &&
 \left.\left. {\tau qv\over\sqrt{v^2\tau^2 +(q^2+v^2)b^2}}\arctan\left(
 {q^2t +v^2(t-\tau)\over q\sqrt{v^2\tau^2 +(q^2+v^2)b^2}}\right)
 \right\}\right|_1^2
  ~.\label{fS} \eea
We are interested in the total change of the distribution function due to
the collapsing texture, i.e. in the limit $t_1\ra -\infty$ and $t_2\ra\infty$.
The logarithmic term in eq. (\ref{fS}), let us call it $L$, diverges in
this limit and the difference,
\[\lim_{\left(\begin{array}{l}t_1\ra -\infty\\t_2\ra\infty
  \end{array}\right)}[L(t_2)-L(t_1)]~,   \]
 crucially depends on how we perform the limit. It can take any value
  from $-\infty$ to $\infty$.
This is due to the fact that we are dealing with an infinite energy solution,
and we certainly have to 'renormalize' our results.
If we would change the energy
momentum tensor  in a way that the texture would be 'born' some time in
the finite past, or if we would compensate it in a consistent way, as we
do it in the expanding universe, this problem would disappear.
A physically intuitive procedure is to introduce a cutoff at some time
$t_2=-t_1\gg b,|\tau|$. With such a cutoff the logarithmic term cancels,
and we obtain
\[ {\cal F}(t,\tau,b,v) = -{\ep\over 2}\frac{d\bar{f}}{dv}
   {\tau qv\over\sqrt{v^2\tau^2 +(q^2+v^2)b^2}}\left.\arctan\left(
 {q^2t +v^2(t-\tau)\over q\sqrt{v^2\tau^2 +(q^2+v^2)b^2}}\right)
\right|_1^2~~.   \]
We then can remove the cutoff and obtain the change of the distribution
function long after the corresponding particles have passed the texture
\be  {\cal F}(t,\tau,b,v) = -{\ep\pi\over 2}\left(\frac{d\bar{f}}{dv}\right)
   {\tau qv\over\sqrt{v^2\tau^2 +(q^2+v^2)b^2}}~. \label{4col} \ee
Introducing our old variables
\bean
 r^2 &=&  (t-\tau)^2v^2/q^2 +b^2    \\
 \mu &=& (t-\tau){v\over qr} ~,  \eean
we find
\be {\cal F}(t,r,\mu,v) = -{\ep\pi\over 2}\left(\frac{d\bar{f}}{dv}\right)
   { q(vt-q\mu r)\over\sqrt{(vt-q\mu r)^2 +r^2(1-\mu^2)(q^2+v^2)}}~.
  \label{fF}  \ee
This result can be inserted in equations (\ref{2Dint}) to (\ref{2Pint})
 to obtain the induced perturbations of the energy momentum tensor.
Here we just perform  the non--relativistic and extremely relativistic limits.
 \vspace{10pt}\\
{\bf Non--relativistic limit:} \\
In the non--relativistic case we have $v\ll q$. The gravitational field of
the texture is strong only for $r\sim |t|$. We can therefore expand
(\ref{fF}) in the small quantity $ vt/(qr)$. The first order approximation
yields
\be {\cal F}(t,r,\mu,v) = (\ep\pi/2)\frac{d\bar{f}}{dv}\!q^2
 [\mu +(vt/qr)(\mu^2-1) + {\cal O}(vt/qr)^2
   ] \; , \label{fnr}  \ee
This leads to
\bea V &=& {2\pi\over \rho}\int{\cal F}v^3\mu dvd\mu = -{\ep\pi\over 2} \\
  D &=& {2\pi\over \rho}\int{\cal F}v^2q dvd\mu = {\ep\pi}t/r~, \eea
the well known late time ($t\gg r$) results of the preceding subsection.
 \vspace{10pt}\\
{\bf Extremely relativistic limit:} \\
In this case we have $v=q$ and therefore
\be {\cal F}(t,r,\mu,v) = -{\ep\pi\over 2}\frac{\bar{f}}{dv}
   { v( t-\mu r)\over\sqrt{(t-\mu r)^2 +2r^2(1-\mu^2)}}~.
  \ee
Integrating ${\cal F}v^3$ over $v$ and dividing by $\rho/4\pi$ yields the
 fractional perturbation of the brightness introduced in Chapter~2,
\be {\cal M}(t,r,\mu) = 2\ep\pi
   { t-\mu r\over\sqrt{(t-\mu r)^2 +2r^2(1-\mu^2)}}~.
  \label{fer}  \ee

Since $\mm$ describes the fluctuations in the energy density radiated in
a given direction $\mu$, the temperature fluctuation is given by
$\Delta T/T = \Delta\mm/4$ (see also the more explicit discussion of this point
below eqn (\ref{2M})). This coincides exactly with eq. (\ref{3SW}) for
photon redshift in the next paragraph.

In a flat, eternal universe the signal from a texture
collapsing at $t=0$, as seen
  from an observer at time $t_o$ and distance $r_o$ would thus be
\be {\Delta T\over T}(\th) = {\pi\ep \over 2}{t_o -r_o\cos\th\over \sqrt{
     (t_o-r_o\cos\th)^2 +2r_o^2\sin^2\th}} ~.  \label{fdT} \ee
A similar calculation using a specific gauge is presented in Durrer et
al. [1992a]. Unlike in the expanding universe  (see next chapter), there
is no horizon present in this calculations. Photons that pass the
texture long before or after collapse, $|t_o|\gg r_o$ are still influenced
by it and yield even a maximum temperature shift,
\[ {\Delta T\over T} = \pm {\pi \over 2} \ep ~.\]
In the expanding universe of finite age we expect, because of the
finite size of the event horizon, $\De T/T$ to achieve a maximum for
$r_o\approx t_o-t_c$ and to vanish for $t_o-t_c \ll r_o$.
The comparison of the flat space result (\ref{fdT}) and the effect of
a compensated texture in the expanding universe is shown in Fig.~8.
\vspace{1cm}

\subsection{  Redshift of photons in the texture metric}

\noindent
Let us first  determine the energy shift which a photon experiences
by passing a texture. Without loss of generality, we set $t_c=0$ in this
paragraph.  If we neglect the distinctive dipole  term and intrinsic
density perturbations, equation (\ref{2deltaE}) leads to
\be \left.{\de E\over E}\right|^f_i =  \int_i^f (\dot{\Phi} -
	\dot{\Psi})d\la + \Psi|_i^f  \; . \label{3deltaE}  \ee
Denoting  the impact parameter of the photon trajectory by $b$  and
the time when the photon passes the texture (the impact time)
by $\tau$, we get $ r^2 = b^2 +  (t-\tau)^2$ . Eq. (\ref{3deltaE}) then
yields after the same ``renormalization procedure which led to (\ref{4col})
\be  \left.{\de E\over E}\right|^f_i = \frac{\ep \tau}{2(\tau^2 +2b^2)^{1/2}}
     [\arctg(\frac{2t-\tau}{\sqrt{\tau^2 +2b^2}})]_i^f  \; . \ee
For $t_f,\; -t_i \gg \tau,\;b$,  we obtain
\be \left.{\de E\over E}\right|^f_i \approx (\ep\pi/2)\frac{\tau}
{(\tau^2 +2b^2)^{1/2}}     \; .
    \label{3SW} \ee
This result was first found by different methods by Turok and Spergel [1990].
 Photons which pass the texture before
it collapses, $\tau < 0$, are redshifted, and photons passing it
after collapse, $\tau > 0$, are blueshifted. This produces a very distinctive
hot spot --- cold spot signal in the microwave sky wherever a texture
has collapsed.

Of course our result is not strictly correct in the expanding universe,
since we have neglected expansion in the calculation of $\Psi_s$ and $\Phi_s$.
 But  the
main contribution to the energy shift comes from times $|t| \le |\tau| + b$.
Therefore, our approximation is reasonable also for the expanding case, if
$|\tau|\le l_H$
 and $b \le l_H$, where $l_H$ denotes the horizon distance at the time of
collapse, $t=0$. On the other hand, by causality  the texture
cannot have a big effect on photon trajectories with $|\tau| > l_H$ or
$b>l_H$. A first approximation to the situation in the expanding
universe is thus
\be \left.\frac{\de E}{E}\right|^f_i = \left\{ \begin{array}{ll}
   {\ep\pi\over 2}\frac{\tau}{\sqrt{\tau^2+2b^2)}} & , \mbox{ for }
   |\tau| < l_H \mbox{ and } b < l_H   \\
    0  & , \mbox{ for }
   |\tau| > l_H \mbox{ or } b > l_H  \; . \end{array} \right.  \ee
\vspace{1cm}

\subsection{Light Deflection}

\noindent
To obtain the light deflection in the gravitational field of a spherically
symmetric texture, we consider (as above) a photon passing the texture
at impact time $\tau$ at a distance $b$ (impact parameter). The trajectory
of this photon is then given by $ x(\la)= (\tau+\la, \la\bm{n}+b\bm{e})$.
Making use of eq. (\ref{2ldefl}), neglecting spherical
abberation due to the relative motion of emitter and observer, we find
\bea \varphi &=& \int_i^f(\Phi-\Psi)_{,i}e^id\la  \nonumber \\
     &\approx& \ep\int_{-\infty}^{\infty}{b\over
  b^2+2\la^2+2\la\tau+\tau^2}d\la   \nonumber  \\
 &=& \ep\pi{b\over\sqrt{2b^2+\tau^2}}~.  \eea
For $\tau=0$, the deflection angle assumes the maximum value,
$\varphi_{\max} = \ep\pi/\sqrt{2}$. For $b=0$ or $\tau\ra\infty$, light
deflection vanishes. This result was first obtained by different
methods in \cite{DHJS}.

In contrast to global monopoles and strings, the deflection angle is not
independent of the impact parameter $b$ of the photon trajectory (except
for photons which pass the texture exactly at collapse time, $\tau =0$).
This leads to a qualitative difference to the lensing caused by
global monopoles and cosmic strings \cite{BV,Vi84}:
If a light source and an observer at distances $L$, respectively $D$
behind, respectively in front of the texture are perfectly aligned
with the texture, the observer sees an Einstein ring with a time dependent
opening angle
\be \beta(t) = \sqrt{2\left({\ep\pi L\over L+D}\right)^2-2\left({t\over D}-1
  \right)^2}~. \ee
This ring opens up at time $t = D-\ep\pi LD/(L+D)$, reaches a maximum
opening angle $\b_{\max} = \sqrt{2}\pi\ep L/(L+D)$ at $t=D$ and   shrinks
back to a point at $t = D+\ep\pi LD/(L+D)$. It exists over a
time span $\De t = 2\pi\ep{DL\over D+L}$. For realistic values of
$\ep \sim  10^{-4}$ the maximum opening angle can become about $1'$.

If source and observer are not perfectly aligned, but deviate by an angle
$\ga<\varphi_{\max}$ from alignment the ring reduces to two points which first
move apart and then together again within the time $\De t$ and with a
maximum separation angle $\b_{\max}$. Unfortunately, even if textures exist,
since late time textures are so rare, the probability of observing this
effect is rather small. There is typically one texture which has collapsed
after $z=4$ and for which the photons that have passed it close to the time
of collapse are visible for us now (see Durrer and Spergel [1991]).
The probability that
there is a quasar behind it with angular separation less than $\b_{\max}
 \sim 10^{-3}$ is even smaller than for global monopoles (see Section~3.4.1).

\chapter{Textures in Expanding Space}
We now want to discuss the effect of collapsing textures in a spatially
flat, expanding  Friedmann Universe. To be specific, we assume a
universe which is dominated by cold
dark matter (CDM) which contributes about 95\% of the total
matter density of the universe. The remaining 5\% are baryonic.
In the next section we write down the system of equations which describes
a spherically symmetric  collapsing texture in this background and the
perturbations it induces in the dark matter, baryons and photons. We have
solved this equations numerically. We then indicate how the single texture
events can be distributed in space and time to obtain a full sky map of
the cosmic microwave background. We  report briefly on the results of
this numerically obtained microwave sky and compare it with the COBE data.
(A more extensive presentation of this numerical work, which was done
in collaboration
with A. Howard and Z.--H. Zhou will be presented elsewhere \cite{DSH}.)

Numerical simulations and analytical estimates have
already shown that the texture scenario leads to early formation of small
objects which are likely to reionize the universe as early
as $z \approx 100$ (see Sect.~4 and Durrer [1990], Spergel et al. [1991]
and Gooding et al. [1991]).
Photons and baryons then are coupled again via Compton scattering of
electrons (see Section 3.2.3).
They remain so until the electrons are too diluted to scatter effectively.
This decoupling time is determined by the distance into the past, at
which the optical depth becomes unity:
\[ \tau(t_{dec}) = \int_{t_0}^{t_{dec}}d\! t a\si_Tn_e = 1 ~. \]
In a flat universe, $\Om =1$ one obtains
$ 1+z_{dec} \approx 100\left({0.05\over \Om_b h_{50}}\right)^{-2/3}$ ,
where $\Om_b$ denotes the density parameter of  baryons, and $h_{50}$
is Hubble's constant in units of $50km/s/Mpc$. This last scattering
surface is of course not as instantaneous as the recombination shell. The
decoupling due to dilution is a rather gradual process, and the thickness
of the last scattering surface is approximately equal to the horizon scale
at $z_{dec}\approx t_{dec}$.

Due to reionization, perturbations smaller than $t_{dec}$ but larger than
the mean free path of the photons are exponentially damped by
photon diffusion. A rough estimate of this effect is given in Section~3.2.3.
Results of a numerical calculation of this damping  are shown
in Fig.~7.

\section{Formalism}

In this section, we  present the equations for calculating the
response of matter and radiation to the collapse of a single spherically
symmetric texture.
 The background metric is given by
\[ ds^2 = a^2(-dt^2 +d\bm{x}^2)  \; .\]

The spherically symmetric ansatz for an ${\bf S}^3$ texture unwinding at
a given time $t=t_c$ is, like in the preceding chapter
\be \bm{\phi} = \eta^2(\sin\chi\sin\th\cos\varphi,\sin\chi\sin\th\sin\varphi,
                       \sin\chi\cos\th, \cos\chi)  \label{2phi} \; , \ee
where $\th \; , \; \varphi$ are the usual polar angles, and $\chi(r,t)$ has
the  properties
\[ \begin{array}{ll}
     \chi(r=0,t<t_c) & = 0  \\
     \chi(r=0,t>t_c) & = \pi  \\
     \chi(r=\infty,t) & = \pi \; .
  \end{array}  \]
In expanding space, the evolution equation (\ref{fchi}) is replaced by
\be \dd_t^2\chi + 2(\dot{a}/a)\dd_t\chi - \dd_r^2\chi -{2\over r}
    \dd_r\chi  = -{\sin 2\chi \over r^2} \; . \label{2chi} \ee
We parametrize the energy momentum tensor of the spherically symmetric
texture field,
\be T_{\mu\nu} = \dd_{\mu}\bm{\phi}\cd\dd_{\nu}\bm{\phi} -
  {1\over 2}g_{\mu\nu}\dd_{\la}\bm{\phi}\cd\dd^{\la}\bm{\phi}
	\label{5Tphi} ~,  \ee
  in terms of  scalar seed perturbations
\[ T_{00} = {\ep \over 4\pi}f_{\rho}/l^2 \; , \;\;\;
   T_{0i} = -{\ep \over 4 \pi} f_{v,i}/l \; , \]
\[ T_{ij} = {\ep \over 4 \pi}[ (f_p/l^2 - {1\over 3}\lap f_{\pi})\de_{ij}
                + f_{\pi},_{ij}]  \; , \]
where $\ep = 16 \pi G\eta^2$. Inserting ansatz (\ref{2phi}) in (\ref{5Tphi}),
one finds
\be f_{\rho}/l^2 = 1/8[(\dd_t\chi)^2 +(\dd_r\chi)^2 + {2\sin^2\chi\over r^2}]
       \label{2fro} \; , \ee
\be f_p/l^2 = 1/8[(\dd_t\chi)^2 -{1\over 3}(\dd_r\chi)^2 -
              {2\sin^2\chi\over 3r^2}]  \label{2fp} \; , \ee
\be f_v/l  = {1\over 4}\int_r^\infty
(\dd_t\chi)(\dd_r\chi)dr \label{2fv} \; , \ee
\be \lap f_{\pi} = {1 \over 4}[(\dd_r\chi)^2 -{\sin^2\chi \over r^2}] +
                   {3\over 4}\int_{\infty}^r[(\dd_r\chi)^2 -
                  {\sin^2\chi \over r^2}]{dr \over r}  \label{2fpi} \; . \ee

The variables
$f_{\rho}/l^2$ and $f_p/l^2$ denote the energy density and isotropic pressure
of the texture field,  $f_v$ is the potential of the velocity field
and $f_\pi$ is the potential for  anisotropic stresses.
 Spherically symmetric perturbations are of
course always of scalar type. Due to the adoption of spherical symmetry,
we loose, e.g., all information about gravitational waves produced during the
collapse.

In addition to the texture, we want to describe dark matter, baryons
and radiation. Since dark matter has zero
pressure, we just need a variable describing its density perturbation.

By $\rho_d$ we denote  the density of  dark matter and $D$ is
its gauge invariant  density perturbation, as defined in Chapter~2.
The evolution of the dark matter fluctuations is governed by
(\ref{2D})  for $c_s^2=w=0$:
\be \ddot{D} + (\dot{a}/a)\dot{D} - 4\pi G\rho_da^2D =
       \ep(f_{\rho} +3f_p)/l^2 \; , \label{5D} \ee

To describe the baryon-- photon system we need two additional variables:
The potential for the baryon velocity $V$ and the perturbation of the
energy integrated photon distribution function $\cal M$.
At times $t\le t_{dec}$, the collisionless Boltzmann equation for the
photons and the equation of motion for the baryons must be modified to
take into account scattering. The dominant effect is non--relativistic
Compton scattering by free electrons.
Let us denote the collision integral which is calculated in Section~3.2.1
by $C(\mm,V)$. As in Chapter~3, \bm{\ep} is the
direction  of the photon momentum. One finds (\ref{BC})
\be C  = a\si_Tn_e[D_g^{(r)} -\mm + 4\ep_i\dd_iV +
	{1\over 2} \ep_{ij}M^{ij}]~, \ee
with
  \[ M^{ij} =
  {3\over 2\pi}\int\mm(\ep)\ep_{ij}d\Om]~;~~ \ep_{ij}\equiv
   \ep_i\ep_j -(1/3)\de_{ij} \]
(We do not have to worry about the position of spatial indices of perturbation
 variables, they are raised and lowered with the Euclidean metric
 $\de_{ij}$ since $k=0$ in this chapter.)

The drag force due to Thomson drag of photons on the matter is given
by (\ref{3drag})
\[ F_i = -{\rho_r\over 4\pi}\int C\ep_id\Om =
   {a\si_Tn_e\rho_r\over 3}(M_i +4l\dd_iV) \; , \]
with $ M_i = (3/4\pi)\int\ep_i{\cal M}d\Om$. Including this drag force
into the equation of motion for the baryons, (\ref{2con2}) for $w=c_s^2=0$,
one obtains (\ref{3bar})
\[ l\dd_i\dot{V} + (\dot{a}/a)l\dd_iV = \dd_i\Psi -
     {a\si_Tn_e\rho_r\over 3\rho_b}(4l\dd_iV + M_i) \;\; \mbox{ or} \]
\be l\lap\dot{V} +(\dot{a}/a)l\lap V = \lap\Psi + a\si_Tn_e
   {\rho_r\over\rho_b}(\dot{M} - {4\over 3}l\lap V) \; . \label{V} \ee
For the last equation we have used the zeroth moment of
Boltzmann's equation, the continuity equation,
\[ \dot{M} +(1/3)\dd_iM^i = 0 \; . \]
In our numerical
computations, we have made the simplifying assumption $n_e = 0$ for $z>z_i$
and  $n_e = n_B$ for $z<z_i$ for some ionization redshift $z_i\approx 200$.

The  evolution of the photons
 is given by the perturbation of Boltzmann's equation,
(\ref{ABs}).
\bea \lefteqn{\dot{\cal M} +  \mu\dd_r\mm + {1-\mu^2\over r}\dd_\mu\mm =}
  \nonumber  \\ &&
   4\mu\dd_r(\Phi-\Psi) +
    a\si_Tn_e[M-\mm -4\mu\dd_rV + 3(\mu^2-1/3)M_2]  \\
    && \mbox{ where }~~ M_2(r,t) = {1\over 2}\int_{-1}^1
  \mm (\mu')(\mu'^2-1/3)d\mu' \; , \label{4Boltz} \eea
where $\mu$ is the direction cosine of the photon momentum in the direction of
{\bf r}.

As in Chapter~4, the photon evolution equation is more
transparent in  characteristic coordinates, $(t,\tau,b)$, where
$b = r\sqrt{1-\mu^2}$ is
 the impact parameter, and $\tau=t-r\mu$ is the impact time.
In these variables (\ref{4Boltz}) simplifies:
\be \dd_t\mm(t,\tau,b) =  4\mu\dd_r(\Phi -\Psi) +C(t,\tau,b)~, \ee
where $C(t,\tau,b)$ is the collision integral above, expressed in terms of
the new variables.

In order to write down the perturbed Einstein equations, we have in principle
to calculate the energy momentum tensor of radiation from \mm. But
since we are only interested in late times where density perturbations
can grow, $\rho_d>\rho_r$,
we may neglect the contribution of radiation to the density
perturbation. The potential $\Phi$ is then determined by the
texture and dark matter perturbations alone:
\be \lap\Phi = -\ep(f_{\rho}/l^2+ 3(\dot{a}/a)f_v/l)  -4\pi Ga^2\rho_dD\;
 . \label{2Phi0}\ee

On the other hand, since dark matter does not give rise to
anisotropic stresses, we have to take into account the contribution
of radiation to the latter.
To calculate the anisotropic  stresses of the photons we recall
the definition of the amplitude of anisotropic stresses, $\Pi$:
\[ \de T_i^j - {1 \over 3}\de T_l^l\de_i^j = p[\Pi_{,i}^{,j} -
     {1\over 3}\lap\Pi\de_i^j] ~.  \]
The anisotropic contributions to the energy momentum perturbations of
the photons are given by
\[ \de T_i^j - {1 \over 3}\de T_l^l\de_i^j = {\rho_r\over 4\pi}
  \int(\ep_i\ep^j-{1\over 3}\de_i^j)\mm d\Om ~.\]
Using these equations and spherical symmetry, one finds
\be \Pi''-\Pi'/r = {9\over 4}\int_{-1}^1(\mu^2-{1\over 3})\mm d\mu
  = {9\over 2}M_2(r,t)  ~.  \label{2Pi}  \ee
This anisotropy and the anisotropy of the texture field contribute to
the sum of the two Bardeen potentials

\be  \lap(\Psi +\Phi) = -2\lap(\ep f_\pi +(4/3)\pi Ga^2\rho_r\Pi)
   ~. \label{5Psi}   \ee

Let us  choose initial conditions that are physically
plausible.  If the phase transition that produced texture occurred
in an initially uniform universe,  causality requires that there
are no geometry fluctuations well outside the horizon \cite{VS}.
This implies that initially, at $t_i\ll t_{c}$, we must require
$\Psi=\Phi = 0$.
We want to compensate as much as possible of the initial texture
fluctuations with an initial dark matter perturbation.
Hence, we use as initial conditions for the density field,
\be D(r,t=t_i) = -{\ep \over 4\pi Ga^2\rho_d} (f_{\rho}/l^2+ 3(\dot{a}/a)f_v/l)
     ~.   \label{2D0}\ee
This initial condition implies that metric fluctuations are induced
by the differences between the texture equation of state, and
the equation of state of the background matter.
This choice yields $\Phi = 0$ at
$t=t_i$, but  not  $\Psi = 0$.  Due to its equation of state the dark
matter cannot compensate anisotropic stresses of the texture. We
thus must compensate them by an initial photon perturbation.
For $\Psi$ to vanish,  we have to require according to (\ref{5Psi})
\[ \Pi = -{3\ep\over 4\pi Ga^2\rho_r}f_\pi~.\]
 This does not lead to a unique initial condition for \mm, but
if we, in addition, require the zeroth and first moments of \mm~ to
vanish (which otherwise would interfere with eq. (\ref{2D0}))
it is reasonable to set
\be \mm(r,\mu,t=t_i) = -{15\ep\over 8\pi Ga^2\rho_r}(\mu^2-{1\over 3})
   (\lap f_\pi -{3\over r}f'_\pi)
  ~. \label{2mm0} \ee
Together with initial conditions for the texture field $\chi(r,t=t_i)$,
the requirements (\ref{2D0}) and (\ref{2mm0}) and the evolution
equations (\ref{2chi}) to (\ref{5Psi}) determine the system which
we have solved numerically.

Figure 9 shows the microwave background fluctuations induced by
the collapse of the texture as a function of $\tau$ for small impact
parameter in the expanding universe for different times.
At $t>t_c$ it is interesting to see the blueshift at $\tau\sim t$ of
the photons which have fallen into the dark matter potential but have
not yet climbed out of it again. (During their way out of the dark matter
potential this blueshift will of course be exactly compensated. This is
also visible in the figure.) Figure~10
shows the microwave background fluctuations induced by texture collapse
as a function of impact parameter.  Note that the temperature fluctuations
are induced only for photons that pass within the event horizon of
the texture.

\section{Textures and the Microwave Sky}

In the previous section, we described how the collapse of a single
texture produces fluctuations in the photon temperature.  In this section,
we sum the contributions of many textures and describe how to
 construct microwave maps of the night sky.

Since COBE observations cover the entire celestial sphere, we construct
a numerical grid consisting of 1 square degree patches.  These patches
are arranged so that they cover equal areas and each patch has
roughly the same shape.

Since the texture fluctuations are in the linear regime, we assume
that the contributions of each texture to the fluctuations at
each point in the sky can be added independently.  We randomly throw
down textures everywhere within the event horizon using the
texture density distribution  \cite{STPR},
\be{dn \over dt} = {1 \over 25} {1 \over t^4}. \ee
We then follow the collapse of each texture and sum their contributions.
We include only textures that collapse after recombination.
Textures that collapse earlier do not contribute significantly to
microwave fluctuations on scales accessible to COBE.

In order to simulate the COBE observations, the
map of the night sky  is smoothed with a Gaussian beam with a
FWHM of 7$^\circ$,
the angular resolution of the DMR detector on COBE \cite{Sm}.
After computing and removing the intrinsic dipole contribution
($10^{-5} - 10^{-4}$) and any monopole fluctuations, we then
compute the microwave quadrupole, the r.m.s. pixel-pixel fluctuations
and other statistics of the microwave sky.

\section{Results}
\def\e0{{\epsilon_0}}
Here we present our  results for the not reionized universe
  ($z_i<z_{dec}$). We just compare the numerical results with the COBE
observations on $\th\ge 10^o$. In the texture scenario, these large
angular scales are not affected by reionization.

The amplitude of the microwave background fluctuations depends
upon the scale of symmetry breaking associated with the texture.
If the symmetry breaking scale is  normalized so that the scenario can
reproduce the amplitude of the galaxy-galaxy correlation function,
we have to set the
value of the dimensionless parameter $\ep = 16\pi G\eta^2$ to
$\ep \approx 5.7 \times 10^{-4} b^{-1}$,
where $b$, the bias factor, is the ratio of the mass-mass correlation
function to the galaxy-galaxy correlation function \cite{GST}.
Note that the definition of $\ep$ in this work is
$2/\pi$ times the definition  used in Gooding et al. [1991].
Comparison of the predictions of the texture model with  observations
of clusters  \cite{Bart}
suggest that $b \approx 2$, a value compatible
with hydrodynamical simulations of texture-seeded galaxy formation
\cite{COST}.  We normalize the microwave background fluctuations
to this value and present our results in units of $\epsilon_0$,
where $\epsilon_0 = \ep/2.8 \times 10^{-4}$.

We numerically performed 100 realizations of the model. For illustration,
a 'COBE map' produced from a typical simulation is shown in Fig.~11.
Averaging over all
realizations, we find an r.m.s. value for the quadrupole moment of
 \[Q = (1.4\pm 1.2)\times 10^{-5}\e0 ~.  \]
  Since only a handful of textures
are the source of most of the large scale fluctuations, the quadrupole
varies significantly from realization to realization.

To compare with the COBE result, we have smoothed our calculations over an
angular scale of $10$ degrees. The average pixel-to-pixel fluctuations
of the smoothed  simulations are
\[(\Delta T/T)_{rms}(10^o) =  (3.8\pm 2.6) \times 10^{-5}\e0\]
The distribution of temperature fluctuations are only mildly non--Gaussian,
the skewness of the distribution is $-4\pm 2$ and
the kurtosis of the distribution is $32\pm 29$. The errors quoted
are  statistical $1\si$ deviations of one hundred realizations.
One example for the pixel distribution of the fluctuations is shown in Fig.~12.

The  results from the COBE differential microwave background
radiometers [DMR]  place strong constraints on CMB fluctuations on scales
larger than $10^o$.
 In this experiment \\  \cite{Co1,Co2} a value of
$Q \approx (0.6\pm 0.2) \times 10^{-5} $ has been found
 for the microwave quadrupole and
\[(\Delta T/T)_{rms}(\th=10^o) =(1.1\pm 0.18)\times 10^{-5}\]
is the result for the fluctuations at a scale of $10$ degrees.
The overall spectrum is compatible with a Harrison--Zel'dovich spectrum:
\[(\De T/T)(\th) \approx 10^{-5}(\th/10^o)^{-(0.1\pm 0.5)/2}~.\]

Our simulations with bias factor $b=2$ are compatible with these results
within one sigma. The $(\De T/T)$ of the simulations is somewhat large
but the scatter is considerable.

We feel that  the adoption of spherical symmetry may lead to underestimates
 of $(\De T/T)$ since contributions due to gravitational waves
and random fluctuations in the scalar field which do not give rise to
texture (i.e., topological winding number) have been neglected in this
approach. If, e.g., $(\De T/T)(\th=10^o)$ is enhanced by a factor of 2 by
these contributions, the COBE measurement is about $2.5\si$ below
the average texture result.

The conclusions from these simulations can thus be put as follows: The
CMB anisotropies from spherically symmetric texture collapse are slightly
high but agree within $1\si$ with the COBE measurements. This is
encouraging. But in the results given above only statistical fluctuations,
(i.e., cosmic variance) have been taken into account. Due to
the uncertainties in modeling the typical texture and the approximations
inherent in modeling the texture as spherically symmetric, these
estimates of the microwave background fluctuations are uncertain
by at least a factor of $\sim 2$ (systematic error),
probably underestimating the true induced
fluctuations. A full $3d$ simulation,  which takes into account
also gravitational waves and non--topological fluctuations of the scalar
field, is necessary to finally decide on the scenario. Such simulations
have now been performed \cite{BR92,PST}. They obtain results which are
higher than those obtained in the spherically symmetric approach by a factor
1.5 to 2.

Smaller scales, $\th\approx 1^o$ lead to somewhat larger fluctuations
and much smaller standard deviations (since many textures
contribute to them):
\[ (\Delta T/T)_{rms}(2^o) =  (4\pm 0.8) \times 10^{-5}\e0\]
If the new measurements which require $(\De T/T)(1^o)< 1.4\times 10^{-5}$
\cite{GSGKSML} are confirmed, we need reionization to damp small scale
fluctuations in the texture scenario.

A  rough estimate of the effects of reionization can be obtained by
just smoothing each texture with a smoothing scale
of about the horizon size at $z_{dec}$,
\[ z_{dec} = 100\left({0.05\over h_{50}\Om_B}\right)^{2/3} ~. \]
This corresponds to an angular scale of
\be  \th_{smooth} = t(z_{dec})/t_0 = (1+z)^{-1/2} \approx 5.7^o ~.\ee
 If the formation of objects leads to reionization prior to $z_{dec}$,
 this would suppress microwave background fluctuations
on  scales $\th\le \th_{dec}\sim 6^o$, but would
not effect the fluctuations on larger scales which are discussed above.

\section{Conclusions}
The texture scenario of large scale structure formation has many
attractive features. Its galaxy galaxy correlation function and the
large scale velocity fields agree better with observations than in the
standard cold dark matter model \cite{GPSTG}. Also a couple of other
statistical
parameters (Mach number, skewness, kurtosis) are in satisfactory
agreement with observations. New simulations which used the COBE quadrupole to
normalize the fluctuations \cite{PST} hint that, in contrary to earlier
results, the scenario may have severe difficulties to reproduce the
very large scale galaxy clustering observed in the infrared \cite{FDSYH},
 just like the standard CDM model.
Due to the steep dark matter potential produced with texture, small
scale structure can form very early. This may lead to early reionization
of the universe. Calculations of the microwave background anisotropies
show that reionization is necessary to reconcile with CMB anisotropies
on small scales (up to about $5^o$). Due to the unique signature
and relatively large amplitude, the CMB anisotropies produced by
textures are one of the most hopeful criterion for confirming or ruling
out this scenario.

Another interesting observational test, lensing of background
quasars by a foreground texture, is very improbable.

This scenario certainly deserves further work.
Especially the investigation of the question if reionization as early
as $z\sim 200 - 100$ is possible and a careful analysis of CMB
anisotropies for the reionized model on angular scales around $1^o$
are important tasks.

\vspace{2cm}

\noindent {\Large\bf Acknowledgement}\\
I thank Dave Spergel, Neil Turok, Norbert Straumann, Slava Mukhanov, Zhi--Hong
Zhou and Markus Meuwly for many stimulating discussions. I'm also  much
indebted to Norbert Straumann, Daniel Wyler and Othmar
Brodbeck for reading parts of the manuscript.\\
This research was supported in parts by the Swiss National Science Foundation.

\newpage

\appendix
\setcounter{equation}{0}
\renewcommand{\theequation}{A\arabic{equation}}

\chapter{The $3+1$ Formalism}
Since we have made use of $3+1$ split of general relativity in deriving
the linear cosmological perturbation equations in this review, we want
to derive the necessary tools in this appendix. A mathematically
rigorous overview is given in Choquet--Bruhat and York [1980] and
Fischer and Marsden [1979]. But  explicit
calculations are missing there, we shall thus be rather detailed.
In this appendix we mainly follow the exposition in Durrer and
Straumann [1988].

\section {Generalities}
We assume that spacetime $({\cal M},g)$ admits a slicing by slices
$\Si_t$,  i.e., there is a diffeomorphism $\phi :{\cal M} \: \ra \:
\Si \times I$, $I\subset {\bf R}$, such that the manifolds $\Si_t =
\phi^{-1}(\Si \times \{ t\})$ are spacelike and the curves
$\phi^{-1}(\{ x\}\times  I)$ are timelike. These curves are what we
call {\em preferred} timelike curves. They define a vector field
$\dd_t$, which can be decomposed into normal and parallel components
relative to the slicing (Figure~13):
\be \dd_t = \al n + \bm{\beta}   \; .  \label{2ddt} \ee
Here $n$ is a unit normal field and \bm{\beta} is tangent to the slices
$\Si_t$. The function $\al$ is the {\em lapse function} and \bm{\beta} is
the {\em shift vector} field.

A coordinate system $\{x^i\}$ on $\Si$ induces natural coordinates on
$\cal M$: $\phi^{-1}(m,t)$ has coordinates $(t,x^i)$ if $m\in \Si$
has coordinates $x^i$. The preferred timelike curves have constant
spatial coordinates. Let us set $\bm{\beta} = \beta^i\dd_i$
 ($\dd_i =\frac{\dd}{\dd x^i}$). From $g(n,\dd_i)=0$ and (\ref{2ddt})
we find
\[ g(\dd_t,\dd_t) = -\al^2+\beta^i\beta_i \:\: ,\: \:
                      g(\dd_t,\dd_i)=\beta_i \: .\]
 In "comoving coordinates" thus
\be g = -(\al^2-\beta^i\beta_i)dt^2 + 2\beta_idx^idt + g_{ij}dx^idx^j
       \label{2metric1} \ee
or
\be g = -\al^2dt^2 + g_{ij}(dx^i + \beta^idt)(dx^j + \beta^jdt)
 \; .  \label{2metric2} \ee
This shows that the forms $dt$ and $dx^i + \beta^idt$ are orthogonal.

The tangent and cotangent spaces of $\cal M$ have two natural
decompositions. One is defined by the slicing
\be T_p({\cal M}) = H_p \oplus V_p \: \: , \label{2TM1}  \ee
where the "horizontal" space $H_p$ consists of the vectors tangent to
the slice through $p$ and the "vertical" sub--space is the 1-dimensional
space spanned by $(\dd_t)_p$ (preferred direction). The dual
decomposition of (\ref{2TM1}) is
\be  T^*_p({\cal M}) = H^*_p \oplus V^*_p \: \: , \label{2T*M1}  \ee
with $H^*_p = \{ \om\in T_p^*({\cal M}) : \langle \om,\dd_t\rangle =0 \}$
and $V^*_p = \{ \om\in T_p^*({\cal M}) : \langle \om,H_p\rangle =0 \}$,
which is spanned by $(dt)_p$.

The metric defines - through the normal field $n$ - yet another
decomposition
\be T_p({\cal M}) = H_p \oplus H^{\perp}_p \: \: , \label{2TM2}  \ee
where $H^{\perp}_p$ is spanned by $n$, and dually
\be T^*_p({\cal M}) = (V^*_p)^{\perp} \oplus V^*_p \: \: , \label{2T*M2}  \ee
Equation (\ref{2ddt}) reflects the fact that in general the two directions
$V_p$ and $H^{\perp}_p$ do not agree. Dually this implies that $H^*_p$ and
$(V_p^*)^{\perp}$ do not coincide. We have for $\om^{\perp} \in
(V_p^*)^{\perp}$ the following decomposition relative to (\ref{2T*M1})
\be \om^{\perp} = \mbox{hor}(\om^{\perp}) +
\langle\om^{\perp},\bm{\beta}\rangle dt  \; . \label{2omperp}  \ee
The decompositions (\ref{2TM1}) to (\ref{2T*M2}) induce two types of
decompositions of arbitrary tensor fields on $\cal M$. We call a tensor
field {\em horizontal} if it vanishes, whenever at least one
argument is $\dd_t$ or $dt$. Relative to a comoving coordinate system
such a tensor has the form
\[\bm{ S} = S^{i_1\cdots i_r}_{j_1\cdots j_s} \dd_{i_1}\otimes\cdots\otimes
        \dd_{i_r}\otimes dx^{j_1}\otimes\cdots\otimes dx^{j_s} \: .\]
This shows that horizontal tensor fields can naturally be identified
with families of tensor fields on $\Si_t$, or with time-dependent
tensor fields on $\Si$ ("absolute" space). We  denote them with
boldface letters (except $\dd_i$ and $dx^i$).

As an often occurring example of a decomposition, we consider a
horizontal p-form \bm{\om} and its exterior derivative $d\bm{\om}$. We
have
\[ d\bm{\om} = \bm{d\om} + dt\wedge\dd_t\bm{\om}  \: \: , \]
where \bm{d\om} is again horizontal. In comoving coordinates \bm{d}
involves only the $dx^i$ ($\bm{ d} =dx^i\wedge\dd_i$) and $\dd_t\bm{\om}$
is the partial time derivative. \bm{d\om} and $\dd_t\bm{\om}$ are
horizontal and can be interpreted as $t$-dependent forms on $\Si$. In
this interpretation \bm{d\om} is just the exterior derivative of
\bm{\om}. Similarly, other differential operators (covariant derivative,
Lie derivative, etc) can be decomposed.
We use two types of bases of vector fields and 1-forms which are adapted
to (\ref{2TM1}) and (\ref{2T*M1}), respectively (\ref{2TM2}) and
(\ref{2T*M2}).
Obviously, the dual pair $\{\dd_{\mu}\}$ and $\{dx^{\mu}\}$ for comoving
coordinates $\{x^{\mu}\}$ are adapted to (\ref{2TM1}) and (\ref{2T*M1}).
On the other hand, equations (\ref{2ddt}) and (\ref{2metric2}) show that
the dual pair
\be  \{ \dd_i,n\} \:\: \mbox{ and } \:\:  \{dx^i+\beta^idt,\al dt\}
       \label{2pair}   \ee
is adapted to (\ref{2TM2}) and (\ref{2T*M2}).

Instead of $\{\dd_i\}$ we  also use an orthonormal horizontal basis
$\{ {\bf e}_i\}$ ($g({\bf e}_i,{\bf e}_j) = \delta_{ij}$), together with
the dual basis $\{ \bm{\vth}^i\}$ instead of $\{dx^i\}$. Then we have the
following two dual pairs, which are constantly used:
\bea \{ {\bf e}_i,\dd_t\} \: \: ,\: \:& \{ \bm{\vth}^i,dt\} \;\;&
     \mbox{ (adapted to slicing),} \label{2pairsl}  \\
    \{ {\bf e}_i,e_0=n\}    & \{\th^{\mu}\}  & \mbox{(adapted to
(\ref{2TM2}) and (\ref{2T*M2})) ,}  \label{2pairorth}   \eea
where the orthonormal tetrad $\{\th^{\mu}\}$ is given by
\be  \th^0 = \al dt \:\: , \:\: \th^i = \bm{\vth}^i +\beta^idt \:\: ,
     \label{2thmu}  \ee
with $\beta^i$ here defined by $\bm{\beta} = \beta^i{\bf e}_i$ .
We note also the relation
\[ e_0 = n = \frac{1}{\al}(\dd_t -\beta^i{\bf e}_i) \;\; .  \]
\vspace{14pt}

\section{ The connection and curvature forms}
 We now  calculate the connection and curvature
forms in the orthonormal basis introduced above.

\subsection{ The connection forms }
  From the first structure equation ,
\[ d\th^{\mu} + \om^{\mu}_{\: \nu}\wedge \th^{\nu} = 0 \; , \]
and the definition of the second fundamental form:
\be  K_{ij} = -n_{i;j} \; , \label{A2ff} \ee
where $n$ denotes the normal field of the slicing, one
finds immediately the Gauss' formulas : (remember $n=e_0$)
\be \om^i_{\: k}({\bf e}_j) = \bm{\om}^i_{\: k}({\bf e}_j) \label{Aomijk} \ee
\be \om^0_{\: i}({\bf e}_j) = -K_{ij} \; . \label{Aom0ij} \ee
We define
\be \dd_{t}\bm{\vth^i} = c^i_{\:j}\bm{\vth^j} \; .  \ee
Then we can calculate the following quantities :
\be \om^0_{\: i}( e_0) = \alpha^{-1} \alpha_{|i} \; ,\label{Aom0i0} \ee
\be \om^i_{\: j}( e_0) = - \alpha^{-1} \bm{\om}^i_{\: j}(\bm{\beta})
    + \frac{1}{2\alpha} (\beta^i_{|j} - \beta^j_{|i} - c^i_{\:j}
    + c^j_{\:i})  \; , \label{Aomij0} \ee
\be K_{ij} =  \frac{1}{2\alpha} (\beta_{i|j} + \beta_{j|i} - c^i_{\:j}
    - c^j_{\:i}) \; ,  \label{Akij} \ee
and thus,
\be \bm{K} = \frac{1}{2\alpha}[L_{\beta}\bm{g}-\dd_{t}\bm{g}]
               \; , \label{AKij} \ee
where the vertical bar $_|$ denotes covariant derivation with respect
to $\bm{g}$ .
Using the general relation
\[ \dd_t(\det\bm{g}) = \tr(\dd_t\bm{g})\det\bm{g} \; , \]
we find from (\ref{AKij})
\be \dd_t\mbox{vol}(\bm{g})=(\bm{\mbox{div}\beta}-\al\mbox{tr}\bm{K})
     \mbox{vol}(\bm{g}) \; . \label{Avol}  \ee

Let's derive (\ref{Aom0i0}), (\ref{Aomij0}) and (\ref{Akij}) briefly .
\[ d\th^0 = d(\alpha dt) = \bm{d} \alpha \wedge dt =
   \alpha_{|i} \bm{\vth}^i \wedge dt =
   \alpha^{-1} \alpha_{|i}\th^i \wedge \th^0 \; . \]
This together with the first structure equation results in (\ref{Aom0i0}).
(\ref{Aomij0}) and (\ref{Akij}) are obtained as follows: From the first
structure equation and (\ref{Aom0ij}) we conclude
\[ \begin{array}{lll}
  i_{{\bf e}_l}i_{ e_0}d\th^i & = & - i_{{\bf e}_l}i_{ e_0}
 ( \om^i_{\: 0} \wedge \th^0 +  \om^i_{\: j} \wedge \th^j) \\
           & = & - (K_{il} +  \om^i_{\: l}( e_0))   \; .
   \end{array} \]
We calculate the left hand side of this equation:
\bean
i_{{\bf e}_j}i_{e_0}d\th^i & = & i_{{\bf e}_j}i_{e_0}d(\bm{\vth}^i +
                              \beta^i dt ) \\
                        & = & i_{{\bf e}_j}i_{e_0}(\bm{d}\bm{\vth}^i +
dt \wedge \dd_{t} \bm{\th}^i + \bm{d} \beta^i \wedge dt ) \\
    & = & \alpha^{-1} i_{{\bf e}_j}(i_{\bm{\beta}}(\bm{\om}^i_{\: l} \wedge
        \bm{\vth}^l) + \dd_{t}\bm{\vth}^i - \bm{d} \beta^i ) \\
    & = & \alpha^{-1}[\om^i_{\: j}(\bm{\beta}) - \om^i_{\: k}({\bf e}_j)
	\beta^k
        - \bm{d}\beta^i({\bf e}_j) + \dd_{t}\bm{\vth}^i({\bf e}_j)] \\
     & = & \alpha^{-1}[\om^i_{\: j}(\bm{\beta})- \beta^i_{\: |j}+ c^i_{\: j}]
  \; .  \eean
The symmetric and antisymmetric contribution of the last identity
yield the formulas (\ref{Akij}) and (\ref{Aomij0}) for $K_{ij}$ and
$ \om^i_{\: j}(e_0)$, respectively.

\subsection{The curvature forms }
We now want to calculate the $3+1$ split of $R_{ij}$, $R_{0j}$ and $G_{00}$.
  From the second structure equation,
\[ \Om^{\mu}_{\: \nu} = d\om^{\mu}_{\: \nu}
                 + \om^{\mu}_{\: \la}\wedge \om^{\la}_{\:\nu} \; , \]
and equations (\ref{Aomijk}) to (\ref{Akij}) one finds immediately
\bea
 \Om^i_{\: j}({\bf e}_k , {\bf e}_l) & = & \bm{\Om}^i_{\: j}({\bf e}_k , {\bf
e}_l) +
                               K^i_{\: k}K_{jl} - K^i_{\: l}K_{jk}
 \mbox{\hspace{1cm} (Gauss)} \label{AGa} \\
 \Om^0_{\: j}({\bf e}_k , {\bf e}_l) & = & K_{jk|l} - K_{jl|k}
\mbox{\hspace{3cm} (Mainardi)} \label{AMa}  \: .
\eea
We need also the normal components of $\Om^0_{\: j}$ . By the second
structure equation we know
\[\Om^i_{\: 0} = -d(K_{ij}\th^j)  + d(\alpha^{-1} \alpha_{|i}\th^0) +
\om^i_{\:l}\wedge (K_{lj}\th^j + \alpha^{-1}\alpha_{|l}\th^0) \; .\]
A straight forward calculation leads to
\[ \Om^i_{\: 0}  =  \alpha^{-1} \alpha_{|ij}(\th^j \wedge \th^0) -
   dK^i_{\:j}\wedge \th^j + K^i_{\:j}(\om^j_{\:l} \wedge \th^l
   - K^j_{\:l}\th^l \wedge \th^0)
                     + \om^i_{\:j}K^j_{\:l}\wedge \th^l \; , \]
which yields (\ref{AMa}) and the normal components of $\Om^i_{\:0}$ :
\be
 \Om^i_{\: 0}({\bf e}_j,e_0) =\alpha^{-1}\alpha^{|i}_{\:|j}+dK^i_{\:j}(e_0)
  - K^i_s\om^s_{\:j}(e_0) + K^{2i}_{\;j}-\om^i_{\:s}(e_0)K^s_{\:j}
 \label{AOm0ij0} \; . \ee
  From equations (\ref{AGa}) to (\ref{AOm0ij0}) we can calculate the Ricci
tensor with the result:
\[R_{\beta \si} = \Om^{\alpha}_{\: \beta}(e_{\alpha},e_{\si}) \; , \]
\be R_{00} = \frac{1}{\al}\bm{\triangle}\al + \alpha^{-1}(\dd_{t}\tr(K) -
              L_{\bm{\beta}}\tr(K_{ij})) + \tr\bm{K}^2 \; . \label{AR00}  \ee
With help of (\ref{AMa}) one finds
\be R_{0i} = (trK)_{|i} -K_{i\: \: |j}^{\:j} \; .\label{AR0i} \ee
For the spatial components we obtain
\[R_{ij}=\Om^0_{\: i}(e_0,{\bf e}_j)+\Om^k_{\:i}({\bf e}_k,{\bf e}_j)\; . \]
Using (\ref{AOm0ij0}) and (\ref{AGa}) for the curvature forms leads to
\be
R_{ij}  =  \bm{R}_{ij} + \tr(K)K_{ij} - 2K^2_{ij} - \al^{-1} \al_{|ij}
           -\al^{-1}(\dd_tK_{ij} - L_{\bm{\beta}}K_{ij})
           + K_{is}\om^s_{\:j}(e_0)
           +K_{js}\om^s_{\:i}(e_0)  \; .          \label{Arij} \ee
With help of (\ref{Aomij0}), (\ref{Akij}) and (\ref{AKij})
one can bring (\ref{Arij}) into the form
\be
 \mbox{hor}(Ricci(g))  =  \bm{Ricci}(\bm{g}) + \tr(\bm{K})\bm{K} - 2\bm{K}^2 -
                \al^{-1}(\dd_{t}\bm{K}-L_{\bm{\beta}}\bm{K})
             - \al^{-1} \bm{Hess}(\al)  \; .             \label{ARij}
\ee
Using (\ref{AR00}) and (\ref{Arij})  we find
\be \begin{array}{lll}
    G_{00}  & =& 1/2(R_{00} + \sum_{i}R_{ii} )   \\
            & =& 1/2[\bm{R} + (\tr(K))^2 - \tr(K^2) ]  \; .
\end{array}      \label{AG00} \ee

\section{The $3+1$ split of hydrodynamics}
Calculations similar to those in the last section lead quite rapidly to
a $3+1$ split of hydrodynamics.

Let us decompose the energy-momentum tensor into horizontal and vertical
components:
\be T = \ep e_0\otimes e_0 + e_0\otimes\bm{S} + \bm{S}\otimes e_0 + \bm{T}
  \; .  \label{5T}  \ee
For an ideal fluid with
\be T=(\rho+p)u\otimes u+pg^{\#} \label{5idfluid} \ee
we find, setting as in special relativity $u=\gamma(e_0+\bm{v})$,
$\gamma =(1-\bm{v^2})^{-1/2}$,
\be \ep =\gamma^2(\rho+p\bm{v^2})   \; , \label{5ep}   \ee
\be \bm{S} = (\rho+p)\gamma^2\bm{v}   \; , \label{5S}   \ee
\be  \bm{T}= (\rho+p)\gamma^2\bm{v}\otimes\bm{v}+p\bm{g^{\#}}
     \; . \label{5T3} \ee
Now we compute $\nabla \cdot T$ for an arbitrary $T$. From
\[ \nabla_{e_0}(\ep e_0\otimes e_0)=L_{\ts e_0}(\ep) e_0\otimes e_0
  +\ep\om^i_{\;0}(e_0)\bm{e}_i\otimes e_0 +
    \ep e_0\otimes\om^i_{\;0}(e_0)\bm{e}_i   \]     and
\[ \nabla_{\bm{e}_k}(\ep e_0\otimes e_0)=L_{\ts \bm{e}_k}(\ep)
    e_0\otimes e_0 +\ep\om^i_{\;0}(\bm{e}_k)\bm{e}_i\otimes e_0 +
    \ep e_0\otimes\om^i_{\;0}(\bm{e}_k)\bm{e}_i   \]
we obtain
\[ \nabla\cdot(\ep e_0\otimes e_0)=L_{\ts e_0}(\ep) e_0
       +\ep\om^i_{\;0}(e_0)\bm{e}_i+\ep\om^i_{\;0}(\bm{e}_i)e_0 \;\;.  \]
In the same manner one finds the other contributions with the result:
 \[ (\nabla\cdot T)^0=L_{\ts e_0}(\ep) +\ep\om^i_{\;0}(\bm{e}_i)
     +\om^0_{\;j}(e_0)S^j+S^k_{\;|k}+\om^0_{\;j}(e_0)S^j+
      \om^0_{\;j}(\bm{e}_i)T^{ij}   \]
Inserting the expressions for the connection forms given in Section~A1,
leads to the following form of the energy equation:
\be  \frac{1}{\al}(\dd_t-\bm{L_{\ts \beta}})\ep =
     -\bm{\nabla\cdot S}-2\bm{\nabla}(\ln\al)\cdot\bm{ S} +
       \ep\tr(\bm{ K}) +\tr(\bm{K\cdot T}) \;\;. \label{5ener}  \ee
Similarely one finds
 \[ (\nabla\cdot T)^i=\om^i_{\;0}(e_0)\ep + L_{\ts e_0}(S^i) +
     [\om^i_{\;0}(\bm{e}_j)+\om^i_{\;j}(e_0)]S^j +\om^j_{\;0}(\bm{e}_j)S^i
      +\om^0_{\;j}(e_0)T^{ji} +T^{ij}_{\;\;|j}   \]
and from this we obtain the momentum conservation
\be  \frac{1}{\al}(\dd_t-\bm{L_{\ts \beta}})\bm{S} =
       -\bm{ \nabla}(\ln\al)\ep + 2\bm{K\cdot S}+\tr(\bm{K})\bm{S}
       -\al^{-1}\bm{\nabla\cdot}(\al\bm{T}) \;\;. \label{5mome}  \ee
This equation is used to derive the vector perturbation equation
(\ref{2dom}) in Chapter~2.

\section{The $3+1$ split of Einsteins field equations}
Here we discuss
the often used $3+1$ split of the gravitational field equations. The
calculation of the curvature forms relative to the basis (\ref{2thmu})
is presented in Section~A2. The reader will note that Cartan's calculus
leads rather quickly to the required results.

We use the notation introduced in the previous section (\ref{5T}) for the
various projections of the energy-momentum tensor $T$ into normal and
horizontal components.
  From equations (\ref{AR0i}), (\ref{ARij}) and (\ref{AG00})
for the Einstein and Ricci
tensors, Einsteins field equations can be written in the form (recall
that boldface letters always refer to the slices $\Si_t$):
\be \bm{ R}+(\tr\bm{ K})^2-\tr\bm{ K}^2 = 16\pi G\ep \; , \label{6G00} \ee
\be \bm{\nabla\cdot K}-\bm{\nabla\cdot}\tr(\bm{ K}) = 8\pi G\bm{S} \; ,
  \label{6Gi0}  \ee
\be \dd_t{\bf K}=\bm{L_{\ts \beta}K}-\mbox{\bf Hess}(\al) +
   \al[\mbox{\bf Ric}(\bm{g}) -2\bm{K\cdot K}+(\tr\bm{K})\bm{K}-8\pi G(
	\bm{T}-{1\over 2}
	\bm{g}(\ep-\tr\bm{T}))]\; . \label{6Gij}  \ee
In addition to (\ref{6G00}), (\ref{6Gi0}) and (\ref{6Gij}) we have the
following relation (Section~A2, equation (\ref{AKij})) between $\bm{g}$
and the second fundamental form $\bf K$:
\be \dd_t\bm{g}=-2\al\bm{K}+\bm{L_{\ts \beta}g} \; . \label{6gij} \ee
Note that this decomposition into constraint equations
(\ref{6G00}), (\ref{6Gi0}) and dynamical equations
(\ref{6Gij}), (\ref{6gij}) involves only horizontal quantities and those
provides the $3+1$ split of the gravitational field equations.

In Chapter~2 we  also use the following consequence of (\ref{6Gij}) and
(\ref{6G00})
\be \dd_t\tr(\bm{ K})=-\bm{\triangle}\al +\bm{L_{\ts \beta}}\tr(\bm{K})
    + \al [\tr(\bm{K}^2)+1/2(\ep+\tr\bm{T})]\; . \label{6dttrK}  \ee
Note that $\dd_t$ and tr do not commute. With (\ref{6gij}) one shows
easily
\be \tr(\dd_t\bm{ K}-\bm{L_{\ts \beta}}\bm{K}) = \dd_t\tr(\bm{ K})
     -\bm{L_{\ts \beta}}\tr(\bm{K}) + 2\al \tr(\bm{K}^2) \; .
\label{6trGij}  \ee
To derive the perturbation equations for vector perturbations we
mainly use (\ref{6Gi0}) and (\ref{6Gij}).
\vspace{14pt}

\section{The $3+1$ split of the Liouville operator for a geodesic spray}
In this section we derive a useful form of the Liouville operator for a
geodesic spray for an arbitrary $3+1$ split.

We start with some generalities. The metric $g$ of the spacetime
manifold ${\cal M}$ defines a natural diffeomorphism between the tangent bundle
$T{\cal M}$ and the cotangent bundle $T^*{\cal M}$, which can be used to pull
back the
natural symplectic form on $T^*{\cal M}$. In terms of natural bundle
coordinates the diffeomorphism is given by $(x^{\mu},p^{\mu})\mapsto
(x^{\mu},p_{\mu}=g_{\mu\nu}p^{\nu})$ and those the induced symplectic
2-form on $T{\cal M}$ is
\be \om = dx^{\mu}\wedge d(g_{\mu\nu}p^{\nu}) \;. \label{4om} \ee
The Lagrangian $L=\frac{1}{2}g_{\mu\nu}p^{\mu}p^{\nu}$ on $T{\cal M}$ defines a
Hamiltonian vector field $X_g$ on $T{\cal M}$, determined by
\[ i_{\ts X_g}\om = dL \;.  \]
In terms of natural bundle coordinates the geodesic spray $X_g$ is given
by
\be X_g = p^{\mu}\dd_\mu -\Gamma^{\mu}_{\;\al\beta}p^{\al}p^{\beta}
	{\dd\over\dd p^\mu} \;,     \label{Xg}  \ee
where $\Gamma^{\mu}_{\;\al\beta}$ are the Christoffel symbols for
$({\cal M},g)$. (For further details see \cite{Ste}.)

The one-particle phase space for particles of mass $m$, i.e., the
sub--bundle $\{p\in T{\cal M}:g(p,p)=-m^2\}$, is invariant under the
geodesic flow
and  we denote the restriction of $X_g$ to the one-particle phase space
also by $X_g$.

Let $f$ be a distribution function on the one-particle phase space. The
Vlasov and Boltzmann equations for $f$ involve the Lie derivative
$L_{\ts X_g}f$. If we consider the spatial components
$p^i$, relative to an orthonormal tetrad $\{e^{\mu}\}$ as independent
variables of $f$, then the Liouville operator $L_{\ts X_g}$ can
be written as
\be L_{\ts X_g}f= p^{\mu}e_{\mu}(f)
  -\om^i_{\;\al}(p)p^{\al}\frac{\dd f}{\dd p^i} \;, \label{4Liou1}\ee
where $\om^{\mu}_{\;\nu}$ are the connection forms relative to the dual
basis $\{\th^{\mu}\}$.

We derive now a more explicit expression of (\ref{4Liou1}) for an
arbitrary $3+1$ slicing. In order to do this, we need the connection
forms relative to the basis $\{\th^{\mu}\}$ calculated in Section~A2 .
They can be expressed in
terms of $\al,\bm{\beta},\bm{\om}^i_{\;j},c^i_{\;j}$.
Using equations (\ref{Aom0i0}), (\ref{Aom0ij}), (\ref{Aomij0}) and
(\ref{Aomijk}) we find
($\bm{p}=p^i\bm{e}_i$, $p^0=\sqrt{\bm{p}^2+m^2}$):
\[\begin{array}{lll}
\om^i_{\;\al}(p)p^{\al}\frac{\dd}{\dd p^i} &=&
\om^i_{\;0}(p)p^0\frac{\dd}{\dd p^i}+\om^i_{\;j}(p)p^j\frac{\dd}{\dd p^i} \\
 &=&[\om^i_{\;0}(e^0)p^0+\om^i_{\;0}(\bm{p})]p^0\frac{\dd}{\dd p^i} +
    [\om^i_{\;j}(e^0)p^0+\om^i_{\;j}(\bm{p})]p^j\frac{\dd}{\dd p^i}  \\
&=& (p^0)^2\al^{-1}\al^{|i}\frac{\dd}{\dd p^i} -
   K^i_{\;j}p^0p^j\frac{\dd}{\dd p^i}+
   \bm{\om}^i_{\;j}(\bm{p})p^j\frac{\dd}{\dd p^i} +
   \om^i_{\;j}(e^0)p^0p^j\frac{\dd}{\dd p^i} \\
&=& (p^0)^2\al^{-1}\al^{|i}\frac{\dd}{\dd p^i} +
    \bm{\om}^i_{\;j}(\bm{p}-\al^{-1}\bm{\beta}p^0)p^j\frac{\dd}{\dd p^i}-
    \frac{p^0}{\al}(\beta_j^{\;|i}-c_j^{\;i})p^j\frac{\dd}{\dd p^i}
\; .  \end{array} \]
Here $K^i_{\;j}$ are the components of the second fundamental form of
$\Si_t$, for which we also use equation (\ref{Akij}) of Section~A2.

This leads to the following useful $3+1$ split of the Liouville
operator:
\be L_{\ts X_g}f = [\frac{p^0}{\al}\dd_t +
 \bm{L}_{\bm{p}-\frac{p^0}{\al}\bm{\beta}}]f - [\bm{\om}^i_{\;j}
   (\bm{p}-\frac{p^0}{\al}\bm{\beta})p^j+(p^0)^2(\ln\al)^{|i} -
     p^0H^i_{\;j}p^j]\frac{\dd f}{\dd p^i} \;\; , \label{4Liou2} \ee
where we have introduced the horizontal tensor field
\be H^i_{\;j}=\al^{-1}(\beta^i_{\;j}-c^i_{\;j}) \;\; . \label{4H} \ee
Equation (\ref{4Liou2}) is  used in Section 2.3.

\section{Glossary}

In this appendix we provide a glossary of the variables used in the text.
For most terms we give a short explanation and refer to the equation or
section where this variable is first used. Usually it is defined there.
If not, this should be a very common variable found, e.g., in most basic
text books on general relativity (like the Christoffel symbols, the
Riemann tensor and so on).
\begin{itemize}
\item[$A$]Perturbation of the $00$ component of the metric, respectively the
	lapse function (2.4),(2.11), Appendix~A.
\item[$B$] Scalar perturbation of the $0i$ component of the metric,
	respectively the shift vector (2.5),(2.11).
\item[$B_i$] Vector perturbation of the $0i$ component of the metric,
	respectively the shift vector (2.7),(2.12).
\item[$B_{ij}$]  Magnetic part of the Weyl tensor (2.27).
\item[$C_{\al\mu\beta\nu}$] $ = R_{\al\mu\beta\nu} -
	(1/2)(g_{\al\beta}R_{\mu\nu}-  +g_{\mu\nu}R_{\beta\al}
   -g_{\mu\beta}R_{\nu\al} -g_{\al\nu}R_{\mu\beta} )
    +{R\over 6}(g_{\al\beta}g_{\mu\nu}-g_{\al\nu}g_{\mu\beta})$,\\
	 Weyl tensor (2.26,27).
\item[$D^{(\al)}$] Gauge invariant density perturbation variable
	for the matter component $\al$ (2.38).
\item[$D_g^{(\al)}$] Gauge invariant density perturbation variable
	for the matter component $\al$ (2.37).
\item[$D_s^{(\al)}$] Gauge invariant density perturbation variable
	for the matter component $\al$ (2.36).
\item[$E_{ij}$] Electrical part of the Weyl tensor (2.26).
\item[$F$] Gauge dependent perturbation variable for the
	distribution function,  paragraph (2.3.1).
\item[${\cal F }^{(S)}$]  Gauge invariant perturbation variable for scalar
	perturbations of the distribution function (2.69).
\item[${\cal F }^{(T)}$]  Gauge invariant perturbation variable for tensor
	perturbations of the distribution function, paragraph (2.3.2).
\item[${\cal F }^{(V)}$]  Gauge invariant perturbation variable for vector
	perturbations of the distribution function, paragraph (2.3.2).
\item[$G$] Newtons constant, $G =6.6720\times 10^{-8}cm^3g^{-1}sec^{-2}$.
\item[$G_{\mu\nu}$] $= R_{\mu\nu} -(1/2)g_{\mu\nu}R$, Einstein tensor.
\item[$H_L$] Trace perturbation  of the spatial part of the metric.
\item[$H_T$] Anisotropic scalar perturbation  of the spatial part
	of the metric (2.6,11).
\item[$H_i$] Anisotropic vector perturbation  of the spatial part
	of the metric (2.8,12).
\item[$H_{ij}$] Anisotropic tensor perturbation  of the spatial part
	of the metric (2.9,13).
\item[$K_{ij}$] Second fundamental form, Section~2.1, Appendix~A.
\item[$L^i$] Spatial components of the vector field $X$ parametrizing
	a gauge transformation, Section~2.1, 2.3.
\item[$L_X$] Lie derivative w.r.t the vector field $X$, Section~2.3.
\item[$M$] Mass used to parametrize the energy momentum tensor of seed
	perturbations, Section~2.5.
\item[$\cal M$] Gauge invariant perturbation variable for the
	energy integrated photon distribution (2.83).
\item[$\cal M$] Spacetime manifold, Appendix~A, Section~2.3
\item[$M_i$] $ = {3\over 4\pi}\int d\Om\ep_i{\cal M}$ The first moment
	of $\cal M$ (3.19).
\item[$M_{ij}$]  $= {3\over 8\pi}\int d\Om\ep_{ij}{\cal M}$ The
	second moment of $\cal M$ (3.17).
\item[$P_\mu^{~\nu} = u_\mu u^\nu +\de_\mu^\nu$] The projection operator
onto the 3--space orthogonal to $u$ (2.30).
\item[$P_m$] The mass bundle, Section~2.3
\item[$R$] The Ricci scalar.
\item[$R= 3\rho_m/4\rho_r$] Parameter used in paragraph~3.2.3.
\item[$\cal R$] Perturbation of the scalar curvature on the slices
	of constant time (2.14).
\item[$R_{\mu\nu}=R_{~~\mu\al\nu}^\al$] The Ricci tensor.
\item[$R_{~~\mu\al\nu}^\beta$] The Riemann tensor
\item[$T$] Temperature of the cosmic background radiation.
\item[$T$] Temporal component of the vector field $X$ parametrizing a gauge
      transformation, Section~2.1, 2.3.
\item[$T{\cal M}$] Tangent space to spacetime, Section 2.3.
\item[$TX$] Tangent vector field associated to the vector
	field $X$, Section~2.3.1
\item[$T_{\mu\nu}^{(sS)}$] Scalar contribution to the energy
	momentum tensor of the seeds (2.117,118,119).
\item[$T_{\mu\nu}^{(sT)}$] Tensor contribution to the energy
	momentum tensor of the seeds (2.122).
\item[$T_{\mu\nu}^{(sV)}$] Vector contribution to the energy momentum
	tensor of the seeds (2.120,121).
\item[$V$] Gauge invariant variable for scalar perturbations of the
velocity field  (2.35).
\item[$V_i$] Gauge invariant variable for vector perturbations
	of the velocity field  (2.42).
\item[$X$] Vector field parametrizing a gauge transformation,
	 Section~2.1, 2.3.
\item[$\dd_\mu = {\dd\over \dd x^\mu}$] Partial derivative (vector field)
\item[$a$] Cosmic scale factor, Section~1.1.
\item[$b$] Impact parameter, Section~3.4, 4.2.2.
\item[$c$] Speed of light, usually set equal to 1 in this text.
\item[$c_s= \sqrt{\dot{p}/\dot{\rho}}$],
	($c_\al=\sqrt{\dot{p}_\al/\dot{\rho}_\al}$) Adiabatic
	sound speed (of matter component $\al$), Section~1.1.
\item[$e_\mu$] Tetrad vector field, Section~2.3, Appendix~A.
\item[$f$] Distribution function of phase space, Section 2.3.
\item[$f_\pi$] Gauge invariant scalar potential parametrizing
	anisotropic stresses of seeds (2.119).
\item[$f_\rho$] Gauge invariant perturbation variable parametrizing
	the energy density  of seeds (2.117).
\item[$f_p$] Gauge invariant perturbation variable parametrizing
	the pressure  of seeds (2.119).
\item[$f_v$] Gauge invariant perturbation variable parametrizing the scalar
	velocity potential  of seeds (2.118).
\item[$g_{\mu\nu}$] Metric of spacetime, Chapter~2.
\item[$h$] Used to parametrize Hubble's constant
	$H_0 =h\times 100{km\over sec Mpc}$, Section 1.1.
\item[$h_{\mu\nu}$] Metric perturbation (2.10).
\item[$\hbar$] Planck's constant,
	$\hbar =1.0546\times 10^{-27}cm^2 g sec^{-1}$,
	usually set equal to 1 in this text.
\item[$k$] Spatial curvature of a Friedmann universe (1.1).
\item[$k$] Comoving wave number, Section~3.2, paragraph~3.4.2.
\item[$k_B$] Boltzmann's constant, $k_B=1.3807\times 10^{-16} erg/K$,
	usually	set equal to 1 in this text.
\item[$l$] Length introduced to keep perturbation variables
	dimensionless, in
	applications it may be set equal to a typical scale of
	perturbations, Section~2.1.\\
\item[$l_H=t$] Comoving size of the horizon, Section~1.3.
\item[$q$] Redshift corrected energy, paragraph~2.3.1.
\item[$t$] Conformal time, Section 1.1.
\item[$t_T$] Conformal Thomson mean free path, Section~3.2.
\item[$u$] Energy velocity field, Section~2.1
\item[$v$] Scalar velocity potential Section~2.1.
\item[$v$] Redshift corrected momentum, paragraph~2.3.1.
\item[$v^i$] Vector peculiar velocity field, Section~2.1.
\item[$w$] Enthalpy, Section 1.1.
\item[$w_i^{(\pi)}$] Gauge invariant vector potential parametrizing
	anisotropic stresses of seeds (2.121).
\item[$w_i^{(v)}$] Gauge invariant vector contribution to the energy
	flow of	seeds (2.120).
\item[$z$]	Cosmological redshift.
\item[$\Ga$]	Gauge invariant entropy perturbation variable, Section~2.1.
\item[$\Ga^\mu_{\nu\la}$] Christoffel symbols, Section~2.3.
\item[$\De$] Laplacian.
\item[$\La$] Cosmological constant (1.1).
\item[$\Pi$]  Gauge invariant scalar potential for anisotropic stresses,
	Section~2.1.
\item[$\Pi_i$] Gauge invariant vector potential for anisotropic stresses,
	Section~2.1.
\item[$\Pi_{ij}$] Gauge invariant tensor contribution to anisotropic
	stresses, Section~2.1.
\item[$\Si$] Three dimensional spatial hypersurface, Appendix~A.
\item[$\Phi$] Gauge invariant scalar potential for geometry
	perturbations (2.24).
\item[$\Psi$] Gauge invariant scalar potential for geometry
	perturbations (2.25).
\item[$\Om^i$] Gauge invariant perturbation variable for the fluid
	vorticity (2.43).
\item[$\Om^{\mu}_{~\nu}$] Curvature 2--form, Appendix~A.
\item[$\al$] Lapse function (2.4), Appendix~A.
\item[$\beta$] Shift vector (2.5), Appendix~A.
\item[$\ga_{ij}$] Metric of a three space of constant curvature,
	Section~1.1.
\item[$\de$] Gauge dependent density perturbation (2.28).
\item[$\bm{\ep}^i$] Spatial unit vector (e.g. denoting photon directions),
	Section~2.3.
\item[$\ep = 4\pi GM^2$] Smallness parameter for the amplitude of seed
	perturbations, paragraph~2.5.2.
\item[$\ep_{ijk}$] Three dimensional totally antisymmetric tensor (2.27).
\item[$\ep_{ij}= \ep_i\ep_j-\ga_{ij}$], Section 3.2.
\item[$\eta$] Symmetry breaking scale (4.1).
\item[$\th^\mu$] Orthonormal tetrad of 1--forms, Appendix~A.
\item[$\vth^i$] Orthonormal triad of 1--forms on the hypersurfaces
	of constant time, Appendix~A.
\item[$\io$] Isomorphism between the perturbed and unperturbed mass bundles,
	Section~2.3.
\item[$\io$] Gauge dependent perturbation variable for the energy integrated
	photon distribution paragraph~(2.3.4).
\item[$\la$] Parameter in the scalar field potential (4.1).
\item[$\la=(1+kr^2/4)^{-1}$] Conformal factor for the metric of a
	3~space of constant curvature, Section~2.3.
\item[$\mu$] Cosine between the photon direction and the radial
	direction, paragraph~3.2.1.
\item[$\pi^\mu$] Orthonormal momentum components Section~2.3.
\item[$\pi^i_j$] Anisotropic stresses (2.31).
\item[$\pi_L$] Gauge dependent pressure perturbation variable (2.31).
\item[$\rho_{(\al)}$, $ \bar{\rho}_{(\al)}$] Background energy density
	of component $\al$.
\item[$\si$] Scalar potential for the shear of the equal time hypersurfaces,
	extrinsic curvature (2.15).
\item[$\si^i$] Vector potential for the shear of the equal time
	hypersurfaces,	extrinsic curvature (2.17).
\item[$\si_T$] Thomson cross section, $\si_T=6.6524\times 10^{-25}cm^2$.
\item[$\tau$] Physical time, Section 1.1.
\item[$\tau$] Impact time (4.30)
\item[$\tau$] Optical depth, Section~5.1.
\item[$\tau_{\mu\nu}$] Stress tensor (2.30).
\item[$\phi$] Scalar field, Chapter~4.
\item[$\chi$] Variable parametrizing spherically symmetric scalar field
	configurations (4.4).
\item[$\om$] Winding number density of the scalar field, Section~4.1.
\item[$\om^{~\mu}_\nu$] Connection forms, Appendix~1.
\item[$\om_{ij}$] Vorticity of a velocity field Section~2.1.

\end{itemize}

\newpage
\vspace*{2cm}
{\LARGE \bf FIGURE CAPTIONS}
\vspace{1cm}\\

{\bf Fig. 1:\  } The spectrum of the cosmic microwave background radiation
as measures by COBE  (Figure  from Mather et al. [1990]).
\vspace{1cm}\\
{\bf Fig. 2:\ } Limits on the CMB anisotropy on different angular scales.
 The COBE result is the only positive detection. All other marks represent
95\% confidence upper limits.
\vspace{1cm}\\
{\bf Fig. 3:\ } The linear perturbation spectrum of hot dark matter, for one
 (2) or three (1) types of massive neutrinos. The fluctuations are heavily
  damped on scales smaller than $\la_{FS}\approx m_{Pl}/m_\nu^2$ (Figure
  from Durrer [1989]).
\vspace{1cm}\\
{\bf Fig. 4:\ } Simulations of structure formation with HDM (top right
 and bottom pictures) compared with the corresponding picture from the
 CfA survey (top left). Triangles are high density regions identified as
 galaxies. One sees that the simulations lead to highly over developed
 large scale structure (Figure from White [1986]).
\vspace{1cm}\\
{\bf Fig. 5:\ } CDM simulations (a) and (b) compared with the CfA survey (c).
  No striking inconsistencies are visible at first sight (Figure from
  Kolb and Turner [1990]).
\vspace{1cm}\\
{\bf Fig. 6:\ } The angular galaxy galaxy correlation function
  as measured by the
  IRAS survey (black dots) compared with the predictions from CDM models
  with $h=0.4$ (black line) and $h=0.5$ (dotted line). The open circles
  and squares are results from an older analysis of the Lick catalogue
  (Figure from Maddox et al. [1990]).
\vspace{1cm}\\
{\bf Fig. 7}\hspace{.3cm} The CMB anisotropy (in units of $10^{-3})$
  from a spherically symmetric texture collapsing at $z=30$ (left) and
$z=200$ (right) respectively as
a function of angular separation from the center of the texture.  This
figure is calculated for a universe which reionizes at $z=200$. It
shows how signals from small scale textures are substantially damped
and  broadened  by photon diffusion.
\vspace{1cm}\\
{\bf Fig. 8:\ } The hot spot---cold spot signal of a spherically symmetric
  collapsing texture in units of $\ep \sim 2.8\times 10^{-4}$ .
  The horizontal variable $\tau = t-r\cos\th$ denotes the 'impact time'
  of a photon arriving at a distance $r$ from the texture at time $t$
traveling with an angle $\th$ with respect to the radial direction.
The hot spot--cold spot is shown for photons with fixed impact parameter
$b=r\sin\th\approx 0.1t_c$ ($t_c$ is the time of texture collapse).
The signal from the  expanding universe  at $t=t_c$, line (1), and
$t=1.5t_c$, line (2), is compared with the flat space result (dashed curve).
The second peak appearing at $t=1.5t_c$ is due to the dark matter potential.
\vspace{1cm}\\
{\bf Fig. 9}\hspace{.3cm}
As Fig. 8 but for different times with time steps $\De t \approx 0.25t_c$.
One sees an outgoing wake of blue shift at $\tau \approx t$. This is caused
by  photons which have fallen into the dark matter potential but have
not yet climbed out of it again. This blueshift will of course be
completely compensated by the redshift these photons will acquire during their
way out of the dark matter potential.
\vspace{1cm}\\
{\bf Fig. 10}\hspace{.3cm}
The CMB perturbation in units of $\ep \sim 2.8\times 10^{-4}$ as a
function of the impact parameter $b$ for fixed $\tau \sim 0.5t_c = 10$.
The signal disappears at an impact parameter $b\sim 1.5t_c  $
($t_c = 20$ in the units chosen).
\vspace{1cm}\\
{\bf Fig. 11}\hspace{.3cm}
A simulated COBE map as it might look in a scenario with texture + CDM.
The color scheme goes from $-4\times 10^{-4}$ (dark blue)  to
$1\times 10^{-4}$ (deep red).  Monopole and dipole
contributions are subtracted in this map.
A description of how the map is produced (in collaboration with D.N.~Spergel
and A. Howard) is given in the text.
\vspace{1cm}\\
{\bf Fig. 12}\hspace{.3cm}
The statistical distribution of microwave anisotropies in the texture
scenario. The number of pixels showing a given anisotropy are counted
for one realization of the CMB sky.
The distribution is slightly non--Gaussian with skewness $\approx -1$
and curtosis $\approx 3$.
\vspace{1cm}\\
{\bf Fig. 13}\hspace{.3cm}
	A $3+1$ slicing of spacetime $\cal M$. The family of immersions
	of $\Si$ into $\cal M$ is  denoted by  $i_t$;~~ $i_t(m) =
                      \phi^{-1}(m,t)$.

\end{document}